\pdfoutput=1

\documentclass[journal]{IEEEtran}
\usepackage{cite}

\ifCLASSINFOpdf
\else
\fi
\usepackage[cmex10]{amsmath,mathtools}
\usepackage{amsfonts,amssymb}

\usepackage{array}
\usepackage{amsfonts}

\usepackage{tikz}
\usetikzlibrary{shapes.geometric, arrows}

\tikzstyle{rect} = [rectangle, minimum width=1.6cm, minimum height=0.6cm, text centered, draw=black, fill=blue!3]
\tikzstyle{arrow} = [thick,->,>=stealth]

\usepackage{algorithm}
\usepackage{algcompatible} 
\usepackage{algpseudocode}

\usepackage{setspace}

\usepackage[latin1]{inputenc} 

\usepackage{graphicx}
\usepackage{epstopdf}
\usepackage[process=auto]{pstool}
\usepackage{psfrag}

\usepackage{wrapfig}

\usepackage{relsize}

\usepackage[implicit=false]{hyperref}

\usepackage{colortbl}

\newcommand{\algorithmfootnote}[2][\footnotesize]{%
  \let\old@algocf@finish\@algocf@finish
  \def\@algocf@finish{\old@algocf@finish
    \leavevmode\rlap{\begin{minipage}{\linewidth}
    #1#2
    \end{minipage}}%
  }%
}

\begin{document}

\bstctlcite{IEEEexample:BSTcontrol}

%
\title{Smartphone-based Vehicle Telematics \\--- A Ten-Year Anniversary}
%
%
%

\pagenumbering{gobble} 
\author{Johan Wahlstr\"{o}m,
      Isaac Skog,~\IEEEmembership{Member,~IEEE}, and Peter H\"{a}ndel,~\IEEEmembership{Senior Member,~IEEE}
      \thanks{J. Wahlstr\"{o}m, I. Skog, and P. H\"{a}ndel are with the ACCESS Linnaeus Center, Dept. of Signal Processing, KTH Royal Institute of Technology, Stockholm, Sweden (e-mail:  \{jwahlst, skog, ph\}@kth.se).}
}

%
%

\markboth{}%
{Shell \MakeLowercase{\textit{et al.}}: Bare Demo of IEEEtran.cls for Journals}
%



\maketitle

\vspace*{-10mm}
\begin{abstract}
Just like it has irrevocably reshaped social life, the fast growth of smartphone ownership is now beginning to revolutionize the driving experience and change how we think about automotive insurance, vehicle safety systems, and traffic research. This paper summarizes the first ten years of research in smartphone-based vehicle telematics, with a focus on user-friendly implementations and the challenges that arise due to the mobility of the smartphone. Notable academic and industrial projects are reviewed, and system aspects related to sensors, energy consumption, cloud computing, vehicular ad hoc networks, and human-machine interfaces are examined. Moreover, we highlight the differences between traditional and smartphone-based automotive navigation, and survey the state-of-the-art in smartphone-based transportation mode classification, driver classification, and road condition monitoring. Future advances are expected to be driven by improvements in sensor technology, evidence of the societal benefits of current implementations, and the establishment of industry standards for sensor fusion and driver assessment.


\end{abstract}


%
\IEEEpeerreviewmaketitle

\section{Introduction}

The \emph{internet of things} (IoT) refers to the concept of connecting everyday physical objects to the existing internet infrastructure. In a potential future scenario, the connected devices could range from household appliances and healthcare equipment, to vehicles and environmental probes \cite{Singh2014}. Two main benefits are expected to result from this increased connectivity. First, IoT enables services based on the monitoring of
devices and their surroundings.
Connected devices can build and send status reports, as well as take appropriate actions based on received commands. Second, IoT will provide massive amounts of data, which can be used for analysis of behavioral patterns, environmental conditions, and device performance.
In the year 2020, the number of connected devices is expected to reach 13 billion, which would be an increase of $350\,\%$ since 2015 \cite{Gartner2014}. According to predictions, the main catalyst for this tremendous growth in IoT will be the smartphone \cite{McLeod2013}.


Smartphones utilize software applications (apps) to unify the capabilities of standard computers with the mobility of cellular phones. Beginning with the birth of the modern smartphone about a decade ago (Nokia N95, first-generation iPhone, etc.), the smartphone has come to define social life in the current age, with almost one and a half billion units sold in 2015 \cite{Gartner2015}. The steadily increasing interest in utilizing the smartphone as a versatile measurement probe has several explanations. For example, smartphones have a large number of built-in sensors, and also offer efficient means of wireless data transfer and social interaction.
The use of smartphones for data collection falls under the general field of \emph{telematics} \cite{Xu2014}, referring to services where telecommunications are employed to transmit information provided by sensors in, e.g., vehicles (\emph{vehicle telematics}) or smart buildings \cite{Jose2015}.

\begin{figure}[t]
\vspace*{-6mm}
\def\svgwidth{3.8in}
\hspace*{-3.5mm}
\scalebox{0.94}{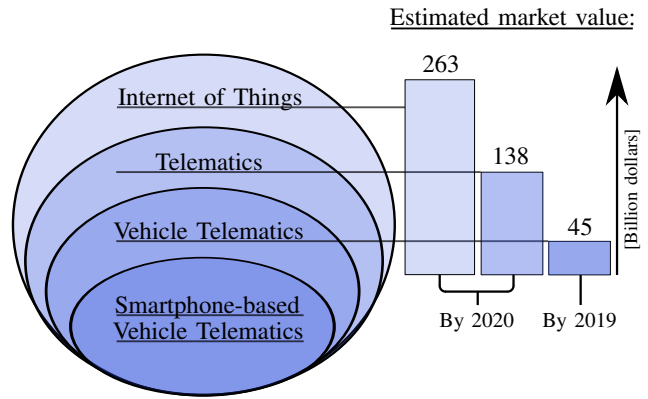}
\caption{\label{venn_figure}A Venn diagram illustrating the relation between the internet of things, telematics, vehicle telematics, and smartphone-based vehicle telematics. Predicted market values are $\$263$ billion (internet of things, by 2020) \cite{Gartner2014}, $\$138$ billion (telematics, by 2020) \cite{IndustryARC2014}, and $\$45$ billion (vehicle telematics, by 2019) \cite{mam2014}.}
\vspace*{-3.5mm}
\end{figure}

Thanks to the ever-growing worldwide smartphone penetration, the vehicle and navigation industry has gained new ways to collect data, which in turn has come to benefit drivers, vehicle owners, and society as a whole.
The immense number of newly undertaken projects, both in academia and in the industry, illustrate the huge potential of \emph{smartphone-based vehicle telematics}, and has laid a steady foundation for future mass market implementations \cite{Engelbrecht2015}. The relation between the IoT, telematics, vehicle telematics, and smartphone-based vehicle telematics is illustrated in Fig. \ref{venn_figure}.

\begin{figure*}
\centering
\hspace*{-2mm}
\begin{tikzpicture}[node distance=1.4cm]
\node (smph) [rect,align = center,xshift=-0.6cm,yshift=0cm] {\underline{Smartphone}\\[0.5ex]Sensing\\Computation};
\node (smph_figure) [align = center,above of=smph,xshift=0cm,yshift=0.45cm]
{\includegraphics[width=.05\textwidth]{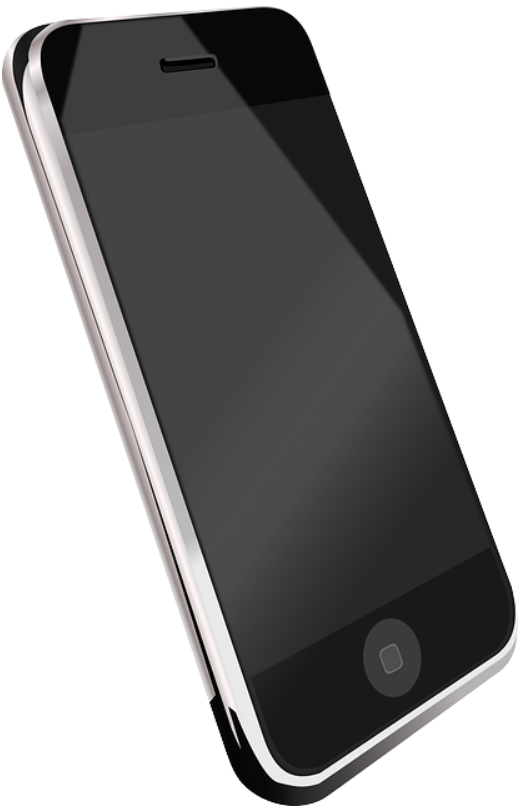}};
\node (title) [draw=none,align = center,above of=smph,xshift=3.5cm,yshift=1.8cm] {\underline{Smartphone-based Vehicle Telematics}};
\node (storage) [rect,align=center,right of=smph,xshift=4.3cm,yshift=0cm] {\underline{Central data storage}\\[0.5ex]Computation};
\node (storage_figure) [align = center,above of=storage,xshift=0cm,yshift=0.4cm]
{\includegraphics[width=.05\textwidth]{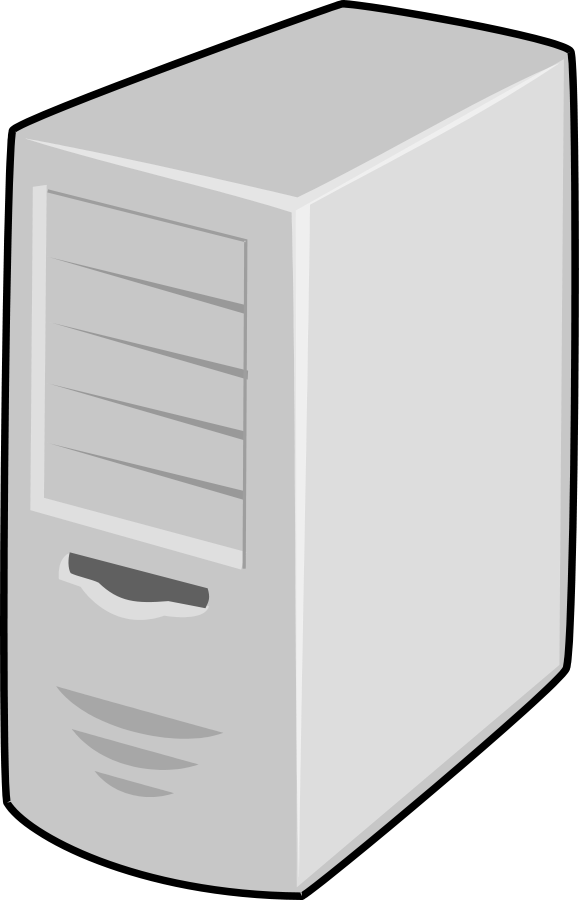}};
\node (comm) [align=center,right of=smph,xshift=1.2cm,yshift=0.45cm] {Wireless \\communication};
\node (comm_figure) [align = center,above of=comm,xshift=0cm,yshift=0cm]
{\includegraphics[width=.06\textwidth]{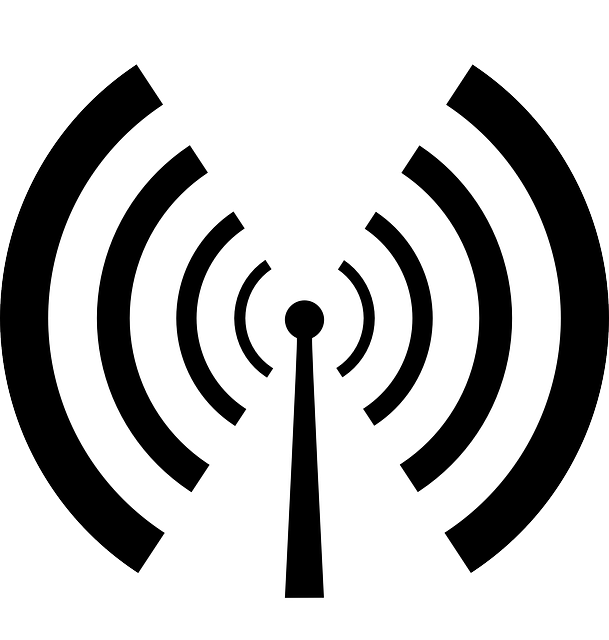}};
\node (compl) [dashed,rect,align = center,below of=smph,xshift=0cm,yshift=-0.35cm] {\\[-0.8ex]Complementary sensors};
\node (human) [rect,align = center,left of=smph,xshift=-2cm,yshift=0cm] {\\[-1.2ex]Human};
\node (vehicle_figure) [align = center,above of=human,xshift=-0.7cm,yshift=-0.1cm]
{\includegraphics[width=.09\textwidth]{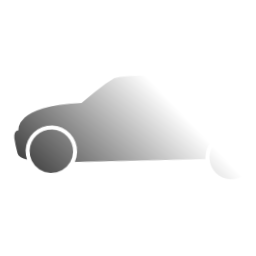}};
\node (human_figure) [align = center,above of=human,xshift=0cm,yshift=0.4cm]
{\includegraphics[width=.06\textwidth]{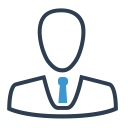}};
\node (HMI) [align = center,left of=smph,xshift=-0.4cm,yshift=0.25cm] {HMI};
\node (business) [draw=black,fill=blue!3,ellipse,align = center,right of=storage,xshift=3.05cm,yshift=-0cm,text height=0.9cm]
{\\\vspace*{-0.9mm}Development\\\vspace*{0mm}and execution of\\\vspace*{-0.7mm}business model};
\node (business_figure) [align = center,above of=business,xshift=0cm,yshift=0.45cm]
{\includegraphics[width=.06\textwidth]{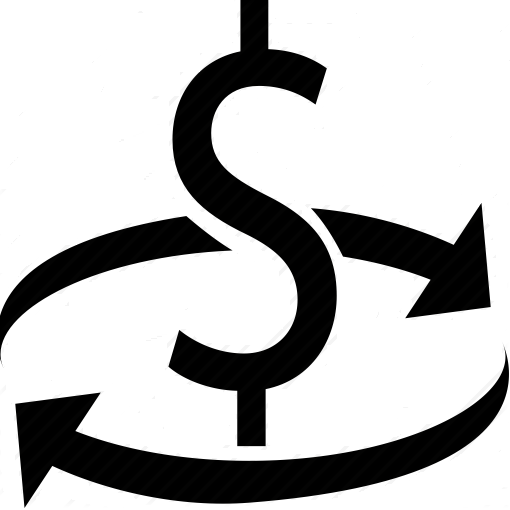}};
\draw [dashed,arrow] (-0.6,-1.35) -- (-0.6,-0.8);
\draw [arrow] (0.55,-0.2) -- (3.45,-0.2);
\draw [arrow] (3.45,0) -- (0.55,0);
\draw [arrow] (-3.1,-0.2) -- (-1.75,-0.2);
\draw [arrow] (-1.75,0) -- (-3.1,0);
\draw [arrow] (6.75,-0.2) -- (7.55,-0.2);
\draw [arrow] (7.55,0) -- (6.75,0);
\end{tikzpicture}
\caption{\label{flow_chart} Process diagram illustrating the information flow of smartphone-based vehicle telematics.}
\vspace*{-1mm}
\end{figure*}
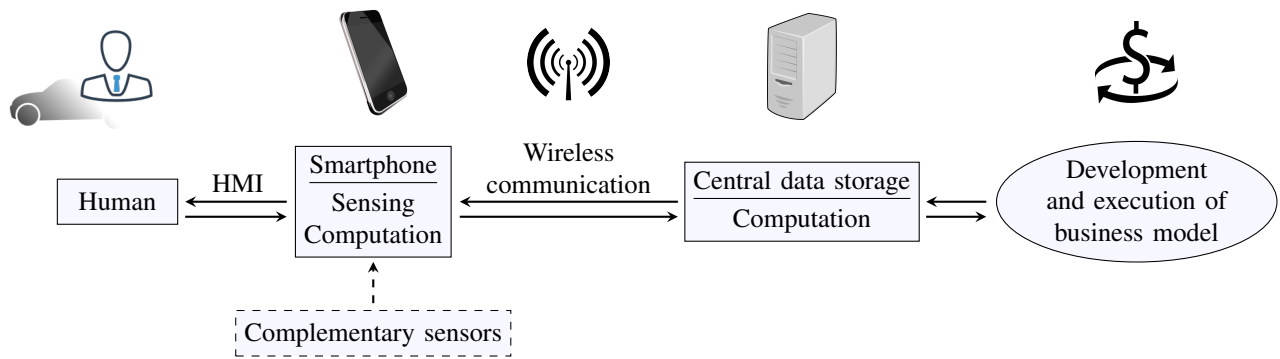

There are several reasons to prefer smartphone-based vehicle telematics over implementations that only utilize vehicle-fixed sensors. Thanks to the unprecedented growth of smartphone ownership, smartphone-based solutions are generally scalable, upgradable, and cheap \cite{Handel2014}.
In addition, smartphones are natural platforms for providing instantaneous driver feedback via audio-visual means, and they enable smooth integration of telematics services with existing social networks \cite{Ekler2015}. Moreover, the replacement and development cycles of smartphones are substantially shorter than those of vehicles, and consequently, smartphones can often offer a shortcut to new technologies \cite{Braun2015}.
At the same time, the logistical demands on the end-user are typically limited to simple app installations.
However, the use of smartphones in vehicle telematics also poses several challenges. Built-in smartphone sensors are generally of poor quality, and have not primarily been chosen or designed with vehicle telematics applications in mind. As a result, the employed estimation algorithms must necessarily rely on statistical noise models, taking the imprecision of the smartphone sensors into account. Further, the smartphone cannot be assumed to be fixed with respect to the vehicle, which complicates the interpretation of data from orientation-dependent sensors such as accelerometers, gyroscopes, and magnetometers \cite{Wahlstrom2015}. Another issue is that of battery drain. An app that continuously collects, stores, process, or transmits data will inevitably consume energy, and hence, there is a demand for solutions that balance constraints on performance and energy efficiency \cite{Tarkoma2014}. Last, we note that the smartphone tends to follow the driver rather than the vehicle. This is often an advantage since it allows the driver to switch vehicle without losing the functionality of smartphone-based telematics services, and to access services and information also when off the road. However, this can also be a disadvantage when, e.g., collecting data for insurance purposes, since automotive insurance policies follow the vehicle rather than the driver in many countries.

\subsection{Conceptual Description of Information Flow}

In what follows, we discuss the information flow of smartphone-based vehicle telematics in more detail. The process is illustrated in Fig. \ref{flow_chart}.
Measurements can be collected from both built-in smartphone sensors and from external, complementary sensor systems. Although vehicle-fixed sensor systems in many cases can offer both higher reliability and accuracy than their smartphone equivalents, they are often omitted to avoid the associated increase in monetary costs and logistical demands. 
Vehicle-fixed sensors are, however, indispensable in the more general field of vehicle telematics where they provide updates on the status of the vehicle's subsystems and describe driver characteristics \cite{Gustafsson2005}. After sensing and processing measurements, data is communicated from the smartphone to a central data storage facility. Vehicle data aggregated at a central point can be used in e.g., traffic state estimation, traffic planning, or comparative driver analysis. Relevant information is sent back to the individual user who can make requests of, e.g., the optimal route in terms of expected travel time to a given location or receive feedback on his driving behavior. The data storage is generally managed by a revenue-generating corporation that supports the services offered to the end-users by financing app development and data analysis \cite{Handel2014_2}. The business model pursued by the corporation relies on the extraction of commercially valuable information from the data storage. In the general case, data is processed both at a local level, in the smartphone, and at a higher level, at the central data storage. Clearly, there is an inherent trade-off between the amount of computations required at a local level and the amount of data that must be sent to the central storage unit \cite{Kjaergaard2011}. (The data set needed for a centralized analysis is often several orders of magnitude smaller in size than the total set of measurements gathered from the smartphones.) 
The amount of computations that are performed directly in the smartphone and at the central storage unit will depend on, e.g., the requested driver feedback and the interests of the service provider. In addition, privacy regulations may require data pre-processing or anonymization at a local level to ensure that the user's privacy is preserved when transmitting data \cite{Hoh2012}.
Telematics is often characterized by the inclusion of a feedback-loop that enables a sensor-equipped end-user to control or change his behavior based on the results of the data analysis. In smartphone-based vehicle telematics, this could be exemplified by a driver who transmits driving data and then receives feedback that can be used to, e.g., improve the safety or fuel-efficiency of his driving \cite{Malikopoulos2013}. By contrast, \textit{telemetry} refers to applications where data is collected from remote locations by one-way telecommunications. Refer to \cite{Reininger2015,Goncalves2014,Handel2014_2} for technical details on the design of a telematics platform.

\subsection{Contributions and Outline}

The aim of this paper is to present an overview of the smartphone-based vehicle telematics field, focusing on the smartphone's role as a measurement probe and an enabler of user-interactive services. In the process, we review recent advances, outline relevant directions for future research, and provide references for further reading. The paper is organized as follows. A selection of academic and industrial projects are discussed in Section \ref{section_proj}. Then, Section \ref{section_system} reviews system aspects, including sensor characteristics, energy efficiency, interfaces for wireless communication, the human-machine interface (HMI) interface, and mobile cloud computing (MCC). Following this, services and applications are covered in Section \ref{section_applications}, which includes discussions on navigation, transportation mode classification, driver behavior classification, and monitoring of road conditions. Finally, the survey is concluded in Section \ref{section_Conclusion}. Table \ref{TableSurveys} provides a list of recommended reviews and other publications relating to the topics of the different sections. The distinguishing features of this survey are that it takes a holistic approach to the field, covers a broad set of applications, and that particular emphasis is placed on the opportunities and challenges that are associated with smartphone-based implementations.

Although the paper's main focus is on smartphone-based sensing, we will also reference a number of studies utilizing other similar sensor setups. However, the choice of references to include is motivated by their relevance for future smartphone-based implementations. For reasons of brevity, we have omitted discussions on topics such as vehicle condition monitoring \cite{Tahat2012}, estimation of fuel consumption \cite{Skog2014}, privacy considerations \cite{Krumm2009b}, accident reconstruction \cite{Thompson2010}, car security \cite{Afzal2014}, and traffic state estimation \cite{Tao2012}.

\begin{table}[t]
\small
\begin{center}
\caption{recommended publications for further reading. \label{TableSurveys}}
\begin{tabular}{l|ll}
\hline
\hline
& & \\ [-2.1ex]
Section & Literature reviews & Other notable studies \\ [-0.2ex]
& & \\ [-2.3ex]
\hline
\multicolumn{1}{l|}{} & \multicolumn{1}{l}{} & \\ [-2ex]
\ref{section_smph_sensors} & \cite{Liu2013,Daponte2013,Lane2010} & \cite{Zandbergen2009} \\
\ref{section_comp_sensors} & \cite{Turker2016,Gustafsson2005} & \cite{Meng2015} \\
\ref{section_energy} & \cite{Tarkoma2014,Perrucci2011,Maloney2012,PerezTorres2016,Wang2014a} & \cite{Carroll2010} \\
\ref{section_cits} & \cite{Lee2010,Lee2014,Araniti2013} & \cite{Salin2012} \\
\ref{section_MCC} & \cite{Fernando2013,Gu2013,Whaiduzzaman2014} & \cite{Yu2013} \\
\ref{section_HMI} & \cite{Williams-Bergen2011,Kinnear2015} & \cite{McEvoy2005,Strayer2013,Strayer2014} \\
\ref{subsection_navigation} & \cite{Quddus2007,Wahlstrom2015_3} & \cite{Wahlstrom2015,Wahlstrom2016} \\
\ref{section_TMC} & \cite{Biancat2014,Gessulat2013} & \cite{Reddy2010,Shin2014} \\
\ref{section_driverbehavior} &
\cite{Weiss2012,Meiring2015,Bolovinou2014,Kaplan2015,Engelbrecht2015,Kalra2014,Handel2014} & \cite{Eren2012,Johnson2011} \\
\ref{section_RCM} & \cite{Chugh2014} & \cite{Mukherjee2016} \\
\multicolumn{3}{l}{} \\ [-2.55ex]
\hline \hline
\multicolumn{3}{l}{}
\vspace*{-8mm}
\end{tabular}
\end{center}
\end{table}




\section{Related academic and industrial projects}
\label{section_proj}

A large number of academic and industrial projects related to smartphone-based vehicle telematics have been conducted.
The projects differ in several aspects, and their value for future implementations will depend on the considered application. As an example, the possibilities for information extraction are, to a large extent, determined by the employed sensors. Fine-grained traffic state estimation generally set high demands on the penetration rate of global navigation satellite systems (GNSS)-equipped probe vehicles, whereas detailed driver behavior profiling requires high rate data from, e.g., accelerometers and gyroscopes. Thus, although there are traffic apps with millions of users, their utility may still be constrained depending on which sensors and sampling rates that are employed.
By contrast, academic projects are often conducted on a smaller scale and in controlled environments, with high-accuracy sensors used to benchmark the performance of smartphone-based implementations. Next, we present a selected number of academically oriented studies, and then go on to discuss industrial projects. A summary of the considered academic projects is provided in Table \ref{Tab3}.

\subsection{Academic Projects}

\emph{CarTel}, introduced at the Massachusetts Institute of Technology in 2006, was a sensing and computing system designed to collect and analyze telematics data. The project resulted in ten published articles addressing topics such as traffic mitigation \cite{Hull2006}, road surface monitoring \cite{Eriksson2008}, and user privacy \cite{Popa2009}.

\emph{Nericell}, initiated by Microsoft in 2007, was a project that specialized in smartphone-based traffic data collection in developing countries. Most of the data was collected in Bangalore, India. The final report included discussions on IMU alignment, braking detection, speed bump detection, and honk detection \cite{Mohan2008}. Earlier, similar projects conducted by Microsoft include the JamBayes and ClearFlow projects, which used data collected in the US to study traffic estimation by means of machine learning \cite{Horvitz2005}.

\begin{table*}[t]
\small
\begin{center}
\caption{significant projects related to smartphone-based vehicle telematics. \label{Tab3}}
\begin{tabular}{l|lllllll}
\hline
\hline
& \multicolumn{1}{l|}{} & \multicolumn{1}{l|}{} & \multicolumn{1}{l|}{} & \multicolumn{1}{l|}{} & \multicolumn{1}{l|}{} & \multicolumn{1}{l|}{} \\ [-2.1ex]
& \multicolumn{1}{l|}{Key} & \multicolumn{1}{l|}{Institution/} & \multicolumn{1}{l|}{Period} & \multicolumn{1}{l|}{Geographical} & \multicolumn{1}{l|}{Data} & \multicolumn{1}{l|}{Primary} & \multicolumn{1}{l}{Main field} \\ [-0.2ex]
& \multicolumn{1}{l|}{} & \multicolumn{1}{l|}{} & \multicolumn{1}{l|}{} & \multicolumn{1}{l|}{} & \multicolumn{1}{l|}{} & \multicolumn{1}{l|}{} \\ [-2.3ex]
Project & \multicolumn{1}{l|}{publ.} & \multicolumn{1}{l|}{company} & \multicolumn{1}{l|}{\scalebox{0.95}{(yy-yy)}} & \multicolumn{1}{l|}{area} & \multicolumn{1}{l|}{quantity} & \multicolumn{1}{l|}{sensors} & \multicolumn{1}{l}{of study} \\
& & & \multicolumn{5}{l}{} \\ [-2.6ex]
\hline
& & & \multicolumn{5}{l}{} \\ [-2.1ex]
CarTel & \cite{Hull2006} & MIT. & 06-11 & Massachusetts, & - & GNSS, & Data management, \\ [-0.2ex]
& & & & US. & & accelerometer.$^{\dag}$ & energy efficiency. \\
\arrayrulecolor{gray}\hline\arrayrulecolor{black} \\ [-2.1ex]
Nericell & \cite{Mohan2008} & Microsoft. & 07-08 & Bangalore, & $622\,[km]$ data & GNSS, & Traffic monitoring \\ [-0.2ex]
& & & & India, and & recorded over & accelerometer, & in developing \\ [-0.2ex]
& & & & Seattle, US. & $27.6\,[h]$. & microphone. & countries. \\
\arrayrulecolor{gray}\hline\arrayrulecolor{black} \\ [-2.1ex]
Mobile & \cite{Bayen2011}, & UC Berkeley, & 08-11 & California and & More than 2000 & GNSS. & Traffic state \\ [-0.45ex]
Millenium & \cite{Herrera2010}$^{\ddagger}$ & Nokia. & & New York, US. & registered drivers. & & estimation. \\
\arrayrulecolor{gray}\hline\arrayrulecolor{black} \\ [-2.1ex]
Movelo & \cite{Handel2014}, & KTH, If P\&C.,  & 10-14 & Stockholm, & More than 1000 & GNSS. & Insurance \\ [-0.2ex]
Campaign & \cite{Handel2014_2} & Movelo. & & Sweden. & registered drivers. & & telematics. \\
\arrayrulecolor{gray}\hline\arrayrulecolor{black} \\ [-2.1ex]
Mobile & \cite{Allstrom2011} & KTH, LiTH, & 11- & Stockholm, & 1500 installed & GNSS.$^{\dag}$ & Traffic state \\ [-0.2ex]
Millenium & & Sweco, & & Sweden. & on-board units. & & estimation. \\ [-0.2ex]
Stockholm & & UC Berkeley. & & & & & \\
\arrayrulecolor{gray}\hline\arrayrulecolor{black} \\ [-2.1ex]
Future & \cite{d'Orey2014}, & University of  & 12- & Porto, Portugal. & 624 installed & GNSS, & Urban traffic \\ [-0.2ex]
Cities & \cite{Rodrigues2014} & Porto. & & & on-board units. & accelerometer. & management. \\
\hline \hline
\multicolumn{8}{l}{} \\ [-2ex]
\multicolumn{8}{l}{\footnotesize{$^{\dag}$ The study only considered vehicle-fixed sensors.}} \\ [-0.2ex]
\multicolumn{8}{l}{\footnotesize{$^{\ddagger}$ Considers the closely related Mobile Century field experiment.}}
\vspace*{-4mm}
\end{tabular}
\end{center}
\end{table*}

\emph{The Mobile Millennium project}, conducted in collaboration between the University of California, Berkeley and Nokia, included one of the first smartphone-based large-scale collections of traffic data. The project resulted in more than forty academic publications, primarily in the field of applied mathematics \cite{Herrera2010}, 
\cite{Hoh2008,Work02010,Hunter2011,Herrera2010_2}.
Several future challenges were identified, all the way from defining the business roles of the involved actors, to delivering efficient driver assistance while minimizing distraction \cite{Bayen2011}. The project was extended by the \emph{Mobile Millennium Stockholm project} in Sweden, in which position data was collected at one-minute intervals from $1500$ GNSS-equipped taxis \cite{Allstrom2012}.

\emph{The Movelo Campaign} introduced a smartphone-based measurement system that focused on applications within insurance telematics. As part of the campaign, If P\&C Insurance launched the commercial pilot \emph{if Safedrive}, in which around $4500\,[h]$ of data was collected from more than 1000 registered drivers \cite{Handel2014,Handel2014_2}. The project was the first large-scale study focusing on local data processing and real-time driver feedback. Published articles discussed, e.g., driver risk assessment \cite{Wahlstrom2015_2}, estimation of fuel consumption \cite{Skog2014}, and business innovation \cite{Ohlsson2015}.

\emph{The Future Cities project}, conducted by the University of Porto, focuses on traffic management in urban environments. The studied research areas include sustainability, mobility, urban planning, as well as information and communication technology. Data have been collected from hundreds of local buses and taxis equipped with vehicle-fixed GNSS receivers and accelerometers. Among other things, the project led to the development of the SenseMyCity app, which can be employed to, e.g., estimate fuel consumption or investigate the possibilities for car sharing \cite{Rodrigues2014}.

\subsection{Commercial Projects}

The market for insurance telematics, i.e., automotive insurances with a premium that is based on driving data, has grown to include more than fifteen million policyholders \cite{Ptolemus2016}. Most insurance telematics programs collect data by the use of in-vehicle sensors or externally installed hardware components, referred to as black boxes, in-vehicle data recorders, or aftermarket devices. The data collection process is often associated with large costs attributed to device installment and maintenance. However, by collecting data from the policyholders' smartphones, it is possible to decrease the logistical costs while at the same time increase driver engagement and transparency \cite{Handel2014}. For example, in 2014, the software provider Vehcon launched the app MVerity, intended as a stand-alone insurance telematics solution, for the mere cost of $\$1$ per year and vehicle \cite{Reuters2014}. As of June 2016, there were thirty-three active smartphone-based insurance telematics programs (including trials) \cite{Ptolemus2016}. The proliferation of smartphones has also had a huge impact on the taxi industry. Specifically, companies such as Uber, Ola Cabs, and Careem are providing apps that connect passengers and drivers based on trip requests, while incorporating dynamic pricing, automated credit card payments, as well as customer and driver ratings. In December 2015, Uber was valued at $\$62.5$ billion \cite{Newcomer2015}. Additional examples of apps specifically designed for vehicle owners include: Waze (acquired by Google in 2013 for $\$1.15$ billion), allowing drivers to share information about, e.g., accidents, traffic jams, and road closures; iWrecked, facilitating easy assemblies of accident reports; Torque, enabling the smartphone to extract data from in-vehicle sensors; FuelLog, logging fuel consumption as well as the maintenance and service costs of the vehicle; Lyft, connecting passengers and drivers for on-demand ridesharing; and iOnRoad (acquired by Harman in 2013), a camera-based collision warning system that sends warning signals at insufficient headway distances.

To summarize the section, a large number of projects related to smartphone-based vehicle telematics have been conducted both in academia and in the industry. The conducted projects differ in the employed sensors, the number of users, the required logistics, and the considered applications. The primary drivers for future projects are expected to be found in applications related to, e.g., insurance telematics, ridesharing, and driver assistance.

\section{System aspects}
\label{section_system}

This section discusses system aspects of smartphone-based vehicle telematics. We review available sensors, battery usage, cooperative intelligent transportation systems (C-ITS), MCC, and the HMI. Many of the covered topics are relevant in a broad range of applications, and hence, this section will serve as a basis for the subsequent section that covers services and applications.

\subsection{Smartphone Sensors}
\label{section_smph_sensors}

In the following, we review the exteroceptive (GNSS, Bluetooth, WiFi-based positioning, cellular positioning, magnetometers, cameras, and microphones) and proprioceptive (accelerometers and gyroscopes) sensors and positioning technologies commonly utilized in smartphones.

\textit{1) Exteroceptive Sensors}: iPhone devices employ a built-in black-box feature called Location Services to provide apps with navigation updates. The updates are obtained by fusing WiFi, cellular, Bluetooth, and GNSS data.
WiFi and cellular positioning are mainly used to aid the initialization of the GNSS receiver
(the stand-alone accuracy of these positioning systems is generally inferior to that of a GNSS). However, the coverage of WiFi and Bluetooth in urban environments is continuously being improved, enabling position fixes in areas where GNSS often is unavailable. Location Services provide three-dimensional position updates with horizontal and vertical accuracy measures, as well as planar speed, planar course, and timestamps. Details on the positioning methods utilized in Locations Services and their associated accuracy can be found in \cite{Zandbergen2009} and \cite{Watzdorf2010}. In contrast to smartphones from the iPhone series, Android devices enable apps to read GNSS messages in the NMEA 1803 standard. The standard includes not only GNSS measurements of position, planar speed, and planar course, but also additional information such as detailed satellite data (refer to \cite{NMEA2008} for details on NMEA 1803). According to reports, raw GNSS measurements, such as pseudoranges and doppler shifts, will eventually become accessible from all new Android phones \cite{ION2016}.

Thanks to its high accuracy and availability, GNSS is often the positioning technology of choice in today's location-based services (LBSs). Nevertheless, the performance of commonly utilized GNSS receivers has been found to be a limiting factor in a number of ITS applications \cite{Zhang2011}. A performance evaluation of GNSS receivers in several commercial smartphones is presented in \cite{Menard2011,Menard2011b}. All of the studied smartphones were found to provide horizontal position measurements accurate to within $10\,[m]$ with a $95\,\%$ probability.
Since comprehensive performance evaluations of GNSS receivers in real-world scenarios often are both time-consuming and costly, simulation-based methodologies are an attractive complement. Using simulations, test scenarios can be repeated in a controlled laboratory setting, with many possibilities for error modeling \cite{Aloi2007}. Obviously, 
the external validity of such experiments
will be limited by the simulator's ability to mimic the contributions of the numerous error sources in real-world GNSS \cite{Kaplan2006}. The standard update rate of smartphone-embedded GNSS receivers is currently $1\,[H\!z]$.

A triad of magnetometers measure the magnetic field in three-dimensions and is in many applications used as a compass, indicating the direction of the magnetic north in the plane that is tangential to the surface of the earth.
In other words, smartphone-embedded magnetometers can often be used to gain information about the orientation of the smartphone.
The accuracy of smartphone-embedded magnetometers has been studied in \cite{Blum2012}, which reported mean absolute errors in the order of $10-30\,[\hspace*{0.2mm}^{\circ}\hspace*{0.2mm}]$ when estimating the direction of the magnetic north during pedestrian walking. In vehicle telematics, the use of magnetometers is further constrained by magnetic disturbances caused by the vehicle engine \cite{Godha2005,Bo2016,Zong2015}. However, in some cases these disturbances can be used to extract valuable information. For example, magnetometers can be placed on the road side to detect and classify passing vehicles based on the magnetic fields that they induce \cite{WahlstromGustafsson2014,Prateek2013}. Unfortunately, studies on how to extend this work to implementations using vehicle-fixed and possible smartphone-embedded magnetometers are scarce.


\begin{table}[t]
\small
\begin{center}
\caption{\hspace*{9mm} smartphone sensors commonly utilized in \newline vehicle telematics applications. \label{Tab4}}
\begin{tabular}{l|l}
\multicolumn{1}{l}{} & \multicolumn{1}{l}{} \\ [-1ex]
\hline
\hline
\multicolumn{1}{l|}{} & \multicolumn{1}{l}{} \\ [-2ex]
Sensor & Measurement \\
\hline
\multicolumn{1}{l}{} & \multicolumn{1}{l}{} \\ [-2ex]
\multicolumn{2}{l}{\emph{Exteroceptive sensors}} \\
\hline
\multicolumn{1}{l|}{} & \multicolumn{1}{l}{} \\ [-2ex]
GNSS & Position, planar speed, and planar course \\
Magnetometer & Magnetic flux density \\
Camera & Visual images \\
Microphone & Audio \\
\hline
\multicolumn{1}{l}{} & \multicolumn{1}{l}{} \\ [-2ex]
\multicolumn{2}{l}{\emph{Proprioceptive sensors}} \\
\hline
\multicolumn{1}{l|}{} & \multicolumn{1}{l}{} \\ [-2ex]
Accelerometer & Specific force (non-gravitational acceleration) \\
Gyroscope & Angular velocity \\
\hline \hline
\multicolumn{2}{l}{}
\vspace*{-4mm}
\end{tabular}
\end{center}
\end{table}

The possibilities for information extraction based on built-in smartphone cameras are continuously increasing as a result of improved image resolutions and increased frame rates. Smartphone cameras can for example be used to aid an inertial navigation system \cite{Zachariah2010}, to detect or track surrounding vehicles \cite{Sivaraman2013}, to estimate the gaze direction of drivers \cite{Chuang2014}, to detect traffic signals \cite{Koukoumidis2011}, or to detect traffic signs \cite{Mathias2013}. Geometric camera calibration is a vast topic encompassing the estimation of intrinsic (describing the internal geometry and optical characteristics of the image sensor) and extrinsic (describing the pose of the camera with respect to an external frame of reference) parameters, as well as the joint calibration of cameras and motion sensors \cite{Salvi2002}. Calibration approaches specifically designed for smartphone-embedded cameras are presented in \cite{Delaunoy2014,Jia2014,Saponaro2013,Skocaj2014,Panahandeh2015}. Generally, camera-based implementations are constrained by their high computational cost, and requirements on, e.g., orientation and visibility. Moreover, they also tend to raise privacy concerns from users. As a result, they are often disregarded in telematics solutions that prioritize reliability and convenience.


Although microphones cannot directly be used for purposes of navigation, they are often used to provide information regarding context and human activities \cite{Lane2010}. Commercial examples include the app Shopkick, which uses inaudible audio signals, unique to different stores, to track users' consumer behavior. In vehicle telematics, microphones have been used for honk detection \cite{Mohan2008}, emergency vehicle detection \cite{Mielke2013}, as well as audio ranging inside of the vehicle \cite{Yang2011}. The use of smartphone-embedded microphones has brought attention to the issue of user privacy, and several related legislations can be expected in the coming years \cite{Gomez-Martin2012}.


\begin{table}[t]
\small
\begin{center}
\caption{\hspace*{9mm}obd measurements commonly utilized in \newline vehicle telematics applications. \label{Tab5}}
\begin{tabular}{l|l}
\multicolumn{1}{l}{} & \multicolumn{1}{l}{} \\ [-1ex]
\hline
\hline
\multicolumn{1}{l|}{} & \multicolumn{1}{l}{} \\ [-2ex]
Measurement & Resolution$^{\dagger}$ \\
\hline
\multicolumn{1}{l|}{} & \multicolumn{1}{l}{} \\ [-2ex]
Vehicle speed & $1\,[km/h]$\\
Engine revolutions per minute & $0.25\,[rev/min]$\\
Throttle position & $0.39\,\%$ \\
Airflow rate & $0.01\,[g/s]$ \\
\hline \hline
\multicolumn{2}{l}{} \\ [-2ex]
\multicolumn{2}{l}{\footnotesize $^{\dagger}$ The resolution gives the smallest difference between} \\ [-0.25ex]
\multicolumn{2}{l}{\footnotesize possible output values. However, note that this is \underline{not} an} \\ [-0.25ex]
\multicolumn{2}{l}{\footnotesize upper bound on the measurement error.}
\vspace*{-4mm}
\end{tabular}
\end{center}
\end{table}

\textit{2) Proprioceptive Sensors}:
A triad of accelerometers and a triad of gyroscopes comprise what is known as a 6-degrees of freedom inertial measurement unit (IMU), which produces three-dimensional measurements of specific force and angular velocity. As shown in \cite{Perlmutter2012}, the cost of smartphone-embedded IMUs has steadily decreased and is expected to continue to do so during the coming years (a smartphone-embedded IMU can be expected to cost less than $\$1$ for the manufacturer). IMUs in smartphones are considered to belong to the lowest grade of inertial sensors (the commercial grade), and can exhibit significant bias, scale factor, misalignment, and random noise errors \cite{Groves2008}. Detailed error analyses of IMUs in ubiquitous smartphone models have been conducted in \cite{Aicardi2014,Chowdhury2014b,Niu2015,Chowdhury2016}. The maximum sample rate of IMUs in smartphones is typically in the order of $20-300\,[H\!z]$. By fusing IMU and GNSS measurements, it is possible to obtain high-rate estimates of the vehicle's position, velocity, and attitude \cite{Groves2008}. In addition, IMUs are commonly used for the detection of, e.g., harsh braking \cite{Johnson2011}, potholes, and speed bumps \cite{Yu2016}. The discussed smartphone sensors are summarized in Table \ref{Tab4}.

\subsection{Complementary Sensors}
\label{section_comp_sensors}

%

In addition to using built-in smartphone sensors, information can also be obtained from vehicle-installed black boxes or from the vehicle's on-board diagnostics (OBD) system. The OBD system is an in-vehicle sensor system designed to monitor and report on the performance of a large number of vehicle components. As of 2001, OBD is mandatory for all passenger cars sold in Europe or in Northern America. OBD data can be sent from an OBD dongle to a smartphone using either WiFi or Bluetooth \cite{Zaldivar2011}. The dongles costs from around $\$10$ (the cost of an OBD dongle used for insurance telematics is typically higher), with the exact price dependent on factors such as smartphone compatibility and data output. Some of the most commonly used OBD measurements are summarized in Table \ref{Tab5}.
The utility of OBD measurements of speed have been compared with standalone smartphone solutions in the context of acceleration-based accident detection \cite{Zaldivar2011}, detection of harsh braking \cite{Handel2014}, maneuver recognition \cite{Sathyanarayana2012}, \cite{Sathyanarayana2013}, and vehicle speed estimation \cite{Chowdhury2014b,Chowdhury2014,Chowdhury2016}. On the one hand, OBD measurements are not subject to multipath errors (unlike GNSS measurements) and only rely on vehicle-fixed sensors (as opposed to smartphone sensors whose readings often are affected by the dynamics of the smartphone with respect to the vehicle). On the other hand, smartphone-based solutions benefit from the high sampling rate of the embedded IMU and can often provide contextual information related to, e.g., the driver's activities prior to driving \cite{Meng2015}. Fusion of OBD and smartphone data for vehicle positioning is discussed in \cite{Walter2013}.


\begin{figure}[t]
\def\svgwidth{\columnwidth}
\hspace*{-2mm}
\scalebox{0.88}{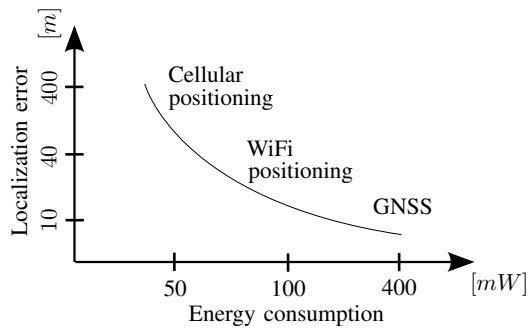}
\caption{\label{tradeoff_figure}The trade-off between localization error \cite{Wang2014a} and energy consumption at $0.1\,[H\!z]$ \cite{Chon2011} for cellular positioning, WiFi positioning, and GNSS.}
\vspace*{-3.5mm}
\end{figure}

\subsection{Energy Consumption}
\label{section_energy}

The energy consumed by technology and hardware in smartphones has been identified as a key factor in the design of new services and applications. Recognizing that the GNSS receiver often can be attributed to a large share of a smartphone's energy consumption, plenty of research has been invested into the design of low-power LBSs.

In \emph{dynamic tracking}, the GNSS receiver is only turned on when GNSS updates are expected to result in a sufficiently large improvement of the navigation accuracy. Dynamic tracking can be roughly categorized into two overlapping branches: \emph{dynamic prediction} and \emph{dynamic selection}. In dynamic prediction, the localization uncertainty is quantified using less power-intensive sensors \cite{Kjaergaard2009}. Accelerometers are for example often used to assess if the target is moving or not. In dynamic selection, the navigation system alternates between GNSS and low-power positioning systems (e.g., WiFi-based) by considering the time and location dependence of 1) the accuracy requirements of the application; and 2) the performance of the positioning technologies. For example, the localization accuracy requirements for turn-by-turn navigation will typically increase as the vehicle approaches an intersection \cite{Hu2015}, and GNSS measurements are known to be unreliable in urban canyons \cite{Paek2010}. By exploiting such variations in required accuracy, the smartphone can dynamically choose the positioning technology that will optimize the trade-off between localization accuracy and energy efficiency (see Fig. \ref{tradeoff_figure}).
Although sampling the GNSS receiver less often typically leads to a decrease in the energy consumed per time, the total energy required per GNSS sample will increase as a result of the cold/warm/hot-start nature of the receiver \cite{Hu2015}.

\begin{table}[t]
\small
\begin{center}
\caption{typical figures of energy consumption in smartphones \newline \cite[p. 127]{Tarkoma2014}.\label{Tab6}}
\begin{tabular}{l|c}
\hline
\hline
\multicolumn{1}{l|}{} & \multicolumn{1}{l}{} \\ [-2ex]
Sensor & Power $[mW]$ \\
\hline
\multicolumn{1}{l|}{} & \multicolumn{1}{l}{} \\ [-2ex]
Magnetometer & \hspace*{1.655mm}$50$ \\
Camera & \hspace*{-1.5mm}$1500$ \\
Microphone & \hspace*{0.1mm}$100$ \\
Accelerometer & \hspace*{1.62mm}$20$ \\
Gyroscope & $150$ \\
\multicolumn{2}{l}{} \\ [-2.55ex]
\hline \hline
\multicolumn{2}{l}{}
\vspace*{-5mm}
\end{tabular}
\end{center}
\end{table}

Several additional measures can be taken to improve the energy efficiency of smartphone-based LBSs. For example, navigation filters are often complemented by digital map-matching. The resulting increase in localization accuracy will depend on the road density (the performance of map-matching algorithms generally deteriorates in areas where the road density is high), and hence, this must be considered in the design of energy-efficient sampling schemes \cite{Alrefaie2013}. Generally, map-matching reduces the error growth during inertial navigation and thereby makes it possible to rely only on low-power inertial sensors for a longer period of time. One option is also to reject GNSS measurements completely. As an example, the CarTel project included a study on map-aided cellular positioning of vehicle-fixed mobile devices \cite{Thiagarajan2011,Aly2015}. An additional alternative is to employ cooperative localization, i.e., to fuse sensor measurements, e.g., GNSS and ranging measurements, from several near-by smartphones. In this way, it is possible to reduce the total number of GNSS updates, and thereby also the energy consumption, while maintaining sufficient localization accuracy \cite{Liu2012}.

For easy reference, Table \ref{Tab6} displays typical figures of energy consumption for the sensors discussed in Section \ref{section_smph_sensors} (the energy consumed by the most commonly employed positioning technologies was displayed in Fig. \ref{tradeoff_figure}). Additional parameters characterizing the energy consumption attributed to transmission, computation, and storage, can be found in \cite{Miettinen2010,Abdelmotalib2012,Friedman2013}.
Since the exact energy characteristics are dependent on many factors, including the device model, sensing specifications, and smartphone settings, the provided figures should be interpreted as order-of-magnitude approximations. As an example, the energy consumption attributed to the IMU will depend on the IMU model, the chosen sampling rate, and eventual duty-cycling schemes \cite{Chon2011}.
It should also be noted that the inertial sensors and magnetometers themselves generally consume less energy than stated in Table \ref{Tab6}. However, since many smartphones use the main processor to directly control the sensors, continuous sensing often incurs a substantial energy overhead. A potential remedy is to let a dedicated low-power processor handle the tasks of duty cycle management, sensor sampling, and signal processing, thereby allowing the main processor to sleep more frequently \cite{Priyantha2010}.

In smartphone-based vehicle telematics, the issue of battery drainage can be mitigated by using a battery charger connected to the vehicle's power outlet (cigarette lighter receptacle). Similarly, several telematics solutions combine sensors, battery, and a charger in a single device that connects to either the OBD port or the cigarette lighter receptacle \cite{Wang2013}.



\subsection{Cooperative Intelligent Transportation Systems}
\label{section_cits}
%

%

\begin{figure}[t]
\def\svgwidth{\columnwidth}
\hspace*{-2mm}
\scalebox{0.88}{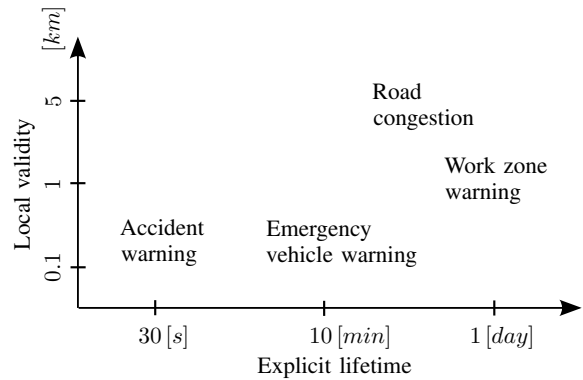}
\caption{\label{validitylifetime_figure}Local validity and explicit lifetime for a number of content types used in applications of C-ITS \cite{Lee2014}.}
\vspace*{-3.5mm}
\end{figure}

The idea of cooperative intelligent transportation systems (C-ITS) is to fuse measurements and information from vehicles, pedestrians, and local infrastructure within a limited geographical area. Studied applications include collision avoidance \cite{Guido2012}, emergency vehicle warning systems \cite{Buchenscheit2009}, and traffic mitigation \cite{Gramaglia2011}. The considered information content can be characterized by its \emph{local validity}, i.e., its spatial scope of utility, and its \emph{explicit lifetime}, i.e., its temporal scope of validity \cite{Lee2014}. Fig. \ref{validitylifetime_figure} displays the local validity and explicit lifetime of several content types typical for applications within C-ITS.
The communication of information in C-ITS is performed over a vehicular ad hoc network (VANET), i.e., a type of mobile ad hoc network (MANET) where vehicles are used as mobile nodes \cite{Lee2010}. VANETs comprise both vehicle-to-vehicle (V2V) and vehicle-to-infrastructure (V2I) communications. Typically, one assumes the employment of the IEEE 802.11p standard for local wireless communication, exclusively dedicated to vehicular environments. However, despite a vast amount of conducted research, the deployment of real-world VANETs has been slow. This is primarily due to the large costs associated with the required communication transceivers, i.e., the costs of setting up local roadside units and installing on-board units in vehicles \cite{Caballero-Gil2013}. Using smartphones in VANETs is appealing both from a cost perspective and because it facilitates the integration of pedestrians, bicyclists, and motorcyclists into the network. Nevertheless, the use of smartphones also brings about some technical difficulties.
Most importantly, commercial smartphones are not equipped with the IEEE 802.11p interface. More so, \cite{Salin2012} concluded that it is currently not feasible, neither from an economical nor from a practical point of view, to make smartphones compliant with the IEEE 802.11p standard. One solution is to make use of the sensing, computing, and storage capabilities of the smartphone, and then use Bluetooth or WiFi to send data from the smartphone to a low-cost in-vehicle 802.11p device, only equipped with the most indispensable communication technology. This 802.11p device can in turn be used to communicate with roadside units or other on-board units \cite{Li2015}, \cite{Barcelos2014}. Implementations have also been proposed where the wireless communication is instead partially or completely based on the IEEE 802.11a/b/g/n standards \cite{Caballero-Gil2013,Tornell2013,Busanelli2013,Djajadi2014}, or on cellular connectivity \cite{Araniti2013,Park2014}. However, the link performance of these implementations is expected to be worse than under the IEEE 802.11p standard. Whether the performance is sufficient for time-critical safety applications in smartphone-based VANETs is still an open question \cite{Vandenberghe2011}. The pros and cons of different wireless technologies in VANETs are reviewed in \cite{Araniti2013}. Refer to \cite{Busanelli2013} and \cite{Silva2014} for details on the implementation of a smartphone-based VANET.

Notable real-world deployments of VANETs include the Cooperative ITS Corridor, a planned vehicular network that will initially be deployed on motorways in Austria, Germany, and the Netherlands. The project's main focus will be on utilizing the IEEE 802.11p interface for safe and environmentally friendly driving \cite{CooperativeITS2013}. In Japan, vehicle drivers are provided with real-time traffic information through the Vehicle Information and Communication System (VICS). Traffic data is collected from road sensors and then sent to vehicles with installed VICS units. As of 2015, the total number of installed VICS units was over $47$ million \cite{Mashiko2016}. Japan also holds a 3.5-hectare ITS Proving Ground in Susono, Shizuoka, where Toyota conducts research on C-ITS. Although large-scale real-world implementations of smartphone-based VANETs have yet to be studied, smaller field tests on V2V communications have been conducted for applications within, e.g., platooning \cite{Berggren2011}.




%
%
%

\subsection{Mobile Cloud Computing}
\label{section_MCC}

\begin{figure}[t]
\def\svgwidth{3.5in}
\hspace*{-3.5mm}
\scalebox{0.95}{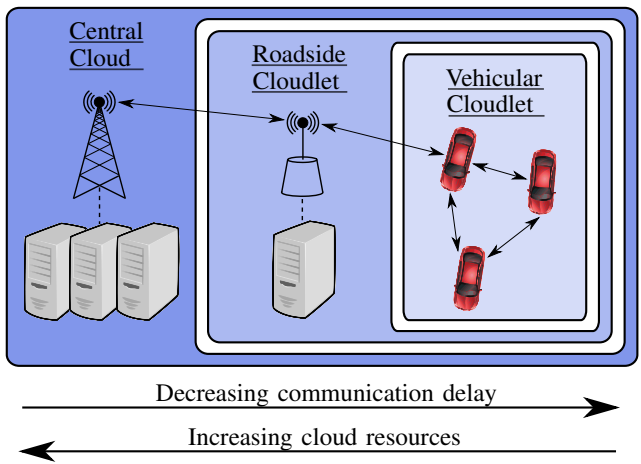}
\caption{\label{MCCfigure}The three-layered vehicle cloud architecture consisting of the vehicle cloudlet, the roadside cloudlet, and the central cloud.}
\vspace*{-3.5mm}
\end{figure}

MCC refers to the concept of offloading computation from mobile nodes to remote hosts. By exploiting resources of "the cloud", mobile users can bypass the computation and storage constraints of their individual devices. Resources can either be provided by a remote cloud, serving all mobile devices, or be provided by a local cloud, utilizing resources from other nearby mobile devices \cite{Fernando2013}. Lately, MCC applications have been discussed in the context of VANETs, giving rise to the term vehicular cloud computing (VCC). Vehicular clouds differ from internet clouds in several aspects. One distinguishing feature derives from the unpredictability of the vehicles' behavior. Vehicles can unexpectedly leave the VANET, and hence, the computation and storage scheme must be made resilient to such events. Moreover, VCC is often performed using so called vehicular cloudlets, i.e., clouds utilizing the resources of surrounding vehicles without requiring an active internet connection. The use of vehicular cloudlets will generally both reduce the transmission cost and increase service availability \cite{Gu2013}. The network architecture proposed in \cite{Yu2013} divides the vehicular cloud into three layers. These are the earlier described vehicular cloudlet, the roadside cloudlet, composed of dedicated local servers and roadside units, and the central cloud (the basic internet cloud)\footnote{The cloud based exclusively on V2V communications, here called the vehicular cloudlet, is sometimes referred to as the vehicular cloud. However, we follow the taxonomy of \cite{Gu2013} and reserve the term vehicular cloud for the overarching network architecture.}. As illustrated in Fig. \ref{MCCfigure}, usage of the added resources provided by the outer layers of the vehicular cloud tends to come at the expense of an increased communications delay.

VCC have been applied in industrial projects conducted by Ford, General Motors, Microsoft, and Toyota. The focus has been on providing navigation services and infotainment systems with features such as voice-controlled traffic updates and in-vehicle WiFi hotspots \cite{Whaiduzzaman2014}.
MCC has also been applied at a sensing level to, e.g., mitigate the large energy consumption of GNSS receivers. In \cite{Liu2012a}, raw GNSS signals were uploaded from a mobile device to the cloud for post-processing. Leveraging the computational capabilities of the cloud and publicly available data such as the satellite ephemeris and earth elevation, the energy efficiency was increased as a result of reducing both the computational cost and the amount of raw GNSS data acquired for each position fix. Obviously, when offloading computations to the cloud, the energy consumption attributed to data transmission will increase. To reduce this effect, a framework for pre-transmission compression of GNSS signals was presented in \cite{Misra2014}.


%
	
%

%

%
%


\subsection{Human-Machine Interface}
\label{section_HMI}

The term HMI encompasses all functionalities that allow a human to interact with a machine. Our main interest is the human-smartphone interface in applications of smartphone-based vehicle telematics. Refer to \cite{Green2008} for details on human-vehicle interfaces.

The design of a human-smartphone interface can be considered to have two objectives \cite{Yamabe2014}. First, the interface must be efficient. In other words, the tasks of control and monitoring should be performed with as little computational and human effort as possible. Second, the interface should be configured to minimize driver distraction. Hence, all driver-smartphone communication should be dynamically integrated with the driving operations and discourage excessive engagement in secondary tasks. As noted in \cite{Watkins2011}, driver distraction due to smartphone usage has become one of the leading factors in fatal and serious injury crashes. For example, having a conversation on your phone while driving can in some situations increase the crash risk by a factor of four \cite{McEvoy2005}, with texting typically being even more distracting \cite{Saiprasert2014}. Four facets of driver distraction can be identified: cognitive distraction, taking your mind off the road; visual distraction, taking your eyes off the road; auditory distraction, focusing your auditory attention on sounds that are unrelated to the traffic environment; and manual distraction, taking your hands off the wheel \cite{Williams-Bergen2011}. While smartphone usage may contribute to each of these types of distractions, the distraction caused by smartphone calls is mainly cognitive and auditory in nature \cite{Strayer2011}. In other words, the use of hands-free equipment do little to alleviate this distraction. In what follows, we discuss the current use of audio and video in human-smartphone interfaces, and describe how the interfaces can be designed to not interfere with the task of driving.

\begin{table}[t]
\small
\begin{center}
\caption{characterization of typical forms of smartphone-related driver distractions \cite{Kinnear2015}. \label{TableDistraction}}
\begin{tabular}{l|cccc}
\hline
\hline
\multicolumn{1}{l|}{} & \multicolumn{1}{l}{} \\ [-2ex]
& \multicolumn{4}{c}{Distraction} \\
\cline{2-5}
\multicolumn{1}{l|}{} & \multicolumn{1}{l}{} \\ [-1.9ex]
Activity & Cognitive & Visual & Auditory & Manual \\
\hline
\multicolumn{1}{l|}{} & \multicolumn{1}{l}{} \\ [-2ex]
Texting & High & High & Low & High \\
Dialing & Medium & High & Low & High \\
Phone call & High & Low & High & Low \\
Voice control & High & Low & Medium & Low \\
\multicolumn{5}{l}{} \\ [-2.55ex]
\hline \hline
\multicolumn{5}{l}{}  \\
\multicolumn{4}{l}{} & \multicolumn{1}{l}{} \\ [-1ex]
\end{tabular}
\vspace*{-8mm}
\end{center}
\end{table}

Audio communication has proven to be an efficient method for relaying information to the driver while minimizing visual and manual distractions. For example, in \cite{Hu2015}, respondents generally thought that vehicular navigation benefited more from turn-by-turn voice guidance than from visual map display. Voice guidance is a common feature in today's navigation apps.
In addition to voice guidance, many driver assistance systems also make use of audio alerts. One example is the {\it iOnRoad} app, which provides warning signals when detecting speeding, insufficient headway, or the crossing of a solid line. Although voice control is an integral part of many existing apps, its use can at times require a significant cognitive workload. Refer to \cite{Strayer2013} and \cite{Strayer2014} for details on how different auditory tasks affect, e.g., reaction times and headway distances. Table \ref{TableDistraction} describes how the activities of texting, dialing a phone number, having a conversation on the phone, and delivering voice commands relate to different forms of distractions.

Most human-smartphone interfaces are, to a large extent, based on visual communication utilizing touch-based operations and virtual keyboards. Obviously, visual display is the most efficient way to deliver complex information such as annotated maps. In addition, displayed information is not intrusive in the same way as, e.g., audio, and therefore, it allows the driver to manually coordinate driving maneuvers and information gathering. The smartphone display may also be used for augmented reality, i.e., to overlay computer-generated graphics onto images of reality. This has been utilized in apps such as iOnRoad, which highlights the vehicle's current lane in a real-time video shown in the smartphone display, and Hudway, which creates a head-up display (HUD), independent of the existing vehicle technology, by reflecting smartphone images in the windshield. The latter technique is particularly valuable in low visibility conditions since it enables the driver to see the curvature of the upcoming road overlaid in sharp lines on the windshield. In general, the increasing prevalence of in-vehicle HUDs is expected to both reduce collisions (thanks to automated obstacle detection) and discourage inattentive driver behavior \cite{Park2013}. By using the Google and Apple standards Android Auto and CarPlay, respectively, any smartphone operation can be directly integrated into the HUD, thereby further reducing visual and manual distractions. However, according to predictions, only $9\%$ of the automobiles on the road in the year 2020 will have a built-in HUD \cite{IHS2013}, and consequently, there will continue to be a high demand for inexpensive and flexible aftermarket solutions.

As should be evident, the level of concentration required for the task of driving varies with both the vehicle's state and the traffic situation. Hence, to satisfy requirements on communication efficiency while minimizing driver distraction, the driver-smartphone interactions should be adjusted with respect to, e.g., the vehicle's speed and location \cite{Matsuyama2014,Li2016b}. This idea is utilized in apps such as DriveID, which enables customized restrictions of smartphone functionality while driving. By using a vehicle-mounted Bluetooth device to position smartphones within the vehicle, the restrictions can be limited to the area around the driver's seat, so that only the driver's smartphone is subject to the restrictions.

\section{Services and applications}
\label{section_applications}

Next, we cover practical applications within smartphone-based vehicle telematics such as navigation, transportation mode classification, the study of driver behavior, and road condition monitoring.

\subsection{Navigation}
\label{subsection_navigation}

As opposed to navigation utilizing vehicle-fixed sensors, smartphone-based automotive navigation is constrained by the fact that the sensor measurements depend not only on the vehicle dynamics, but also on the orientation, position, and movements of the smartphone relative to the vehicle. In the following, we describe what this means for a designer of a smartphone-based automotive navigation system, and review methods for estimating the orientation and position of the smartphone with respect to the vehicle. Subsequently, we discuss supplementary information sources and how these can be used to improve the navigation solution.

\begin{table}[t]
\small
\begin{center}
\caption{publications discussing smartphone-to-vehicle alignment. \label{TableS2B}}
\begin{tabular}{l|l}
\hline
\hline
\multicolumn{1}{l|}{} & \multicolumn{1}{l}{} \\ [-2ex]
Sensors & Publications \\
\hline
\multicolumn{1}{l|}{} & \multicolumn{1}{l}{} \\ [-2ex]
Accelerometer & \cite{Dai2010,Paefgen2012,Banerjee2014,Xianping2011,Osafune2016,Li2016} \\
Accelerometer, GNSS & \cite{Almazan2013,Mohan2008,Hong2014,Li2012,Sharma2016,Chowdhury2016,Pfriem2014} \\
Accelerometer, GNSS, Magnetometer & \cite{Wahlstrom2015_3,Bhoraskar2012,Zong2015,Ghose2016,Zhao2013}\\
Gyroscope & \cite{Woo2016} \\
IMU & \cite{Wang2013,Yu2016,He2014} \\
IMU, GNSS & \cite{Wahlstrom2015,Wahlstrom2016,Wallin2013,Larsdotter2014} \\
IMU, Magnetometer & \cite{Kang2014,Bruwer2015} \\
IMU, GNSS, Magnetometer & \cite{Promwongsa2014,Khalegi2015} \\
\multicolumn{2}{l}{} \\ [-2.55ex]
\hline \hline
\multicolumn{2}{l}{}  \\
\multicolumn{1}{l}{} & \multicolumn{1}{l}{} \\ [-1ex]
\end{tabular}
\vspace*{-8mm}
\end{center}
\end{table}

\textit{1) Smartphone-based Vehicle Navigation and Smartphone-to-Vehicle Alignment}: Low-cost automotive navigation is often performed by employing a GNSS-aided inertial navigation system (INS), fusing measurements from a GNSS receiver and an IMU. Typically, a GNSS-aided INS will provide three-dimensional estimates of the vehicle's position, velocity, and attitude (orientation) \cite{Groves2008}. The estimates have the same update rate as the IMU measurements; refer to \cite{Niu2012,Gikas2016,Forster2012} for examples of smartphone-based GNSS-aided INSs. Since the measurements originate from sensors within the smartphone, a smartphone-based navigation system will provide estimates of the smartphone's, rather than of the vehicle's, dynamics and position. For the position and velocity estimates, the difference is typically negligible (when considering applications such as driver navigation and speed compliance) as long as the smartphone remains fixed to the vehicle. However, the attitude of the smartphone may obviously be very different from the attitude of the vehicle. To construct estimates of the vehicle's attitude based on estimates of the smartphone's attitude, one typically needs some knowledge of the smartphone-to-vehicle orientation. In \cite{Niu2012}, the smartphones were fixed to the vehicle during the entire field test, and the smartphone-to-vehicle orientation was estimated using measurements from a vehicle-fixed tactical-grade IMU that had already been aligned to the vehicle frame. In most practical applications though, measurements from IMUs that have been pre-aligned to the vehicle will not be available.
\begin{figure*}[t]
\def\svgwidth{7.18in}
\hspace*{-0.5mm}
\scalebox{0.94}{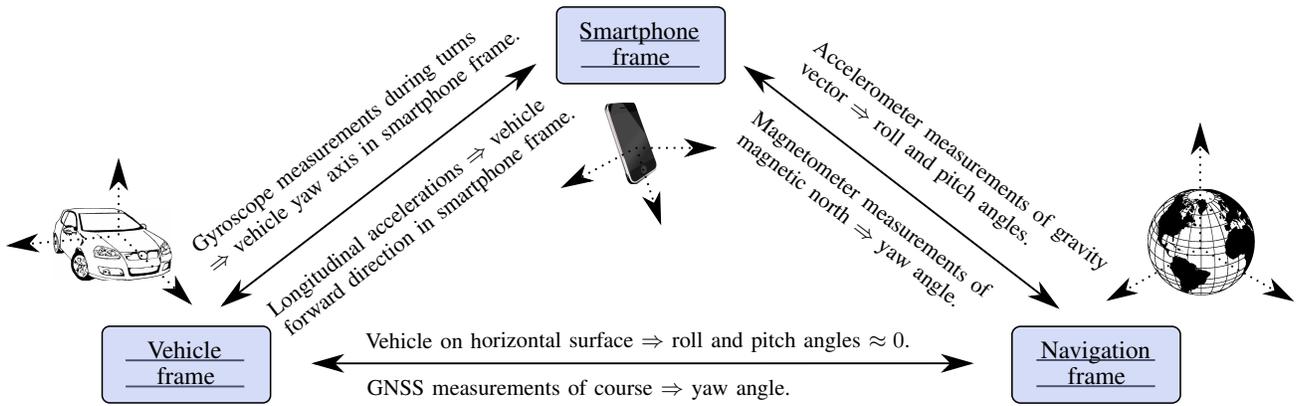}
\caption{\label{S2V_figure}The three coordinate frames of interest in smartphone-based automotive navigation and methods that can be applied to infer their relation.}
\vspace*{-3.5mm}
\end{figure*}
As a consequence, several stand-alone solutions for smartphone-to-vehicle alignment (estimation of the smartphone-to-vehicle orientation) have been proposed. In Table \ref{TableS2B}, publications discussing smartphone-to-vehicle alignment are categorized based on the employed sensors. Often, the estimation will use IMU measurements and exploit the fact that the vehicle's velocity is roughly aligned with the forward direction of the vehicle frame \cite{Wahlstrom2015}. Moreover, the smartphone's roll and pitch angles can be estimated from accelerometer measurements during zero acceleration \cite{Li2012}, the smartphone's yaw angle can be estimated from magnetometer measurements \cite{Paefgen2012}, the vehicle's roll and pitch angles can in many cases be assumed to be approximately zero \cite{Wahlstrom2015_3}, and the vehicle's yaw angle can be estimated from GNSS measurements of course (or consecutive GNSS measurements of position)\footnote{Obviously, estimates of the smartphone's and the vehicle's attitude with respect to some chosen navigation frame can be directly transformed into estimates of the smartphone-to-vehicle orientation.} \cite{Khalegi2015}. Alternatively, the smartphone-to-vehicle orientation can be estimated by applying principal component analysis to 1) accelerometer measurements \cite{Larsdotter2014}: most of the variance is (due to forward acceleration and braking) concentrated along the longitudinal direction of the vehicle frame, and most of the remaining variance is concentrated along the lateral direction of the vehicle frame (with large accelerations in this direction associated with vehicle turns); or 2) gyroscope measurements \cite{Woo2016}: most of the variance is concentrated along the vehicle's yaw axis (with large angular velocities in this direction associated with vehicle turns), and most of the remaining variance is concentrated along the vehicle's pitch axis (with large angular velocities in this direction associated with changes in the inclination of the road). The three coordinate frames and the methods to infer their relation are summarized in Fig. \ref{S2V_figure}.

Once the smartphone-to-vehicle orientation has been estimated, the IMU measurements can be rotated to the vehicle frame, thereby making it possible to estimate, e.g., the acceleration in the vehicle's forward direction directly from the IMU measurements (see Section \ref{section_driverbehavior}). Although smartphone-to-vehicle alignment can normally be performed with good precision when the smartphone is fixed with respect to the vehicle (the errors presented in \cite{Wahlstrom2015} were in the order of $2\,[\hspace*{0.2mm}^\circ\hspace*{0.2mm}]$ for each Euler angle), the process is complicated by the fact that the smartphone-to-vehicle orientation may change at any time during a trip. Moreover, the accelerations and angular velocities experienced by the smartphone when it is picked up by a user are often orders of magnitude larger than those associated with normal vehicle dynamics \cite{Wahlstrom2015_3}. Therefore, the vehicle dynamics observed in measurements collected by smartphone-embedded IMUs tend to be obscured by the dynamics of the hand movements whenever the smartphone is set in motion with respect to the vehicle. Moreover, the high dynamics of the human motions may excite, e.g., scale factor errors, thereby rendering continued inertial navigation all the more difficult. Solutions presented in the literature have discarded the IMU measurements during detected periods of smartphone-to-vehicle movements, and then re-initialized the inertial navigation once the smartphone is detected to be fixed with respect to the vehicle again \cite{Mohan2008}. Obviously, GNSS data can still be used for position and velocity estimation, even when the IMU measurements are discarded. However, stand-alone GNSS navigation suffers from both its low update rate and its inability to provide attitude estimates.

\begin{table*}[t]
\small
\begin{center}
\caption{publications on smartphone-to-vehicle positioning and driver/passenger classification. \label{Tab2}}
\begin{tabular}{l|lllllll}
\hline
\hline
& & & & \multicolumn{3}{|l|}{} & \\ [-2.1ex]
& \multicolumn{3}{c}{Sensors} & \multicolumn{3}{|c|}{Classification features} & \\ [-0.2ex]
& & & & \multicolumn{3}{|l|}{} & \\ [-2.3ex]
\arrayrulecolor{black}\cline{2-4}\cline{5-7} \\ [-2.1ex]
Pub. & IMU & mag.$^{\dag}$ & mic.$^{\dag}$ & \multicolumn{1}{|l}{Human motion} & Vehicle dynamics &  \multicolumn{1}{l|}{Other} & Assumptions/requirements \\
& & & \multicolumn{5}{l}{} \\ [-2.6ex]
\hline
& & & \multicolumn{5}{l}{} \\ [-2.1ex]
\cite{Chu2014}, & x & x & x & Vehicle entry, & & Audio ranging & The smartphone is in the \\[-0.2ex]
\cite{Pote2016} & & & & pedal pressing, & & based on & pocket or handbag of the \\[-0.2ex]
& & & & seat belt fastening. & & turn signals. & driver. \\
\arrayrulecolor{gray}\hline\arrayrulecolor{black} \\ [-2.1ex]
\cite{Bo2016} & x & x & & Vehicle entry. & Pothole crossing. & Recognition of &  The smartphone is in the \\[-0.2ex]
& & & & & & magnetic field & pocket of the driver.\\[-0.2ex]
& & & & & & from engine. & \\
\arrayrulecolor{gray}\hline\arrayrulecolor{black} \\ [-2.1ex]
\cite{Wang2013}$^{\ddagger}$ & x & & & & Centripetal & & Data is available from one \\[-0.2ex]
& & & & & acceleration. & & smartphone and from OBD \\[-0.2ex]
& & & & & & & or additional smartphones. \\
\arrayrulecolor{gray}\hline\arrayrulecolor{black} \\ [-2.1ex]
\cite{He2014} & x & & & & Centripetal & & Data is available from \\ [-0.2ex]
& & & & & acceleration, & & at least two smartphones. \\[-0.2ex]
& & & & & pothole crossing. & & \\
\arrayrulecolor{gray}\hline\arrayrulecolor{black} \\ [-2.1ex]
\cite{Wahlstrom2016} & x & & & & Any dynamics. & & Data is available from \\ [-0.2ex]
& & & & & & & at least two smartphones. \\
\arrayrulecolor{gray}\hline\arrayrulecolor{black} \\ [-2.1ex]
\cite{Yang2011}, & & & x & & & Audio ranging. & Vehicle-installed Bluetooth \\[-0.2ex]
\cite{Yang2012} & & & & & & & hands-free system. \\
\arrayrulecolor{gray}\hline\arrayrulecolor{black} \\ [-2.1ex]
\cite{Feld2010}$^{\ast}$ & & & x & & & Audio ranging. & Four vehicle-installed \\[-0.2ex]
& & & & &  & & microphones available. \\ [-0.2ex]
& & & & &  & & Access to a training set\\ [-0.2ex]
& & & & &  & & for voice recognition.\\
\hline \hline
\multicolumn{8}{l}{} \\ [-1.9ex]
\multicolumn{8}{l}{\footnotesize{$^{\dag}$ Abbreviations of magnetometer (mag) and microphone (mic).}} \\ [-0.2ex]
\multicolumn{8}{l}{\footnotesize{$^{\ddagger}$ The study only estimates the smartphone's lateral position in the vehicle frame.}} \\ [-0.2ex]
\multicolumn{8}{l}{\footnotesize{$^{\ast}$ Although the work was motivated by customization of personal devices such as smartphones, the sensing system functions}} \\ [-0.2ex]
\multicolumn{8}{l}{\footnotesize{independently of the devices.}}
\vspace*{-2mm}
\end{tabular}
\end{center}
\end{table*}

\textit{2) Smartphone-to-Vehicle Positioning}: The problem of estimating the smartphone's position with respect to the vehicle is closely related to the problem of driver/passenger classification, i.e., determining whether a smartphone belongs to the driver or a passenger of the vehicle. Usually, it is assumed that the smartphone is placed in the vicinity of its owner, and hence, smartphone-to-vehicle positioning makes it possible to assess whether the owner of a given smartphone is also the driver of the vehicle. While neither of these problems are critical to vehicle positioning, driver/passenger classifications are of use in several applications, including distracted driving solutions \cite{Wang2013}, and insurance telematics \cite{Chu2014}. A large number of classification features have been considered. One alternative is to study human motions as measured by the smartphone. Studied motions have included vehicle entries \cite{Bo2016}, seat-belt fastening, and pedal pressing \cite{Chu2014}. The dynamics of the two former motions will typically differ depending on which side of the vehicle the user entered from. Similarly, the detection of pedal pressing always indicates that the user sat in the driver's seat. While these features possess some predictive power, they are susceptible to variations in the motion behavior of each individual. In addition, they are constrained by the assumption that the user carries the smartphone in his pocket. A second option is to utilize that the specific force, measured by accelerometers, will vary depending on where in the vehicle the accelerometers have been placed \cite{He2014}. The effect is most clearly seen during high-dynamic events such as when passing a pothole or during heavy cornering \cite{Wahlstrom2016}. Typically, the measurements from a smartphone-embedded accelerometer must be compared with measurements from an additional smartphone (that has a different position in the vehicle) or from the OBD system\footnote{Smartphone-to-vehicle positioning based on comparison of smartphone-based IMU measurements with OBD measurements of vehicle speed hinges on the assumption that the vehicle trajectory can be approximated with a circle \cite{Wang2013}.}.
Since the difference in the specific force measured by two accelerometers only depends on their relative position, and not on their absolute positions, methods of this kind can only be used for absolute positioning within the vehicle when the absolute position of one of the sensor nodes is already known \cite{Wahlstrom2016}. A third option is to utilize audio ranging. Proposed frameworks have used smartphone-embedded microphones to recognize, e.g., the vehicle's turn signals \cite{Chu2014}, or high frequency beeps sent through the car's stereo system \cite{Yang2011}, \cite{Yang2012}. The latter positioning systems are in general very accurate. However, the required vehicle infrastructure (Bluetooth hands-free systems) can only be found in high-end vehicles, and thus, implementations in older vehicle models necessitate expensive aftermarket installations \cite{Wang2013}. A fourth option is to estimate the relative position of two devices from their GNSS measurements. However, since the errors normally are heavily correlated in time and often exceed the typical device distance, convergence can be expected to be slow \cite{Wahlstrom2015_3}. A fifth option is to utilize vehicle-installed technology to directly identify the driver and the passengers, without making use of any smartphone sensors. If needed, the data or the results from the driver/passenger classification can then be communicated to all personal devices in the vehicle and be used to, e.g., customize their settings based on the design of distracted driving solutions. This of course assumes that each smartphone already has been associated with its owner's identifying characteristics. One way to implement a system of this kind is to record speech using directional microphones attached to each seat \cite{Feld2010}. Each smartphone can then compare the recorded audio with its own unique pre-stored voice signature, and can thereby identify the seat of the smartphone owner. Some additional approaches to driver/passenger classification are to identify drivers from their driving characteristics, so called driver fingerprinting \cite{Zhang2016}, to position smartphones in the vehicle using Bluetooth or near field communication \cite{Chu2014}, to use smartphone cameras \cite{Paruchuri2015}, or to detect texting while driving based on texting characteristics \cite{He2015}. A selection of publications related to driver/passenger classification are summarized in Table \ref{Tab2}.

\textit{3) Supplementary information sources}: Several supplementary information sources can be employed to improve the navigation estimates. These include road maps, road features, and dynamic constraints.
By using the added information to aid an INS, it is often possible to reduce the dependence on GNSS availability. Next, we discuss the three mentioned information sources and how they can be integrated with sensor measurements. Map-matching (MM) algorithms use location-based attributes (speed limits, restrictions on travel directions, etc.) and spatial road maps together with sensor measurements to infer locations, links (arcs between nodes in a road map), or the path of a traveling vehicle \cite{Quddus2007}. The algorithms are broadly categorized as online or offline, with the former primarily being used for turn-by-turn navigation, and the latter being used as, e.g., input in traffic analysis models. Although GNSS has been established as the dominant source of input data to map-matching algorithms, cellular positioning has been employed in several instances \cite{Dong2013}. In addition, IMU or odometer measurements can be used to increase the estimation rate and bridge GNSS gaps. Since many MM algorithms have been developed for navigation products using data from dedicated GNSS receivers, they often have to be modified to cope with the low density and poor performance of GNSS data from smartphone-embedded receivers \cite{Bierlaire2013}. Today, digital road maps are created using both professional methods based on, e.g., satellite imagery, and using crowdsourcing by collecting GNSS traces from smartphone-equipped drivers. The gains of crowdsourcing have been magnified by the ubiquity of smartphones, thereby motivating many navigation service providers to replace the commercial maps used in their products with crowdsourced open license alternatives. One of the most popular crowdsourcing platforms is OpenStreetMap (OSM), which benefits from a large community of contributing users enabling rapid updates to geographical changes \cite{Haklay2008}. However, just as many other crowdsourced databases, OSM, to some extent, suffers from logical inconsistency (duplicated lines, dead-ended one-way-roads, etc.) \cite{Hashemi2015}, and incomplete coverage (missing or simplified objects) \cite{Haklay2010}.

Artificial position measurements can be obtained by detecting road features associated with map locations. Typically, IMUs are used for the detection. Previously studied road features within smartphone-based automotive navigation include traffic lights \cite{Hu2013}, bridges \cite{Aly2015}, tunnels \cite{Bo2013b}, speed bumps \cite{Tan2014}, potholes \cite{Yu2016}, and turns \cite{Hu2015}. As demonstrated in \cite{Aly2016}, driving events such as turns, lane changes, and potholes can be used for smartphone-based lane-level positioning. The problem of building a map of road features was considered in \cite{Hu2013}, which used GNSS traces collected from smartphones to detect traffic lights or stop signs at intersections. In the future, we can expect more implementations to make use of the vast research on simultaneous localization and mapping (SLAM) conducted by the robotics community.

Constraints on the vehicle dynamics are employed to reduce the dimension of possible navigation solutions, and thereby improve the estimation accuracy. For example, the estimated altitude of the vehicle can be constrained when the vehicle is expected to travel on an approximately flat surface \cite{Godha2005b}, the vehicle's speed and angular velocity can be constrained to comply with the minimal turning radius \cite{Niu2010}, and the estimated velocity can be assumed to be aligned with the forward direction of the vehicle frame \cite{Wahlstrom2015}.

Last, we note that measurements from smartphone-embedded motion sensors and positioning technologies often display signs of pre-processing, which can affect the navigation capabilities of the device. One potential reason for this is that the sensor measurements have been filtered through built-in estimation filters, designed to, e.g., mitigate sensor failures or increase accuracy. As a result, the measurements can may display signs of latency or temporal error correlation \cite{Marti2014}. Unfortunately, the transparency of built-in estimation filters is typically low.

\subsection{Transportation Mode Classification}
\label{section_TMC}

As opposed to in-vehicle sensors, smartphone-embedded sensors can be used to collect data not only on automotive driving, but also on pedestrian activities, bus rides, train rides, etc. Hence, for those smartphone-based vehicle telematics applications where the primary interest lies in data collected while traveling in a specific vehicle type, accurate transportation mode classification is an essential capability. For example, in insurance telematics and participatory bus arrival time prediction systems, the primary interest typically lies in car trips \cite{Wahlstrom2015_3} and bus trips \cite{Zhou2014b}, respectively. Smartphone-based transportation mode classification have also been employed for general trip reconstruction, which may be used for health monitoring or in studies of travel behavior \cite{Reddy2010}. While it would be theoretically possible to let the smartphone users manually label their trips or start and stop the data collection, the resulting burden would discourage many people from using the app to begin with \cite{Handel2014_2}.

In the days before the breakthrough of the smartphone, mobile-based transportation mode classification only utilized cellular positioning \cite{Sohn2006,Anderson2006}. Today, the classification primarily relies on GNSS receivers and accelerometers. The classification methods can be divided into heuristic rule-based approaches and machine-learning methods \cite{Bolbol2012}, \cite{Stenneth2012}. In some cases, the algorithm will in a first stage attempt to detect the time points when the user switches from one mode to another, and in a second stage classify each segment of data between two such identified switches. Obviously, the accuracy of the classification is in these cases constrained by the accuracy of the mode-switch detection \cite{Bolbol2012}. Generally, it is comparatively easy to separate motorized modes from non-motorized, and hence, many studies will first make a coarse classification by attempting to detect all motorized data segments \cite{Hemminki2013,Thiagarajan2010}. The classification benefits from that pedestrians and bicyclists have more freedom to change their direction than automobiles \cite{Lari2015}, and that the typical speeds achieved when, say, riding in passenger cars and walking, tend to be very different \cite{Assemi2016}.

There are several measurement features that can be used to separate different motorized modes. For example, the acceleration and braking frequency of disconnected vehicles that move alongside other traffic (i.e., cars, buses and trams) tends to be vastly different from that of vehicles moving independently of other traffic (i.e., trains) \cite{Hemminki2013}. Similarly, \cite{Shin2014} illustrated that the distributions of mean (as taken over a complete trip) absolute accelerations can be expected to be very different for trains, trams, cars, and buses. Moreover, cars are more prone to be involved in quick driving maneuvers than larger vehicles \cite{Hemminki2013}, often reach higher speeds on the freeway \cite{Assemi2016}, and does typically not, as opposed to public transport, permit walking inside the vehicle. Of course, public transport also tends to make a larger number of stops per driven kilometer. Furthermore, if information on timetables, bus routes, or rail maps are available, this will clearly simplify the identification of the corresponding transportation modes \cite{Thiagarajan2010,Montoya2015}. It may also be possible to make use of semantic information, such as home and work location, and information on behavioral patterns \cite{Shin2014}. Likewise, one may use that if the last recorded car trip ended at a given location, it is likely that the next recorded car trip starts at the same location. In \cite{Gessulat2013}, it is also proposed that one could use prior probabilities on different transportation modes that are dependent on the current day of the week, the current month, or the weather. For example, people are more likely to walk longer distances in clear weather than they are when it is raining.

Proposals on how to reduce the energy consumed by smartphones collecting data for transportation mode classification have included sparse GNSS sampling \cite{Bolbol2012}, implementations only utilizing inertial sensors \cite{Eftekhari2016,Hemminki2013}, and an increased use of cellular positioning. In the last case, one option is to let changes in the cellular position estimates indicate when the user moves from indoor to outdoor settings. A GNSS fix does then only have to be attempted when the user is expected to be outdoors \cite{Reddy2010}. Last, we note that it is possible to label trips made with a specific vehicle by installing dedicated vehicle-fixed devices that can communicate with the smartphone. Previous devices used for this purpose have included iBeacons \cite{Li2016c}, smart battery chargers \cite{Handel2014_2}, or near-field communication tags mounted on a smartphone cradle \cite{Handel2014_2}.

\subsection{Driver Behavior Classification}
\label{section_driverbehavior}

\begin{figure}[t]
\def\svgwidth{3.38in}
\hspace*{-4.5mm}
\scalebox{0.94}{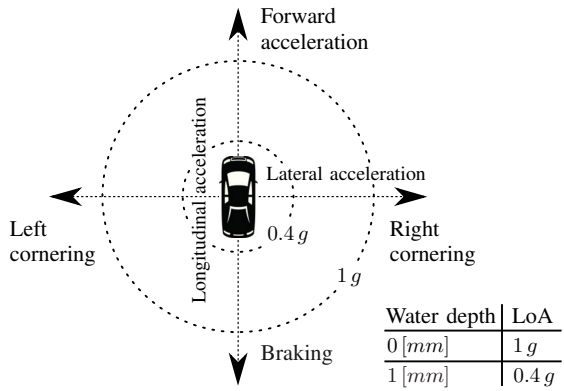}
\caption{\label{frictionfigure}Friction circle (also known as the ellipse of adherence) illustrating the limits of adhesion (LoA) on a vehicle's acceleration in terms of gravity $g\approx 9.8\,[m/s^2]$ when driving at a velocity of $50\,[km/h]$ on a horizontal surface with worn tires and a tread depth of $1.6\,[mm]$ \cite{Girling1996}.}
\vspace*{-3.5mm}
\end{figure}

The literature on driver behavior classification is vast and encompasses a wide range of sensors and algorithms. Here, the primary focus is on smartphone-based implementations.
What separates smartphone-based methods from other methods is that the former primarily relies on GNSS, accelerometer, gyroscope, and magnetometer data, whereas the latter more often also will make use of OBD data extracted from the controller area network (CAN) bus, vehicle-mounted cameras, electroencephalograms (EEGs), or other high-end sensors \cite{Kaplan2015,Engelbrecht2015}.

Assessments of driver safety are usually performed by detecting driving events that are perceived as dangerous or aggressive. Two of the most commonly studied categories of driving events, which also are of particular interest in insurance telematics \cite{Progressive2012}, are harsh acceleration\footnote{Here, we use harsh acceleration to refer to both aggressive forward acceleration and harsh braking.} and harsh cornering. Today, it is widely accepted that drivers who frequently engage in harsh acceleration and harsh cornering also tend to be involved in more accidents \cite{Klauer2009}, \cite{Osafune2016}, get more traffic tickets \cite{Hong2014}, and drive in a less eco-friendly manner \cite{Yamakado2009}. The intimate connection between harsh acceleration and harsh cornering is made clear by the fact that the study of harsh cornering usually is considered to be synonymous with the study of horizontal (longitudinal and lateral) acceleration. As shown in \cite{Wahlstrom2015_2}, this can be motivated by a kinematic analysis of vehicle rollovers and tire slips. Often, the resulting constraints on a vehicle's acceleration are illustrated using a friction circle as displayed in Fig. \ref{frictionfigure}.
\begin{figure}[t]
\def\svgwidth{3.38in}
\hspace*{-3.5mm}
\scalebox{0.94}{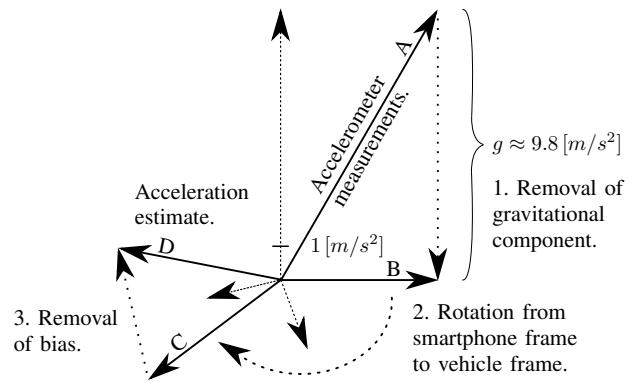}
\caption{\label{accrot_figure}Illustration of how to construct acceleration estimates in the vehicle frame (D) from smartphone-based accelerometer measurements (A) by 1) removing the gravitational component; 2) rotating the measurements to the vehicle frame; and 3) compensating for bias. The order of the three operations may be modified.}
\vspace*{-3.5mm}
\end{figure}
Although both longitudinal and lateral acceleration estimates can be obtained by differentiating GNSS measurements \cite{Handel2014,Saiprasert2013b,Saiprasert2014b}, the higher sampling rate of accelerometers generally makes them more suitable for the detection of short-lived acceleration events \cite{Saiprasert2015}, \cite{Bruwer2015b}. As illustrated in Fig. \ref{accrot_figure}, the computation of acceleration estimates in the vehicle frame from smartphone-based accelerometer measurements can be carried out in three steps \cite{Bruwer2015}. First, the gravitational component needs to be removed from the measurements. To do this, one must estimate the attitude of the smartphone with respect to the earth (using, e.g., a GNSS-aided INS). Since the gravitational component is larger than typical vehicle accelerations, an improper removal may significantly distort the acceleration estimates. Second, the measurements need to be rotated from the smartphone frame to the vehicle frame. This requires an estimate of the smartphone-to-vehicle orientation (see Section \ref{subsection_navigation}). Third, one needs to compensate for measurement errors. Often, it will be sufficient to consider bias and noise terms. One solution that can be used to simultaneously estimate the smartphone's attitude, the smartphone-to-vehicle orientation, and the sensor biases was presented in \cite{Wahlstrom2015}. The impact of random noise can be mitigated by the use of low-pass filters\footnote{Most of a passenger vehicle's horizontal accelerations will be in the frequency spectrum below $2\,[H\!z]$ \cite{Skog2009}.}. Although much of the literature on smartphone-based detection of harsh acceleration and harsh cornering has been focused on the estimation of horizontal acceleration, several alternative methods have been proposed. For example, harsh cornering may be detected by using magnetometer \cite{Castignani2015} or gyroscope \cite{Johnson2011} measurements to estimate the vehicle's angular velocity; in \cite{Stoichkov2013}, harsh braking and cornering was detected by thresholding the vehicle's pitch and roll angles, respectively; and the authors of \cite{Chen2015} and \cite{Bergasa2014} used microphones to detect whether turn signals were employed during critical steering maneuvers (if not, the maneuver was classified as careless).

\begin{table*}[t]
\small
\begin{center}
\caption{publications on smartphone-based driver safety classification. \label{Tab1}}
\begin{tabular}{l|llllllllll}
\hline
\hline
& \multicolumn{4}{l|}{} & \multicolumn{2}{c|}{} & \multicolumn{3}{l|}{} & \multicolumn{1}{l|}{} \\ [-2.3ex]
& \multicolumn{4}{c|}{Smartphone sensors} & \multicolumn{2}{c|}{External sensors} & \multicolumn{3}{l|}{Driving events$^{\dag}$} & \multicolumn{1}{l|}{} \\ [-2.1ex]
& \multicolumn{4}{c|}{} & \multicolumn{2}{c|}{} & \multicolumn{3}{l|}{}  & \multicolumn{1}{l|}{} \\ [-0.4ex]
& \multicolumn{4}{l|}{} & \multicolumn{2}{l|}{} & \multicolumn{3}{l|}{} & \multicolumn{1}{l|}{} \\ [-2.6ex]
\cline{2-5}\cline{6-7}\cline{8-10} \\ [-2.2ex]
& & & & \multicolumn{1}{c|}{} & & \multicolumn{1}{l|}{} & \multicolumn{3}{l|}{} & \multicolumn{1}{l|}{} \\[-2.4ex]
Publications & GNSS & Acc.$^{\dag}$ & Gyro.$^{\dag}$ & \multicolumn{1}{c|}{Mag.} & OBD & \multicolumn{1}{l|}{IMU} & HA & HC & \multicolumn{1}{l|}{LC} & \multicolumn{1}{l|}{Detection/classification method} \\
& \multicolumn{8}{l}{} & \\ [-2.5ex]
\hline
& \multicolumn{8}{l}{} & \\ [-2.05ex]
\cite{Chowdhury2015,Chowdhury2015b,Banerjee2016} & x & & & & & & x & & & Thresholding. \\
\cite{Vaiana2014} & x & & & & & & x & x & & Thresholding. \\
\cite{Wahlstrom2014,Wahlstrom2015_2} & x & & & & & & & x & & Thresholding. \\
\cite{Osafune2016} & x & x & & & & & x & x & & Thresholding/SVM. \\
\cite{Akhtar2014} & x & x & & & & & x & & x & Thresholding. \\
\cite{Chen2015} & x & x & x & & & & & x & x & Thresholding. \\
\cite{Daptardar2015} & x & x & x & & & & x & x & x & Thresholding/hidden Markov model. \\
\cite{Engelbrecht2014,Engelbrecht2015b} & x & x & x & & & & & x & & Thresholding/DTW/NBC. \\
\cite{Bergasa2014} & x & x & x & & & & x & x & & Thresholding. \\
\cite{Pholprasit2015} & x & x & x & x & & & x & x & x & Thresholding/DTW. \\
\cite{Bruwer2015b} & x & x & x & x & & & x & x & x & Thresholding. \\
\cite{Castignani2013,Castignani2015} & x & x & & x & & & x & x & x & Thresholding. \\
\cite{Saiprasert2013b,Saiprasert2014b,Saiprasert2015} & x & x & & x & & & x & x & x & Thresholding/DTW. \\
\cite{Buscarino2014} & & x & & & & & x & & & Thresholding. \\
\cite{Li2016} & & x & & & & & x & x & x & Thresholding. \\
\cite{Zhao2013} & & x & & & & & x & x & & Thresholding. \\
\cite{Fazeen2012} &  & x & & & & & x & & x & Thresholding. \\
\cite{Dai2010} & & x & & & & & x & & & Thresholding. \\
\cite{Kalra2014b} & & x & & & & & x & x & & Thresholding. \\
\cite{Chigurupati2012} & & x & & & & & & x & x & Thresholding. \\
\cite{Chakravarty2013} & & x & & & & & x & x & & Thresholding. \\
\cite{Kang2014} & & x & x & & & & x & x & x & Thresholding. \\
\cite{Stoichkov2013} & & x & x & x & & & x & x & x & Thresholding. \\
\cite{Khedkar2015} & & x & x & & x & & x & x & x & Thresholding. \\
\cite{Ylizaliturri-Salcedo2015} & & x & & & x & & x & x & x & Thresholding. \\
\cite{Ouyang2016} & & & x & & & & & x & x & Thresholding. \\ [-0.3ex]
\cite{AbuAli2015}$^{\ddagger}$ & & & & & x & & x & & & Thresholding. \\ [-0.3ex]
\cite{Carmona2015}$^{\ddagger}$ & & & & & x & x & x & x & x & Thresholding. \\
\cite{Hosseinioun2015} & x & x & & & & & x & x & & Various machine learning methods. \\
\cite{Tchankue2013} & x & x & x & x & & & x & x & x & Various machine learning methods. \\
\cite{Chaovalit2013} & x & x & & x & & & x & x & & Symbolic aggregate approximation. \\
\cite{Castignani2015b} & x & x & & x & & & x & x & & Maximum likelihood estimation. \\
\cite{Hong2014} & x & x & & & x & x & x & x & & NBC. \\
\cite{Woo2016} & x & & x & & & & x & x & & SVM (to classify maneuvers). \\ [-0.05ex]
\cite{Bhoyar2013} & & x & & & & & x & x & x & Pattern matching. \\
\cite{Antoniou2014}$^*$ & & x & & & & & & & & k-means clustering. \\ [-0.05ex]
\cite{Meseguer2013}$^*$ & & x & & & x & & & & & Artificial neural network. \\
\cite{Chen2015b} & & x & x & x & & & x & x & x & SVM (to classify maneuvers). \\
\cite{Eren2012} & & x & x & x & & & x & x & x & DTW/Bayesian classification.\\
\cite{Johnson2011} & & x & x & x & & & x & x & x & DTW/k-nearest neighbors. \\
\cite{Tecimer2013} & & x & x & x & & & x & x & & Various machine learning methods. \\ [-0.3ex]
\cite{Ly2013}$^{\ddagger}$ & & & & & x & x & x & x & & SVM, k-means clustering. \\
\hline \hline
\multicolumn{11}{l}{}
\\ [-2ex]
\multicolumn{11}{l}{\footnotesize{The table only specifies the sensors that was used for the detection of aggressive acceleration, cornering, and lane changing.}} \\ [-0.2ex]
\multicolumn{11}{l}{\footnotesize{$^{\dag}$ Abbreviations: Accelerometers (Acc.), gyroscopes (Gyro.), harsh acceleration (HA), harsh cornering (HC), and lane changing (LC).}} \\ [-0.2ex]
\multicolumn{11}{l}{\footnotesize{$^{\ddagger}$ Although no smartphone sensors are employed, the study is motivated by smartphone-based driver classification.}} \\ [-0.2ex]
\multicolumn{11}{l}{\footnotesize{$^{*}$ The algorithm did not detect individual driving events per se, but rather provided a general characterization of the driving style.}}
\vspace*{-0mm}
\end{tabular}
\end{center}
\end{table*}

Another category of driving events that are of interest in driver assessments is lane changing. In similarity with cornering events, lane changes are characterized by pronounced lateral accelerations. To separate these two types of events using sensor measurements, one may use that 1) the vehicle's yaw angle will be approximately the same before and after completing a lane change along a straight road \cite{Kang2014}; and 2) during a lane change, the accumulated displacement along the direction perpendicular to the road is approximately equal to the lane width \cite{Chen2015}. In general, neither of these statements will hold for cornering events. When position measurements and map information are available, the separation obviously becomes easier. Sudden or frequent lane changes are often associated with aggressive driving \cite{Kang2014}, and have been reported to account for up to $10$\% of all vehicle crashes \cite{Lee2004}.

Two dominant approaches to the detection of aggressive driving events have emerged. In some studies, these will naturally extend to an overall driver risk assessment, whereas others only discuss the classification of individual driving events. The first alternative approach to the detection of driving events is to use thresholding or some type of fuzzy inference \cite{Kang2014}. In other words, one would construct rule-based methods based on the validity of a given set of statements regarding the measurements. For example, a harsh braking could be considered to have been detected whenever the estimated longitudinal acceleration falls below (assuming that negative values correspond to decelerations) some predetermined threshold. The second alternative is to study the temporal regularities of the measurements or related test statistics. For example, a harsh braking could be considered to have been detected whenever the IMU measurements over a specified time window are sufficiently similar to some predetermined and manually classified reference templates. A popular method used to assess the similarity between two driving events is dynamic time warping (DTW) \cite{Johnson2011,Eren2012,Engelbrecht2015b,Saiprasert2015}. DTW is a time-scale invariant similarity measure, and hence, it enables correct event classifications despite variations in event length. A comparison of DTW and a threshold-based method was conducted in \cite{Saiprasert2013b,Saiprasert2015}, indicating a performance difference in favor of DTW. However, in \cite{Engelbrecht2015b}, DTW came up short against a naive Bayesian classifier (NBC) that used driving event features to classify events as aggressive or normal. In addition to NBC \cite{Hong2014}, driver safety classifications have also been based on machine learning approaches such as artificial neural networks \cite{Meseguer2013}, support vector machines (SVMs) \cite{Ly2013}, and k-nearest neighbors \cite{Tecimer2013}. A summary of studies on smartphone-based classifications of driver safety is provided in Table \ref{Tab1}. Notable related studies using black boxes or in-vehicle data recorders instead of smartphones can be found in \cite{Joubert2016} and \cite{Toledo2008}.

Harsh acceleration, cornering, and lane changing are far from the only driving events of interest in smartphone-based studies of driver safety. For example, swerving or weaving are important events that are similar in nature to lane changes \cite{Dai2010,Bergasa2014,Chigurupati2012}, and similarly, speeding has been considered in many related studies \cite{Whipple2009,Castignani2013,Chigurupati2012,Castignani2015,Carmona2015,Ghose2016,Bruwer2015b,AbuAli2015}. Although speeding detection is straightforward if GNSS measurements are available, IMU-only solutions to speed estimation are of interest for reasons of energy-efficiency, availability, and privacy. These solutions can be based on the relation between the vehicle's speed and the vehicle vibrations measured by inertial sensors \cite{Chakravarty2014,Lindfors2016}, or use that the vehicle speed is inversely proportional to the time that elapses between when the front and rear wheel passes a given road anomaly \cite{Bruwer2015b,Yu2016}. Moreover, the driver behavior can be further characterized by using accelerometers to estimate the current gear \cite{Handel2009}.

Smartphone-embedded cameras can be used to assess driver safety by tracking both the road \cite{Bergasa2014}, as well as the behavior of the driver \cite{Chuang2014}. More so, it is possible to combine these capabilities by rapidly switching between the smartphone's front and rear-facing cameras \cite{Singh2012}. This idea was used in \cite{You2013} to detect drowsy and inattentive driving, tailgating, lane weaving, and harsh lane changes. In \cite{Chen2015}, camera-free and camera-based methods for the detection of lane changing were compared in different weather conditions. The camera-free methods outperformed the camera-based both in terms of performance and computational cost.

\subsection{Road Condition Monitoring}
\label{section_RCM}

Poor road surface conditions can cause damage to vehicles, increase maintenance costs and fuel consumption, reduce driving comfort, and may even increase the risk of accidents \cite{Douangphachanh2014thesis}. In Britain alone, potholes cause more than $£1$ million worth of damages to cars every day \cite{Tecimer2013}. Due to the high costs associated with identifying and analyzing road deteriorations using specialized vehicles and equipment, the collection of crowdsourced data from smartphones in passenger vehicles has emerged as an attractive alternative \cite{Seraj2016}. Typically, related studies will threshold the standard deviation of measurements from smartphone-embedded accelerometers to detect potholes, cracks, speed bumps, or other road anomalies (with accompanying position estimates used to mark their location) \cite{Strazdins2011,Mednis2011,Kang2014,Ghose2012,Astarita2012,Kalra2014b,Darawade2016,Li2016,Mahajan2015b,Toke2016,Kulkarni2016,Akhtar2014,Yu2016}. Nonetheless, many variations to this approach have been presented.

Considering the sensor setup, we note that accelerometers in many instances been complemented with magnetometers \cite{Tecimer2013,Lanjewar2016,Tonde2015} and gyroscopes \cite{Seraj2016,Tecimer2013,Aly2015}. Other smartphone-based implementations have used 1) GNSS measurements of speed to attempt to remove the speed dependency in the accelerometer features describing a given anomaly \cite{Perttunen2011}; 2) microphones to record pothole-induced sound signals \cite{Mednis2010}; and 3) OBD suspension sensors to measure pothole-induced compressions \cite{AbuAli2015}. Although cameras have been popular in the general field of pothole detection \cite{Rajamohan2015,Orhan2013}, camera-based approaches are often considered too computationally expensive for smartphone-based implementations \cite{Wang2015}. Moving on to the road assessment, we note that there has been several proposals on how to increase the granularity of detection algorithms that just aim to identify road anomalies in general. For example, \cite{Seraj2016} used a SVM to differentiate between road-wide anomalies, such as speed bumps and railroad crossings, and one-sided potholes. As noted in \cite{Eriksson2008}, one-sided potholes tend to have a larger effect on the vehicle's pitch angle than road-wide anomalies. Moreover, studies which have aimed to detect potholes have proposed to threshold the speed to reject apparent anomalies caused by curb ramps (typically, curb ramps are approached at low speeds) \cite{Eriksson2008}. In \cite{Vittorio2014}, it was hinted that it is possible to estimate both the depth and length of potholes from accelerometer measurements. This information could then be used to evaluate the priority of specific repairs. Since many drivers will try to avoid driving over potholes, \cite{Seraj2015} noted that repeated instances of swerving at a given position indicates the presence of a pothole.

Although speed bumps are typically not in need of repairs, efforts have still been directed towards detecting and mapping them \cite{Jain2012,Mukherjee2016}. One of the motivations for this is the creation of early warning systems that could give the driver sufficient time to slow down also in, e.g., low visibility conditions. The estimation of bump height was previously discussed in \cite{Fazeen2012}. Instead of detecting individual anomalies, it is also possible to assess the overall road roughness
\cite{Alessandroni2014}, \cite{Tai2010}. Specifically, \cite{Douangphachanh2014thesis} and \cite{Forslof2015} studied how accelerometer measurements relate to the established international roughness index (IRI). Last, we note that while most algorithms for road condition monitoring rely on thresholding techniques, some studies have instead employed linear predictive coding \cite{Alessandroni2014}, SVMs \cite{Seraj2016,Perttunen2011,Tai2010,Mohamed2015}, k-means clustering \cite{Bhoraskar2012,Mahajan2015}, decision-tree classifiers \cite{Aly2015}, or Bayesian networks \cite{Tecimer2013}. Remaining challenges include 1) the development of standardized methods for the detection, classification, and characterization of road anomalies; 2) how to efficiently build a map of anomalies and reject spurious detections; and 3) how to model and consider the effect of vehicle suspension \cite{Mukherjee2016}.

\begin{figure}[t]
\centering
\def\svgwidth{2.2in}
\hspace*{-6mm}
\scalebox{0.94}{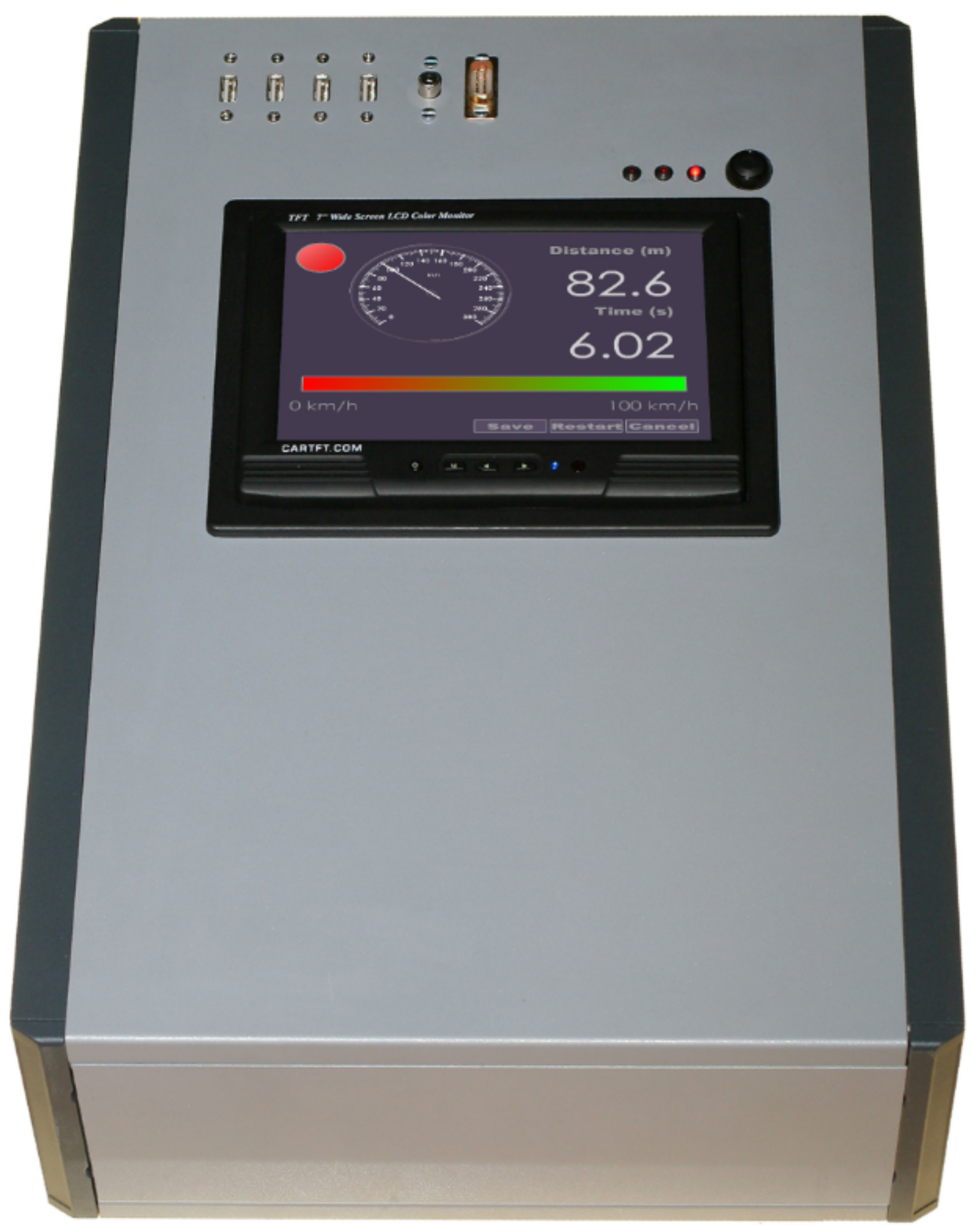}
\caption{\label{prototypefig}Historical flashback: Pre-smartphone era telematics device developed in 2006 with dimensions of $45\times 32\times 13\,[cm]$ \cite{Skog2006}. The device was equipped with a touch-screen, a 2G GSM modem, a USB-connected camera, a GNSS receiver, and an IMU \cite{Skog2006b}. The display shows the user interface of a non-intrusive vehicle performance application \cite{Handel2010}. }
\vspace*{-3.5mm}
\end{figure}

\section{Summary, historical perspective, and outlook}
\label{section_Conclusion}

The internet of things describes a world where personal devices, vehicles, and buildings are able to communicate, gather information, and take action based on digital data and human input. Some of the most promising applications are found within the field of smartphone-based vehicle telematics, which also was the topic of this survey. As demonstrated by services such as Uber and Waze, the idea of smartphone-based vehicle telematics is to employ the sensing, computation, and communication capabilities of smartphones for the benefit of primary (drivers), secondary (passengers), and tertiary users (bystanders, pedestrians, etc.).
The field has been particularly driven by the rapid improvement in smartphone technology seen over the last ten years. This development can be exemplified by the continuous addition of new sensors to devices from the iPhone series, which now come with a GPS-receiver (introduced in the iPhone 3G, in 2008), a digital compass (iPhone 3GS, 2009), a three-axis gyroscope (iPhone 4, 2010), and a GLONASS-receiver (iPhone 4S, 2011), and so forth. For comparison, Fig. \ref{prototypefig} shows a pre-smartphone research platform developed at KTH in 2006.

To begin with, the present study described the information flow that is currently utilized by, e.g., smartphone-based insurance telematics and navigation providers, and emphasized the characteristic feedback-loop that enables users to change their behavior based on the analysis of their data. Notable academic and industrial projects were reviewed, and system aspects such as embedded and complementary sensors, energy-efficiency, and cloud computing were examined. Moreover, it was concluded that while the smartphone potentially can function as an enabler for low-cost implementations of, e.g., VANETs and distracted driving solutions, there are often technical difficulties that have to be overcome due to the non-dedicated nature of the device. Since the smartphone-embedded sensors are not fixed with respect to the vehicle, we reviewed methods to estimate the smartphone's orientation and position with respect to some given vehicle frame. Similarly, we presented methods for labeling smartphone data according to transportation mode. Specifically, we noted that data collected from cars can be distinguished from public transport data by considering driving characteristics, timetables, bus routes, rail maps, and personal information. Further, publications on smartphone-based driver classification were categorized based on the used sensors, considered driving events, and applied classification methods. Last, methods for road condition monitoring were reviewed. While smartphones offer a cheap, scalable, and easily implementable alternative to current road monitoring methods, several methodological challenges still remain.

The abundance of smartphones in the traffic is expected to catalyse new advances in, e.g., crowdsourced traffic monitoring, platooning, ridesharing, car security, insurance telematics, and infotainment. However, although the breadth of the literature is continuously expanding, widespread adoption has so far been limited to a narrow set of applications. One of the most important reasons for this is that potential driving forces are often left unconvinced of the monetary or social value of specific ventures. For example, since few industrial partners have detailed data describing both driving behavior and the associated insurance claims over longer periods of time, the value of collecting driving data for insurance companies is still somewhat obscured. Similarly, there is a need to establish and validate standards for commonly encountered challenges such as transportation mode classification, smartphone-to-vehicle alignment, driver event detection, and driver scoring. Currently, the literature on these topics is very scattered, with many articles detailing ideas that have already been published elsewhere.
Moreover, future implementations may also benefit from improvements in sensor technology, communication standards, and road maps.
While some speculate that the increasing number of vehicles being equipped with factory-installed telematics systems will eventually make smartphone-based solutions obsolete, smartphones will continue to offer scalable and user-friendly telematics platforms for many years to come.





\ifCLASSOPTIONcaptionsoff
  \newpage
\fi



\bibliographystyle{IEEEtran}
\bibliography{refs}

\begin{thebibliography}{100}
\providecommand{\url}[1]{#1}
\csname url@samestyle\endcsname
\providecommand{\newblock}{\relax}
\providecommand{\bibinfo}[2]{#2}
\providecommand{\BIBentrySTDinterwordspacing}{\spaceskip=0pt\relax}
\providecommand{\BIBentryALTinterwordstretchfactor}{4}
\providecommand{\BIBentryALTinterwordspacing}{\spaceskip=\fontdimen2\font plus
\BIBentryALTinterwordstretchfactor\fontdimen3\font minus
  \fontdimen4\font\relax}
\providecommand{\BIBforeignlanguage}[2]{{%
\expandafter\ifx\csname l@#1\endcsname\relax
\typeout{** WARNING: IEEEtran.bst: No hyphenation pattern has been}%
\typeout{** loaded for the language `#1'. Using the pattern for}%
\typeout{** the default language instead.}%
\else
\language=\csname l@#1\endcsname
\fi
#2}}
\providecommand{\BIBdecl}{\relax}
\BIBdecl

\bibitem{Singh2014}
D.~Singh, G.~Tripathi, and A.~Jara, ``A survey of internet-of-things: Future
  vision, architecture, challenges and services,'' in \emph{IEEE World Forum on
  Internet of Things}, Seoul, Korea, Mar. 2014, pp. 287--292.

\bibitem{Gartner2014}
J.~Rivera and R.~van~der Meulen, ``Gartner says 4.9 billion connected "things"
  will be in use in 2015,'' Nov. 2014, {G}artner.

\bibitem{McLeod2013}
J.~McLeod, ``How the smart phone is driving the internet-of-things,'' Aug.
  2013, {K}ilopass Technol. Inc.

\bibitem{Gartner2015}
V.~Woods and R.~van~der Meulen, ``Gartner says worldwide smartphone sales grew
  9.7 percent in fourth quarter of 2015,'' Feb. 2016, {G}artner.

\bibitem{Xu2014}
Z.~Xu, Z.~Chen, and H.~Nie, ``Handheld computers: Smartphone-centric wireless
  applications,'' \emph{IEEE Microwave Mag.}, vol.~15, no.~2, pp. 36--44, Mar.
  2014.

\bibitem{Jose2015}
A.~C. Jose and R.~Malekian, ``Smart home automation security: {A} literature
  review,'' \emph{Smart Comput. Review}, vol.~5, no.~4, pp. 269--285, Aug.
  2015.

\bibitem{IndustryARC2014}
``Global consumer telematics systems market,'' Tech. Rep., Oct. 2014,
  {IndustryARC}.

\bibitem{mam2014}
``Automotive telematics market,'' Tech. Rep., Aug. 2014, {M}arkets and markets,
  Report Code: AT 2658.

\bibitem{Engelbrecht2015}
J.~Engelbrecht, M.~J. Booysen \emph{et~al.},
  ``\BIBforeignlanguage{English}{Survey of smartphone-based sensing in vehicles
  for intelligent transportation system applications},''
  \emph{\BIBforeignlanguage{English}{IET Intell. Transport. Syst.}}, vol.~9,
  no.~10, pp. 924--935, Dec. 2015.

\bibitem{Handel2014}
P.~Händel, I.~Skog \emph{et~al.}, ``Insurance telematics: {O}pportunities and
  challenges with the smartphone solution,'' \emph{IEEE Intell. Transport.
  Syst. Mag.}, vol.~6, no.~4, pp. 57--70, Oct. 2014.

\bibitem{Ekler2015}
P.~Ekler, T.~Balogh \emph{et~al.}, ``Social driving in connected car
  environment,'' in \emph{Proc. 21th IEEE Int. Conf. European Wireless},
  Budapest, Hungary, May 2015, pp. 136--141.

\bibitem{Braun2015}
P.~Braun, ``Why does in-car tech lag behind phone tech? {I}t's trickier than
  you think,'' Feb. 2015, digitaltrends.com.

\bibitem{Wahlstrom2015}
J.~Wahlström, I.~Skog, and P.~Händel, ``{IMU} alignment for smartphone-based
  automotive navigation,'' in \emph{Proc. 18th IEEE Int. Conf. Inf. Fusion},
  Washington, DC, Jul. 2015, pp. 1437--1443.

\bibitem{Tarkoma2014}
S.~Tarkoma, M.~Siekkinen \emph{et~al.}, \emph{Smartphone Energy
  Consumption}.\hskip 1em plus 0.5em minus 0.4em\relax Cambridge University
  Press, 2014.

\bibitem{Gustafsson2005}
F.~Gustafsson, ``Statistical signal processing for automotive safety systems,''
  in \emph{Proc. IEEE 13th Workshop Statistical Signal Process.}, Novosibirsk,
  Russia, Jul. 2005, pp. 1428--1435.

\bibitem{Handel2014_2}
P.~Händel, J.~Ohlsson \emph{et~al.}, ``Smartphone-based measurement systems for
  road vehicle traffic monitoring and usage-based insurance,'' \emph{IEEE Syst.
  J.}, vol.~8, no.~4, pp. 1238--1248, Dec. 2014.

\bibitem{Kjaergaard2011}
M.~B. Kjaergaard, S.~Bhattacharya \emph{et~al.}, ``Energy-efficient trajectory
  tracking for mobile devices,'' in \emph{Proc. 9th Int. Conf. Mobile Syst.,
  Appl., Services}, Washington, DC, Jun. 2011, pp. 307--320.

\bibitem{Hoh2012}
B.~Hoh, T.~Iwuchukwu \emph{et~al.}, ``Enhancing privacy and accuracy in probe
  vehicle-based traffic monitoring via virtual trip lines,'' \emph{IEEE Trans.
  Mobile Comput.}, vol.~11, no.~5, pp. 849--864, May. 2012.

\bibitem{Malikopoulos2013}
A.~Malikopoulos and J.~Aguilar, ``An optimization framework for driver feedback
  systems,'' \emph{IEEE Trans. Intell. Transport. Syst.}, vol.~14, no.~2, pp.
  955--964, Jun. 2013.

\bibitem{Reininger2015}
M.~Reininger, S.~Miller \emph{et~al.}, ``A first look at vehicle data
  collection via smartphone sensors,'' in \emph{Proc. IEEE Symp. Sensors
  Appl.}, Zadar, Croatia, Apr. 2015.

\bibitem{Goncalves2014}
J.~Goncalves, J.~Goncalves \emph{et~al.}, ``Smartphone sensor platform to study
  traffic conditions and assess driving performance,'' in \emph{Proc. IEEE Int.
  Conf. Intell. Transport. Syst.}, Qingdao, China, Oct. 2014, pp. 2596--2601.

\bibitem{Tahat2012}
A.~Tahat, A.~Said \emph{et~al.}, ``{Android-based} universal vehicle diagnostic
  and tracking system,'' in \emph{Proc. 16th IEEE Int. Symp. Consum.
  Electron.}, Harrisburg, PA, Jun. 2012, pp. 137--143.

\bibitem{Skog2014}
I.~Skog and P.~Händel, ``Indirect instantaneous car-fuel consumption
  measurements,'' \emph{IEEE Trans. Instrum. Meas.}, vol.~63, no.~12, pp.
  3190--3198, Dec. 2014.

\bibitem{Krumm2009b}
J.~Krumm, ``A survey of computational location privacy,'' \emph{Personal
  Ubiquitous Comput.}, vol.~13, no.~6, pp. 391--399, Aug. 2009.

\bibitem{Thompson2010}
C.~Thompson, J.~White \emph{et~al.}, ``\BIBforeignlanguage{English}{Using
  smartphones to detect car accidents and provide situational awareness to
  emergency responders},'' in \emph{\BIBforeignlanguage{English}{Mobile
  Wireless Middleware, Operating Syst., Appl.}}\hskip 1em plus 0.5em minus
  0.4em\relax Springer, Jun. 2010, vol.~48, pp. 29--42.

\bibitem{Afzal2014}
H.~Afzal and V.~Maheta, ``Low cost smart phone controlled car security
  system,'' in \emph{Proc. IEEE Int. Conf. Ind. Technol.}, Busan, South Korea,
  Feb. 2014, pp. 670--675.

\bibitem{Tao2012}
S.~Tao, V.~Manolopoulos \emph{et~al.}, ``Real-time urban traffic state
  estimation with {A-GPS} mobile phones as probes,'' \emph{J. Transport.
  Technol.}, vol.~2, no.~1, pp. 22--31, Nov. 2012.

\bibitem{Liu2013}
M.~Liu, ``A study of mobile sensing using smartphones,'' \emph{Int. J. Distrib.
  Sensor Netw.}, vol.~9, no.~3, Mar. 2013.

\bibitem{Daponte2013}
P.~Daponte, L.~D. Vito \emph{et~al.}, ``State of the art and future
  developments of measurement applications on smartphones,'' \emph{Meas.},
  vol.~46, no.~9, pp. 3291--3307, Nov. 2013.

\bibitem{Lane2010}
N.~Lane, E.~Miluzzo \emph{et~al.}, ``A survey of mobile phone sensing,''
  \emph{IEEE Commun. Mag.}, vol.~48, no.~9, pp. 140--150, Sep. 2010.

\bibitem{Zandbergen2009}
P.~A. Zandbergen, ``Accuracy of i{P}hone locations: {A} comparison of assisted
  {GPS}, {WiFi} and cellular positioning,'' \emph{Trans. GIS}, vol.~13, pp.
  5--25, Jun. 2009.

\bibitem{Turker2016}
G.~F. Türker and A.~Kutlu, ``Survey of smartphone applications based on
  {OBD-II} for intelligent transportation systems,'' \emph{Int. J. Eng.
  Research Appl.}, vol.~6, no.~1, pp. 69--73, Jan. 2016.

\bibitem{Meng2015}
R.~Meng, C.~Mao, and R.~R. Choudhury, ``Driving analytics: Will it be {OBD}s or
  smartphones?'' \emph{Proc. Driver Analytics Workshop}, 2015.

\bibitem{Perrucci2011}
G.~Perrucci, F.~Fitzek, and J.~Widmer, ``Survey on energy consumption entities
  on the smartphone platform,'' in \emph{Proc. 73rd IEEE Int. Conf. Veh.
  Technol.}, Budapest, Hungary, May 2011.

\bibitem{Maloney2012}
S.~Maloney and I.~Boci, ``Survey: Techniques for efficient energy consumption
  in mobile architectures,'' in \emph{Power (mW)}, vol.~16, 2012, pp. 7--35.

\bibitem{PerezTorres2016}
R.~Perez-Torres, C.~Torres-Huitzil, and H.~Galeana-Zapien, ``Power management
  techniques in smartphone-based mobility sensing systems: A survey,''
  \emph{Accepted in Pervasive Mobile Comput.}

\bibitem{Wang2014a}
L.~Wang, Y.~Cui \emph{et~al.}, ``\BIBforeignlanguage{English}{Energy efficiency
  on location based applications in mobile cloud computing: {A} survey},''
  \emph{\BIBforeignlanguage{English}{Comput.}}, vol.~96, no.~7, pp. 569--585,
  Jul. 2014.

\bibitem{Carroll2010}
A.~Carroll and G.~Heiser, ``An analysis of power consumption in a smartphone,''
  in \emph{Proc. USENIX Technical Conf.}, Boston, MA, Jun. 2010, pp. 271--285.

\bibitem{Lee2010}
U.~Lee and M.~Gerla, ``A survey of urban vehicular sensing platforms,''
  \emph{Comput. Netw.}, vol.~54, no.~4, pp. 527--544, Mar. 2010.

\bibitem{Lee2014}
E.~Lee, E.-K. Lee \emph{et~al.}, ``Vehicular cloud networking: {Architecture}
  and design principles,'' \emph{IEEE Commun. Mag.}, vol.~52, no.~2, pp.
  148--155, Feb. 2014.

\bibitem{Araniti2013}
G.~Araniti, C.~Campolo \emph{et~al.}, ``{LTE} for vehicular networking: {A
  survey},'' \emph{IEEE Commun. Mag.}, vol.~51, no.~5, pp. 148--157, May 2013.

\bibitem{Salin2012}
H.~Salin, ``Gap analysis in cooperative systems within intelligent
  transportation systems,'' Master's thesis, KTH Royal Institute of Technology,
  2012.

\bibitem{Fernando2013}
N.~Fernando, S.~W. Loke, and W.~Rahayu, ``Mobile cloud computing: {A} survey,''
  \emph{Future Generation Computer Systems}, vol.~29, no.~1, pp. 84--106, Jan.
  2013.

\bibitem{Gu2013}
L.~Gu, D.~Zeng, and S.~Guo, ``Vehicular cloud computing: {A} survey,'' in
  \emph{Proc. IEEE Globecom Workshops}, Atlanta, GA, Dec. 2013, pp. 403--407.

\bibitem{Whaiduzzaman2014}
M.~Whaiduzzaman, M.~Sookhak \emph{et~al.}, ``A survey on vehicular cloud
  computing,'' \emph{J. Netw. Comput. Appl.}, vol.~40, pp. 325--344, Apr. 2014.

\bibitem{Yu2013}
R.~Yu, Y.~Zhang \emph{et~al.}, ``Toward cloud-based vehicular networks with
  efficient resource management,'' \emph{IEEE Netw.}, vol.~27, no.~5, pp.
  48--55, Sep. 2013.

\bibitem{Williams-Bergen2011}
E.~Williams-Bergen, J.~Hedlund \emph{et~al.}, ``Distracted driving - {What}
  research shows and what states can do,'' Governors Highway Safety
  Association, Tech. Rep., Jul. 2011.

\bibitem{Kinnear2015}
N.~Kinnear and A.~Stevens, ``The battle for attention; {Driver} distraction -
  {A} review of recent research and knowledge,'' Institute of Advanced
  Motorists, Tech. Rep., Dec. 2015.

\bibitem{McEvoy2005}
S.~P. McEvoy, M.~R. Stevenson \emph{et~al.}, ``Role of mobile phones in motor
  vehicle crashes resulting in hospital attendance: a case-crossover study,''
  \emph{BMJ}, vol. 331, no. 7514, pp. 428--430, Aug. 2005.

\bibitem{Strayer2013}
D.~L. Strayer, J.~M. Cooper \emph{et~al.}, ``Measuring cognitive distraction in
  the automobile,'' AAA - Foundation for Traffic Safety, Tech. Rep., Jun. 2013.

\bibitem{Strayer2014}
D.~L. Strayer, J.~Turrill \emph{et~al.}, ``Measuring cognitive distraction in
  the automobile {II}: Assessing in-vehicle voice-based interactive
  technologies,'' AAA - Foundation for Traffic Safety, Tech. Rep., Oct. 2014.

\bibitem{Quddus2007}
M.~A. Quddus, W.~Y. Ochieng, and R.~B. Noland, ``Current map-matching
  algorithms for transport applications: State-of-the art and future research
  directions,'' \emph{Transport. Research Part C: Emerg Technol.}, vol.~15,
  no.~5, pp. 312--328, Oct. 2007.

\bibitem{Wahlstrom2015_3}
J.~Wahlstr\"{o}m, I.~Skog, and P.~H\"{a}ndel, ``Driving behavior analysis for
  smartphone-based insurance telematics,'' in \emph{Proc. 2nd Workshop on
  Physical Analytics}, Florence, Italy, May 2015, pp. 19--24.

\bibitem{Wahlstrom2016}
J.~Wahlström, I.~Skog \emph{et~al.}, ``{IMU}-based smartphone-to-vehicle
  positioning,'' \emph{Accepted in IEEE Trans. Intell. Vehicles}.

\bibitem{Biancat2014}
J.~Biancat, C.~Brighenti, and A.~Brighenti, ``Review of transportation mode
  detection techniques,'' \emph{Endorsed Trans. Ambient Syst.}, vol.~1, no.~4,
  Oct. 2014.

\bibitem{Gessulat2013}
S.~Gessulat, ``A review of real-time models for transportation mode
  detection,'' Free University of Berlin, Tech. Rep., Mar. 2013.

\bibitem{Reddy2010}
S.~Reddy, M.~Mun \emph{et~al.}, ``Using mobile phones to determine
  transportation modes,'' \emph{ACM Trans. Sen. Netw.}, vol.~6, no.~2, pp.
  13:1--13:27, Feb. 2010.

\bibitem{Shin2014}
D.~Shin, D.~Aliaga \emph{et~al.}, ``Urban sensing: Using smartphones for
  transportation mode classification,'' \emph{Comput., Environment, Urban
  Syst.}, vol.~53, pp. 76--86, Sep. 2014.

\bibitem{Weiss2012}
J.~Weiss and J.~Smollik, ``Beginner's roadmap to working with driving behavior
  data,'' \emph{Casualty Actuarial Soc. E-Forum}, vol.~2, 2012.

\bibitem{Meiring2015}
G.~A.~M. Meiring and H.~C. Myburgh, ``A review of intelligent driving style
  analysis systems and related artificial intelligence algorithms,''
  \emph{Sensors}, vol.~15, no.~12, pp. 30\,653--30\,682, Dec. 2015.

\bibitem{Bolovinou2014}
A.~Bolovinou, A.~Amditis, and F.~Bellotti, ``Driving style recognition for
  co-operative driving: A survey,'' in \emph{Proc. 6th Int. Conf. Adaptive
  Self-Adaptive Syst. Appl.}, Venice, Italy, May. 2014, pp. 73--78.

\bibitem{Kaplan2015}
S.~Kaplan, M.~A. Guvensan \emph{et~al.}, ``Driver behavior analysis for safe
  {Driving: A Survey},'' \emph{IEEE Trans. Intell. Transport. Syst.}, vol.~16,
  no.~6, pp. 3017--3032, Dec. 2015.

\bibitem{Kalra2014}
N.~Kalra and D.~Bansal, ``Analyzing driver behavior using smartphone sensors: A
  survey,'' \emph{Int. J. Electron. Electr. Eng.}, vol.~7, no.~7, pp. 697--702,
  Jan. 2014.

\bibitem{Eren2012}
H.~Eren, S.~Makinist \emph{et~al.}, ``Estimating driving behavior by a
  smartphone,'' in \emph{Proc. IEEE Symp. Intell. Vehicles}, Alcalá de Henares,
  Spain, Jun. 2012, pp. 234--239.

\bibitem{Johnson2011}
D.~Johnson and M.~Trivedi, ``Driving style recognition using a smartphone as a
  sensor platform,'' in \emph{Proc. IEEE Conf. Intell. Transport. Syst.},
  Washington, DC, Oct. 2011, pp. 1609--1615.

\bibitem{Chugh2014}
G.~Chugh, D.~Bansal, and S.~Sofat, ``Road condition detection using smartphone
  sensors: {A} survey,'' \emph{Int. J. Electron. Electrical Eng.}, vol.~7,
  no.~6, pp. 595--602, 2014.

\bibitem{Mukherjee2016}
A.~Mukherjee and S.~Majhi, ``Characterisation of road bumps using
  smartphones,'' \emph{European Transport Research Review}, vol.~8, no.~2, Apr.
  2016.

\bibitem{Hull2006}
B.~Hull, V.~Bychkovsky \emph{et~al.}, ``{CarTel: A} distributed mobile sensor
  computing system,'' in \emph{Proc. 4th ACM SenSys}, Boulder, CO, Nov. 2006,
  pp. 125--138.

\bibitem{Eriksson2008}
J.~Eriksson, L.~Girod \emph{et~al.}, ``The pothole patrol: Using a mobile
  sensor network for road surface monitoring,'' in \emph{Proc. 6th Int. Conf.
  Mobile Syst. Appl. Services}, Breckenridge, CO, Jun. 2008, pp. 29--39.

\bibitem{Popa2009}
R.~A. Popa, H.~Balakrishnan, and A.~J. Blumberg, ``{VPriv: Protecting} privacy
  in location-based vehicular services,'' in \emph{Proc. 18th Conf. USENIX
  Security Symp.}, Montreal, QC, Aug. 2009, pp. 335--350.

\bibitem{Mohan2008}
P.~Mohan, V.~N. Padmanabhan, and R.~Ramjee, ``Trafficsense: Rich monitoring of
  road and traffic conditions using mobile smartphones,'' Microsoft Research,
  Tech. Rep., Apr. 2008.

\bibitem{Horvitz2005}
E.~Horvitz, J.~Apacible \emph{et~al.}, ``Prediction, expectation, and surprise:
  Methods, designs, and study of a deployed traffic forecasting service,'' in
  \emph{Proc. 21st Int. Conf. Uncertainty Artificial Intell.}, Edinburgh,
  Scotland, Jul. 2005, pp. 275--283.

\bibitem{Bayen2011}
A.~M. Bayen, J.~Butler \emph{et~al.}, ``Mobile millennium final report,''
  University of California, Tech. Rep., Sep. 2011.

\bibitem{Herrera2010}
J.~C. Herrera, D.~B. Work \emph{et~al.}, ``Evaluation of traffic data obtained
  via {GPS}-enabled mobile phones: The mobile century field experiment,''
  \emph{Transport. Research Part C: Emerg. Technol.}, vol.~18, no.~4, pp.
  568--583, Aug. 2010.

\bibitem{Allstrom2011}
A.~Allström, J.~Archer \emph{et~al.}, ``Mobile millennium {S}tockholm,'' in
  \emph{Proc. 2nd Conf. Models Technol. Intell. Transport. Syst.}, Leuven,
  Belgium, Jun. 2011.

\bibitem{d'Orey2014}
P.~d'Orey and M.~Ferreira, ``{ITS} for sustainable mobility: A survey on
  applications and impact assessment tools,'' \emph{IEEE Trans. Intell.
  Transport. Syst.}, vol.~15, no.~2, pp. 477--493, Apr. 2014.

\bibitem{Rodrigues2014}
J.~G. Rodrigues, A.~Aguiar, and J.~Barros, ``{SenseMyCity}: Crowdsourcing an
  urban sensor,'' \emph{CoRR}, vol. arXiv:1412.2070, Dec. 2014.

\bibitem{Hoh2008}
B.~Hoh, M.~Gruteser \emph{et~al.}, ``Virtual trip lines for distributed
  privacy-preserving traffic monitoring,'' in \emph{Proc. 11th Int. Conf.
  Mobile Systems, Appl., Services}, Breckenridge, CO, Jun. 2008, pp. 15--28.

\bibitem{Work02010}
D.~B. Work, S.~Blandin \emph{et~al.}, ``A traffic model for velocity data
  assimilation,'' \emph{{Appl. Mathematics Research eXpress}}, vol. 2010,
  no.~1, pp. 1--35, Apr. 2010.

\bibitem{Hunter2011}
T.~Hunter, T.~Moldovan \emph{et~al.}, ``Scaling the mobile millennium system in
  the cloud,'' in \emph{Proc. 2nd ACM Symp. Cloud Comput.}, Cascais, Portugal,
  Oct. 2011, pp. 28:1--28:8.

\bibitem{Herrera2010_2}
J.~C. Herrera and A.~M. Bayen, ``Incorporation of {Lagrangian} measurements in
  freeway traffic state estimation,'' \emph{Transport. Research Part B:
  Methodological}, vol.~44, no.~4, pp. 460--481, May 2010.

\bibitem{Allstrom2012}
A.~Allström, D.~Gundlegård, and C.~Rydergren, ``Evaluation of travel time
  estimation based on {LWR-v and CTM-v: A case study in Stockholm},'' in
  \emph{Proc. 15th IEEE Int. Conf. Intell. Transport. Syst.}, Anchorage, AK,
  Sep. 2012, pp. 1644--1649.

\bibitem{Wahlstrom2015_2}
J.~Wahlström, I.~Skog, and P.~Händel, ``Detection of dangerous cornering in
  {GNSS}-data-driven insurance telematics,'' \emph{IEEE Trans. Intell.
  Transport. Syst.}, vol.~16, no.~6, pp. 3073--3083, Dec. 2015.

\bibitem{Ohlsson2015}
J.~Ohlsson, P.~Händel \emph{et~al.}, ``\BIBforeignlanguage{English}{Process
  innovation with disruptive technology in auto insurance: Lessons learned from
  a smartphone-based insurance telematics initiative},'' in
  \emph{\BIBforeignlanguage{English}{BPM - Driving Innovation in a Digital
  World}}.\hskip 1em plus 0.5em minus 0.4em\relax Springer, 2015, pp. 85--101.

\bibitem{Ptolemus2016}
``The state of usage based insurance today,'' Jun. 2016, {P}tolemus -
  Consulting Group.

\bibitem{Reuters2014}
J.~Fuller, ``Vehcon, inc. announces verified mileage solution for \$1 per
  vehicle,'' \emph{Reuters}, Sep. 2014.

\bibitem{Newcomer2015}
E.~Newcomer, ``Uber raises funding at \$62.5 billion valuation,'' Dec. 2015,
  {Bloomberg}.

\bibitem{Watzdorf2010}
S.~von Watzdorf and F.~Michahelles, ``Accuracy of positioning data on
  smartphones,'' in \emph{Proc. 3rd Int. Workshop Location Web}, Tokyo, Japan,
  Nov. 2010, pp. 2:1--2:4.

\bibitem{NMEA2008}
``{NMEA} 0183 - {S}tandard for interfacing marine electronic devices, version
  4.0,'' National Marine Electronic Association, Tech. Rep., 2008.

\bibitem{ION2016}
ION, ``{Google's Android OS} to provide raw {GNSS} measurements,'' \emph{The
  Institute of Navigation - Quarterly Newletter}, vol.~26, no.~3, p.~7, 2016.

\bibitem{Zhang2011}
J.~Zhang, F.-Y. Wang \emph{et~al.}, ``Data-driven intelligent transportation
  systems: A survey,'' \emph{IEEE Trans. Intell. Transport. Syst.}, vol.~12,
  no.~4, pp. 1624--1639, Dec. 2011.

\bibitem{Menard2011}
T.~Menard, J.~Miller \emph{et~al.}, ``{Comparing the GPS capabilities of the
  Samsung Galaxy S, Motorola Droid X, and the Apple iPhone for vehicle tracking
  using FreeSim-Mobile},'' in \emph{Proc. 14th IEEE Int. Conf. Intell.
  Transport. Syst.}, Washington, DC, Oct. 2011, pp. 985--990.

\bibitem{Menard2011b}
T.~Menard and J.~Miller, ``{Comparing the GPS capabilities of the iPhone 4 and
  iPhone 3G for vehicle tracking using FreeSim-Mobile},'' in \emph{Proc. IEEE
  Intell. Vehicles Symp.}, Baden-Baden, Germany, Jun. 2011, pp. 278--283.

\bibitem{Aloi2007}
D.~Aloi, M.~Alsliety, and D.~Akos, ``A methodology for the evaluation of a
  {GPS} receiver performance in telematics applications,'' \emph{IEEE Trans.
  Instrum. Meas.}, vol.~56, no.~1, pp. 11--24, Feb. 2007.

\bibitem{Kaplan2006}
E.~D. Kaplan and C.~J. Hegarty, \emph{Understanding GPS - Principles and
  Applications}, 2nd~ed.\hskip 1em plus 0.5em minus 0.4em\relax Artech House,
  2006.

\bibitem{Blum2012}
J.~R. Blum, D.~G. Greencorn, and J.~R. Cooperstock, ``Smartphone sensor
  reliability for augmented reality applications.'' in \emph{MobiQuitous}, vol.
  120, Beijing, China, Dec. 2012, pp. 127--138.

\bibitem{Godha2005}
S.~Godha, M.~G. Petovello, and G.~Lachapelle, ``Performance analysis of {MEMS}
  {IMU}/{HSGPS}/magnetic sensor integrated system in urban canyons,'' in
  \emph{Position, Location and Navigation Group}, Long Beach, CA, Sep. 2005,
  pp. 1977--1990.

\bibitem{Bo2016}
C.~Bo, X.~Jian \emph{et~al.}, ``Detecting driver's smartphone usage via
  non-intrusively sensing driving dynamics,'' \emph{Accept in IEEE Internet of
  Things J.}

\bibitem{Zong2015}
X.~Zong, X.~Wen, and T.~Zhang, ``Establishment of coordinate relationship with
  accelerometer,'' in \emph{Proc. Int. Conf. Intell. Syst. Research
  Mechatronics Eng.}, Zhengzhou, China, Apr. 2015, pp. 1279--1291.

\bibitem{WahlstromGustafsson2014}
N.~Wahlström and F.~Gustafsson, ``Magnetometer modeling and validation for
  tracking metallic targets,'' \emph{IEEE Trans. Signal Process.}, vol.~62,
  no.~3, pp. 545--556, Feb. 2014.

\bibitem{Prateek2013}
G.~V. Prateek, K.~Nijil, and K.~V.~S. Hari, ``Classification of vehicles using
  magnetic field angle model,'' in \emph{Proc. 4th IEEE Int. Conf. Intell.
  Syst., Modelling Simulation}, Bangkok, Thailand, Jan. 2013, pp. 214--219.

\bibitem{Zachariah2010}
D.~Zachariah and M.~Jansson, ``Camera-aided inertial navigation using epipolar
  points,'' in \emph{Proc. IEEE Symp. Position Location Navigation}, Indian
  Wells, CA, May 2010, pp. 303--309.

\bibitem{Sivaraman2013}
S.~Sivaraman and M.~Trivedi, ``Looking at vehicles on the road: A survey of
  vision-based vehicle detection, tracking, and behavior analysis,'' \emph{IEEE
  Trans. Intell. Transport. Syst.}, vol.~14, no.~4, pp. 1773--1795, Dec. 2013.

\bibitem{Chuang2014}
M.-C. Chuang, R.~Bala \emph{et~al.}, ``Estimating gaze direction of vehicle
  drivers using a smartphone camera,'' in \emph{Proc. IEEE Int. Conf. Computer
  Vision Pattern Recognition Workshops}, Columbus, OH, Jun. 2014, pp. 165--170.

\bibitem{Koukoumidis2011}
E.~Koukoumidis, L.-S. Peh, and M.~R. Martonosi, ``{SignalGuru: Leveraging}
  mobile phones for collaborative traffic signal schedule advisory,'' in
  \emph{Proc. 9th Int. Conf. Mobile Syst., Appl., Services}, Bethesda, MD, Jun.
  2011, pp. 127--140.

\bibitem{Mathias2013}
M.~Mathias, R.~Timofte \emph{et~al.}, ``Traffic sign recognition - {How} far
  are we from the solution?'' in \emph{Proc. IEEE Int. Conf. Neural Netw.},
  Dallas, TX, Aug. 2013.

\bibitem{Salvi2002}
J.~Salvi, X.~Armangué, and J.~Batlle, ``A comparative review of camera
  calibrating methods with accuracy evaluation,'' \emph{Pattern Recognition},
  vol.~35, no.~7, pp. 1617--1635, Jul. 2002.

\bibitem{Delaunoy2014}
A.~Delaunoy, J.~Li \emph{et~al.}, ``Two cameras and a screen: How to calibrate
  mobile devices?'' in \emph{Proc. 2nd Int. Conf. 3D Vision}, Tokyo, Japan,
  Dec. 2014, pp. 123--130.

\bibitem{Jia2014}
C.~Jia and B.~L. Evans, ``Online camera-gyroscope autocalibration for cell
  phones,'' \emph{IEEE Trans. Image Process.}, vol.~23, no.~12, pp. 5070--5081,
  Dec. 2014.

\bibitem{Saponaro2013}
P.~Saponaro and C.~Kambhamettu, ``Towards auto-calibration of smart phones
  using orientation sensors,'' in \emph{Proc. IEEE Conf. Comput. Vision and
  Pattern Recognition Workshops}, Portland, OR, Jun. 2013, pp. 20--26.

\bibitem{Skocaj2014}
D.~Skocaj, D.~Tabernik \emph{et~al.}, ``Evaluation of performance of smart
  mobile devices in machine vision tasks,'' in \emph{Comput. Vision Winter
  Workshop}, Krtiny, Czech Republic, Feb. 2014.

\bibitem{Panahandeh2015}
G.~Panahandeh, M.~Jansson, and P.~Händel, ``Calibration of an {IMU}-camera
  cluster using planar mirror reflection and its observability analysis,''
  \emph{IEEE Trans. Instrum. Meas.}, vol.~64, no.~1, pp. 75--88, Jan. 2015.

\bibitem{Mielke2013}
M.~Mielke and R.~Bruck, ``Smartphone application for automatic classification
  of environmental sound,'' in \emph{Proc. 20th Int. Conf. Mixed Design of
  Integr. Circuits Syst.}, Gdynia, Poland, Jun. 2013, pp. 512--515.

\bibitem{Yang2011}
J.~Yang, S.~Sidhom \emph{et~al.}, ``Detecting driver phone use leveraging car
  speakers,'' in \emph{Proc. 17th Int. Conf. Mobile Comput. Netw.}, Las Vegas,
  NV, Sep. 2011, pp. 97--108.

\bibitem{Gomez-Martin2012}
L.~E. Gomez-Martin, ``Smartphone usage and the need for consumer privacy
  laws,'' \emph{Pittsburgh J. Technol. Law, Policy}, vol.~12, pp. 217--237,
  2012.

\bibitem{Perlmutter2012}
M.~Perlmutter and L.~Robin, ``High-performance, low cost inertial {MEMS}: A
  market in motion!'' in \emph{Proc. IEEE/ION Symp. Position Location and
  Navigation}, Myrtle Beach, SC, Apr. 2012, pp. 225--229.

\bibitem{Groves2008}
P.~D. Groves, \emph{Principles of GNSS, inertial, and multisensor integrated
  navigation systems}, 1st~ed.\hskip 1em plus 0.5em minus 0.4em\relax Artech
  House, 2008.

\bibitem{Aicardi2014}
I.~Aicardi, P.~Dabove \emph{et~al.}, ``Sensors integration for smartphone
  navigation: {Performances} and future challenges,'' \emph{Int. Archives of
  the Photogrammetry, Remote Sensing and Spatial Information Sciences}, vol.
  XL-3, pp. 9--16, Aug. 2014.

\bibitem{Chowdhury2014b}
A.~Chowdhury, T.~Chakravarty, and P.~Balamuralidhar, ``A novel approach to
  improve vehicle speed estimation using smartphone's {INS}/{GPS} sensors,'' in
  \emph{Proc. 8th Int. Conf. Sens. Technol.}, Liverpool, UK, Sep. 2014, pp.
  441--446.

\bibitem{Niu2015}
X.~Niu, Q.~Wang \emph{et~al.}, ``Using inertial sensors in smartphones for
  curriculum experiments of inertial navigation technology,'' \emph{Educ.
  Sci.}, vol.~5, no.~1, pp. 26--46, Mar. 2015.

\bibitem{Chowdhury2016}
A.~Chowdhury, A.~Ghose \emph{et~al.}, \emph{Smart Sensors, Measurement and
  Instrumentation}.\hskip 1em plus 0.5em minus 0.4em\relax Springer, Jul. 2016,
  vol.~16, ch. An Improved Fusion Algorithm For Estimating Speed From
  Smartphone's {INS/GPS} Sensors, pp. 235--256.

\bibitem{Yu2016}
J.~Yu, H.~Zhu \emph{et~al.}, ``{SenSpeed: Sensing} driving conditions to
  estimate vehicle speed in urban environments,'' \emph{IEEE Trans. Mobile
  Comput.}, vol.~15, no.~1, pp. 202--216, Jan. 2016.

\bibitem{Zaldivar2011}
J.~Zaldivar, C.~Calafate \emph{et~al.}, ``Providing accident detection in
  vehicular networks through {OBD-II} devices and {A}ndroid-based
  smartphones,'' in \emph{Proc. 36th. IEEE Int. Conf. Local Comput. Netw.},
  Bonn, Germany, Oct. 2011, pp. 813--819.

\bibitem{Sathyanarayana2012}
A.~Sathyanarayana, S.~Sadjadi, and J.~Hansen, ``Leveraging sensor information
  from portable devices towards automatic driving maneuver recognition,'' in
  \emph{Proc. 15th IEEE Int. Conf. Intell. Transport. Syst.}, Anchorage, AK,
  Sep. 2012, pp. 660--665.

\bibitem{Sathyanarayana2013}
A.~Sathyanarayana, S.~O. Sadjadi, and J.~Hansen, ``Automatic driving maneuver
  recognition and analysis using cost effective portable devices,'' \emph{SAE
  Int. J. Passeng. Cars - Electron. Electr. Syst}, vol.~6, no.~2, pp. 467--477,
  Apr. 2013.

\bibitem{Chowdhury2014}
A.~Chowdhury, T.~Chakravarty, and P.~Balamuralidhar, ``Estimating true speed of
  moving vehicle using smartphone-based {GPS} measurement,'' in \emph{Proc.
  IEEE Int. Conf. Syst., Man Cybernetics}, San Diego, CA, Oct. 2014, pp.
  3348--3353.

\bibitem{Walter2013}
O.~Walter, J.~Schmalenstroeer \emph{et~al.}, ``Smartphone-based sensor fusion
  for improved vehicular navigation,'' in \emph{Proc. 10th IEEE Positioning
  Navigation Commun.}, Dresden, Germany, Mar. 2013.

\bibitem{Chon2011}
Y.~Chon, E.~Talipov \emph{et~al.}, ``Mobility prediction-based smartphone
  energy optimization for everyday location monitoring,'' in \emph{Proc. 9th
  ACM Conf. Embedded Netw. Sensor Syst.}, Seattle, WA, Nov. 2011, pp. 82--95.

\bibitem{Kjaergaard2009}
M.~B. Kj{\ae}rgaard, J.~Langdal \emph{et~al.}, ``{EnTracked: Energy}-efficient
  robust position tracking for mobile devices,'' in \emph{Proc. 7th Int. Conf.
  Mobile Syst., Appl., Services}, Kraków, Poland, Jun. 2009, pp. 221--234.

\bibitem{Hu2015}
S.~Hu, L.~Su \emph{et~al.}, ``Experiences with {eNav: A} low-power vehicular
  navigation system,'' in \emph{Proc. ACM Int. Conf. Pervasive Ubiquitous
  Comput.}, Osaka, Japan, Sep. 2015, pp. 433--444.

\bibitem{Paek2010}
J.~Paek, J.~Kim, and R.~Govindan, ``Energy-efficient rate-adaptive {GPS}-based
  positioning for smartphones,'' in \emph{Proc. 8th Int. Conf. Mobile Systems,
  Appl., Services}, San Francisco, CA, Jun. 2010, pp. 299--314.

\bibitem{Alrefaie2013}
M.~Alrefaie, I.~Carreras \emph{et~al.}, ``Map matching accuracy: Energy
  efficient location sampling using smartphones,'' in \emph{Proc. 16th IEEE
  Int. Conf. Intell. Transport. Syst.}, The Hague, the Netherlands, Oct. 2013,
  pp. 2243--2248.

\bibitem{Thiagarajan2011}
A.~Thiagarajan, L.~Ravindranath \emph{et~al.}, ``Accurate, low-energy
  trajectory mapping for mobile devices,'' in \emph{Proc. 8th USENIX Conf.
  Netw. Syst. Design Implementation}, Boston, MA, Mar. 2011, pp. 267--280.

\bibitem{Aly2015}
H.~Aly and M.~Youssef, ``{semMatch: R}oad semantics-based accurate map matching
  for challenging positioning data,'' in \emph{Proc. 23rd Int. Conf. Advances
  in Geographic Inf. Syst.}, Bellevue, WA, 2015, pp. 5:1--5:10.

\bibitem{Liu2012}
K.~Liu, Q.~Huang \emph{et~al.}, ``Less energy higher accuracy: Smartphone
  localization via social collaboration,'' University of Florida, Tech. Rep.,
  2012.

\bibitem{Miettinen2010}
A.~P. Miettinen and J.~K. Nurminen, ``Energy efficiency of mobile clients in
  cloud computing,'' in \emph{Proc. 2nd USENIX Conf. Hot Topics Cloud Comput.},
  Boston, MA, 2010.

\bibitem{Abdelmotalib2012}
A.~Abdelmotalib and Z.~Wu, ``Power consumption in smartphones (hardware
  behaviourism),'' \emph{Int. J. Comput. Sci. Issues}, vol.~9, no.~3, pp.
  161--164, May 2012.

\bibitem{Friedman2013}
R.~Friedman, A.~Kogan, and Y.~Krivolapov, ``On power and throughput tradeoffs
  of {WiFi} and {Bluetooth} in smartphones,'' \emph{IEEE Trans. Mobile
  Comput.}, vol.~12, no.~7, pp. 1363--1376, Jul. 2013.

\bibitem{Priyantha2010}
B.~Priyantha, D.~Lymberopoulos, and J.~Liu, ``Little rock: Enabling energy
  efficient continuous sensing on mobile phones,'' Microsoft Research, Tech.
  Rep., Feb. 2010.

\bibitem{Wang2013}
Y.~Wang, J.~Yang \emph{et~al.}, ``Sensing vehicle dynamics for determining
  driver phone use,'' in \emph{Proc. 11th Int. Conf. Mobile Systems, Appl.,
  Services}, Taipei, Taiwan, Jun. 2013, pp. 41--54.

\bibitem{Guido2012}
G.~Guido, A.~Vitale \emph{et~al.}, ``Estimation of safety performance measures
  from smartphone sensors,'' in \emph{Procedia - Social and Behavioral Sci.},
  vol.~54, Paris, France, Sep. 2012, pp. 1095--1103.

\bibitem{Buchenscheit2009}
A.~Buchenscheit, F.~Schaub \emph{et~al.}, ``A {VANET}-based emergency vehicle
  warning system,'' in \emph{Proc. IEEE Veh. Netw. Conf.}, Tokyo, Japan, Oct.
  2009.

\bibitem{Gramaglia2011}
M.~Gramaglia, P.~Serrano \emph{et~al.}, ``New insights from the analysis of
  free flow vehicular traffic in highways,'' in \emph{Proc. IEEE Int. Symp.
  World of Wireless, Mobile and Multimedia Netw.}, Lucca, Italy, Jun. 2011.

\bibitem{Caballero-Gil2013}
P.~Caballero-Gil, C.~Caballero-Gil, and J.~Molina-Gil, ``Design and
  implementation of an application for deploying vehicular networks with
  smartphones,'' \emph{Int. J. Distrib. Sensor Netw.}, vol.~9, no.~12, Dec.
  2013.

\bibitem{Li2015}
X.~Li, Y.~Feng \emph{et~al.}, ``When smart phone meets vehicle: A new on-board
  unit scheme for {VANETs},'' in \emph{Proc. IEEE Int. Conf. Pervasive Intell.
  Comput.}, Liverpool, U.K., Oct. 2015, pp. 1095--1100.

\bibitem{Barcelos2014}
V.~P. Barcelos, T.~C. Amarante \emph{et~al.}, ``Vehicle monitoring system using
  {IEEE 802.11p device and Android} application,'' in \emph{Proc. IEEE Int.
  Symp. Comput. Commun.}, Funchal, Portugal, Jun. 2014.

\bibitem{Tornell2013}
S.~Tornell, C.~Calafate \emph{et~al.}, ``Evaluating the feasibility of using
  smartphones for {ITS} safety applications,'' in \emph{Proc. 77th IEEE Int.
  Conf. Veh.Technol.}, Dresden, Germany, Jun. 2013.

\bibitem{Busanelli2013}
S.~Busanelli, F.~Rebecchi \emph{et~al.}, ``Cross-network information
  dissemination in vehicular ad hoc networks ({VANETs}): {E}xperimental results
  from a smartphone-based testbed,'' \emph{Future Internet}, vol.~5, no.~3, pp.
  398--428, Aug. 2013.

\bibitem{Djajadi2014}
A.~Djajadi and R.~Putra, ``Inter-cars safety communication system based on
  {Android} smartphone,'' in \emph{Proc. IEEE Int. Conf. Open Syst.}, Subang,
  Malaysia, Oct. 2014, pp. 12--17.

\bibitem{Park2014}
Y.~Park, J.~Ha \emph{et~al.}, ``A feasibility study and development framework
  design for realizing smartphone-based vehicular networking systems,''
  \emph{IEEE Trans. Mobile Comput.}, vol.~13, no.~11, pp. 2431--2444, Nov.
  2014.

\bibitem{Vandenberghe2011}
W.~Vandenberghe, I.~Moerman, and P.~Demeester, ``On the feasibility of
  utilizing smartphones for vehicular ad hoc networking,'' in \emph{Proc. 11th
  Int. Conf. ITS Telecommun.}, St. Petersburg, Russia, Aug. 2011, pp. 246--251.

\bibitem{Silva2014}
R.~Silva, S.~Noguchi \emph{et~al.}, ``Standards for cooperative intelligent
  transportation systems: a proof of concept,'' in \emph{Proc. 10th Int. Conf.
  Telecommun.}, Paris, France, Jul. 2014, pp. 35--40.

\bibitem{CooperativeITS2013}
``Cooperative {ITS} corridor joint deployment,'' Ministry of infrastructure and
  the environment of the {N}etherlands, Tech. Rep., Jun. 2013.

\bibitem{Mashiko2016}
T.~Mashiko, ``Autonomous driving technology and {ITS},'' in \emph{Proc. 8th
  Int. Conf. ETSI ITS Workshop}, Sophia Antipolis, France, Mar. 2016.

\bibitem{Berggren2011}
F.~S. Berggren, ``Traffic flow enhancement through guidance for driver based on
  {GPS} aided smartphones,'' Master's thesis, KTH Royal Institute of
  Technology, Jun. 2011.

\bibitem{Liu2012a}
J.~Liu, B.~Priyantha \emph{et~al.}, ``Energy efficient {GPS} sensing with cloud
  offloading,'' in \emph{Proc. 10th ACM Conf. Embedded Netw. Sensor Syst.},
  Toronto, ON, Nov. 2012, pp. 85--98.

\bibitem{Misra2014}
P.~K. Misra, W.~Hu \emph{et~al.}, ``Energy efficient {GPS} acquisition with
  sparse-{GPS},'' in \emph{Proc. 13th Int. Symp. Inf. Process. in Sensor
  Netw.}, Piscataway, NJ, Nov. 2014, pp. 155--166.

\bibitem{Green2008}
P.~Green, J.~Sullivan \emph{et~al.}, ``Integrated vehicle-based safety systems
  {(IVBSS)}: Human factors and driver-vehicle interface {(DVI)} summary
  report,'' U.S. Department of Transport., National Highway Traffic Safety
  Administration, Tech. Rep., Jan. 2008.

\bibitem{Yamabe2014}
T.~Yamabe and R.~Kiyohara, ``A study of on-vehicle information devices using a
  smartphone,'' in \emph{Proc. IEEE 38th Int. Conf. Comput. Softw. Appl.
  Workshops}, Taichung, Taiwan, Jul. 2014, pp. 561--566.

\bibitem{Watkins2011}
M.~Watkins, I.~Amaya \emph{et~al.}, ``Autonomous detection of distracted
  driving by cell phone,'' in \emph{Proc. 14th IEEE Int. Conf. Intell.
  Transport. Syst.}, Washington, DC, Oct. 2011, pp. 1960--1965.

\bibitem{Saiprasert2014}
C.~Saiprasert, S.~Supakwong \emph{et~al.}, ``Effects of smartphone usage on
  driver safety level performance in urban road conditions,'' in \emph{Proc.
  IEEE Int. Conf. Electr. Eng/Electron., Comput., Telecommun. Inf. Technol.},
  Nakhon Ratchasima, Thailand, May 2014.

\bibitem{Strayer2011}
D.~L. Strayer, J.~M. Watson, and F.~A. Drews, \emph{The Psychology of Learning
  and Motivation}, Dec. 2011, vol.~54, ch. Cognitive Distraction While
  Multitasking in the Automobile, pp. 33--58.

\bibitem{Park2013}
H.~S. Park, M.~W. Park \emph{et~al.}, ``In-vehicle {AR-HUD} system to provide
  driving-safety information,'' \emph{ETRI J.}, vol.~35, no.~6, pp. 1038--1047,
  Dec. 2013.

\bibitem{IHS2013}
``{OEM} automotive electronics - {World},'' Apr. 2013, {IHS Automotive}.

\bibitem{Matsuyama2014}
S.~Matsuyama, T.~Yamabe \emph{et~al.}, ``Intelligent user interface of
  smartphones for on-vehicle information devices,'' \emph{Procedia Comput.
  Sci.}, vol.~35, pp. 1635--1643, Sep. 2014.

\bibitem{Li2016b}
Y.~Li, G.~Zhou \emph{et~al.}, ``Determining driver phone use leveraging
  smartphone sensors,'' \emph{Accepted in Multimedia Tools Appl.}

\bibitem{Dai2010}
J.~Dai, J.~Teng \emph{et~al.}, ``Mobile phone based drunk driving detection,''
  in \emph{Proc. 4th Int. Conf. Pervasive Comput. Technol. for Healthcare},
  Munich, Germany, Mar. 2010.

\bibitem{Paefgen2012}
J.~Paefgen, F.~Kehr \emph{et~al.}, ``Driving behavior analysis with
  smartphones: insights from a controlled field study,'' in \emph{Proc. 11th
  Conf. Mobile and Ubiquitous Multimedia}, Luleå, Sweden, Dec. 2012, pp.
  36:1--8.

\bibitem{Banerjee2014}
D.~Banerjee, N.~Banerjee \emph{et~al.}, \emph{Mobile and Ubiquitous Syst.:
  Comput., Netw., Services}.\hskip 1em plus 0.5em minus 0.4em\relax Springer,
  Sep. 2014, vol. 131, ch. How's My Driving? A Spatio-Semantic Analysis of
  Driving Behavior with Smartphone Sensors, pp. 653--666.

\bibitem{Xianping2011}
F.~Xianping, M.~Yugang, and Y.~Guoliang, ``A driving behavior retrieval
  application for vehicle surveillance system,'' \emph{Int. J. Modern Educ.
  Comput. Sci.}, vol.~2, pp. 44--50, Apr. 2011.

\bibitem{Osafune2016}
T.~Osafune, T.~Takahashi \emph{et~al.}, ``Analysis of accident risks from
  driving behaviors,'' \emph{Int. J. Intell. Transport. Syst. Research}, Sep.
  2016.

\bibitem{Li2016}
Y.~Li, F.~Xue \emph{et~al.}, ``A driving behavior detection system based on a
  smartphone's built-in sensor,'' \emph{Accepted in Int. J. Commun. Syst.}

\bibitem{Almazan2013}
J.~Almazan, L.~Bergasa \emph{et~al.}, ``Full auto-calibration of a smartphone
  on board a vehicle using {IMU }and {GPS} embedded sensors,'' in \emph{Proc.
  IEEE 4th Symp. Intell. Veh.}, Gold Coast City, Australia, Jun. 2013, pp.
  1374--1380.

\bibitem{Hong2014}
J.-H. Hong, B.~Margines, and A.~K. Dey, ``A smartphone-based sensing platform
  to model aggressive driving behaviors,'' in \emph{Proc. 32nd ACM Conf. on
  Human Factors in Comput. Syst.}, Toronto, ON, Apr. 2014, pp. 4047--4056.

\bibitem{Li2012}
K.~Li, M.~Lu \emph{et~al.}, ``\BIBforeignlanguage{English}{Personalized driving
  behavior monitoring and analysis for emerging hybrid vehicles},'' in
  \emph{\BIBforeignlanguage{English}{Pervasive Comput.}}\hskip 1em plus 0.5em
  minus 0.4em\relax Springer, Jun. 2012, vol. 7319, pp. 1--19.

\bibitem{Sharma2016}
H.~Sharma, S.~Naik \emph{et~al.}, ``S-road assist: Road surface conditions and
  driving behavior analysis using smartphones,'' in \emph{Proc. 2015 Int. Conf.
  Connected Vehicles Expo}, Shenzhen, China, Oct. 2015, pp. 291--296.

\bibitem{Pfriem2014}
M.~Pfriem and F.~Gauterin, ``Employing smartphones as a low-cost multi sensor
  platform in a field operational test with electric vehicles,'' in \emph{Proc.
  47th IEEE Int. Conf. Syst. Sci.}, Manoa, HI, Jan. 2014, pp. 1143--1152.

\bibitem{Bhoraskar2012}
R.~Bhoraskar, N.~Vankadhara \emph{et~al.}, ``Wolverine: Traffic and road
  condition estimation using smartphone sensors,'' in \emph{Proc. 4th IEEE Int.
  Conf. Commun. Syst. Netw.}, Bangalore, India, Jan. 2012.

\bibitem{Ghose2016}
A.~Ghose, A.~Chowdhury \emph{et~al.}, ``An enhanced automated system for
  evaluating harsh driving using smartphone sensors,'' in \emph{Proc. 17th Int.
  Conf. Distrib. Comput. Netw.}, Singapore, Singapore, Jan. 2016, pp.
  38:1--38:6.

\bibitem{Zhao2013}
H.~Zhao, H.~Zhou \emph{et~al.}, ``{Join driving: A} smart phone-based driving
  behavior evaluation system,'' in \emph{Proc. IEEE Int. Conf. Global Commun.},
  Atlanta, GA, Dec. 2013, pp. 48--53.

\bibitem{Woo2016}
C.~Woo and D.~Kulic, ``Manoeuvre segmentation using smartphone sensors,'' in
  \emph{Proc. IEEE Symp. Intell. Vehicles}, Göteborg, Sweden, Jun. 2016, pp.
  572--577.

\bibitem{He2014}
Z.~He, J.~Cao \emph{et~al.}, ``Who sits where? {I}nfrastructure-free in-vehicle
  cooperative positioning via smartphones,'' \emph{Sensors}, vol.~14, no.~7,
  pp. 11\,605--11\,628, Jun. 2014.

\bibitem{Wallin2013}
J.~Wallin and J.~Zachrisson, ``Sensor fusion in smartphones with application to
  car racing performance analysis,'' Master's thesis, Linköping University,
  Jun. 2013.

\bibitem{Larsdotter2014}
R.~Larsdotter and D.~Jaller, ``Automatic calibration and virtual alignment of
  {MEMS}-sensor placed in vehicle for use in road condition determination
  system,'' Master's thesis, Chalmers University of Technol., Jan. 2014.

\bibitem{Kang2014}
L.~Kang, Z.~Liu, and S.~Banerjee, ``How's my driving: Sensing driving
  behaviours by using smartphones,'' University of Wisconsin-Madison, Tech.
  Rep., 2014.

\bibitem{Bruwer2015}
F.~Bruwer and M.~Booysen, ``Vehicle acceleration estimation using smart-phone
  based sensors,'' in \emph{Southern African Transport Conf.}, Pretoria, South
  Africa, Jul. 2015.

\bibitem{Promwongsa2014}
N.~Promwongsa, P.~Chaisatsilp \emph{et~al.}, ``Automatic accelerometer
  reorientation for driving event detection using smartphone,'' in \emph{Proc.
  13th ITS Asia Pacific Forum}, Auckland, New Zealand, Apr. 2014.

\bibitem{Khalegi2015}
B.~Khaleghi, A.~El-Ghazal \emph{et~al.}, ``Opportunistic calibration of
  smartphone orientation in a vehicle,'' in \emph{Proc. IEEE Int. Symp. World
  of Wireless, Mobile Multimedia Netw.}, Boston, MA, Jun. 2015.

\bibitem{Niu2012}
X.~Niu, Q.~Zhang \emph{et~al.}, ``Using inertial sensors of i{P}hone 4 for car
  navigation,'' in \emph{Proc. IEEE/ION Position, Location Navig. Symp.},
  Myrtle Beach, SC, Apr. 2012, pp. 555--561.

\bibitem{Gikas2016}
V.~Gikas and H.~Perakis, ``Rigorous performance evaluation of smartphone
  {GNSS/IMU} sensors for {ITS} applications,'' \emph{Sensors}, vol.~16, no.~8,
  Aug. 2016.

\bibitem{Forster2012}
M.~Forster, R.~Frank, and T.~Engel, ``Evaluation of sensors in modern
  smartphones for vehicular traffic monitoring,'' University of Luxembourg,
  Tech. Rep., 2012.

\bibitem{Chu2014}
H.~Chu, V.~Raman \emph{et~al.}, ``I am a smartphone and {I} know my user is
  driving,'' in \emph{Proc. 6th Int. Conf. Commun. Syst. Netw.}, Bangalore,
  India, Jan. 2014.

\bibitem{Pote2016}
C.~Pote, D.~Bhargude \emph{et~al.}, ``Car driver detection and accident
  prevention system,'' \emph{Int. Eng. Research J.}, vol.~2, no.~2, pp.
  625--629, Apr. 2016.

\bibitem{Yang2012}
J.~Yang, S.~Sidhom \emph{et~al.}, ``Sensing driver phone use with acoustic
  ranging through car speakers,'' \emph{IEEE Trans. Mobile Comput.}, vol.~11,
  no.~9, pp. 1426--1440, Sep. 2012.

\bibitem{Feld2010}
M.~Feld, T.~Schwartz, and C.~Muller, ``\BIBforeignlanguage{English}{This is me:
  Using ambient voice patterns for in-car positioning},'' in
  \emph{\BIBforeignlanguage{English}{Ambient Intell.}}\hskip 1em plus 0.5em
  minus 0.4em\relax Springer, Nov. 2010, pp. 290--294.

\bibitem{Zhang2016}
C.~Zhang, M.~Patel \emph{et~al.}, ``Driver classification based on driving
  behaviors,'' in \emph{Proc. 21st Int. Conf. Intell. User Interfaces}, Sonoma,
  CA, Mar. 2016, pp. 80--84.

\bibitem{Paruchuri2015}
V.~Paruchuri and A.~Kumar, ``Detecting driver distraction using smartphones,''
  in \emph{Proc. 29th IEEE Int. Conf. Advanced Inf. Netw. Appl.}, Gwangiu,
  South Korea, Mar. 2015, pp. 468--475.

\bibitem{He2015}
J.~He, A.~Chaparro \emph{et~al.}, ``Mutual interferences of driving and texting
  performance,'' \emph{Comput. Human Behavior}, vol.~52, pp. 115--123, Nov.
  2015.

\bibitem{Dong2013}
H.~Dong, M.~Wu \emph{et~al.}, ``Urban residents travel analysis based on mobile
  communication data,'' in \emph{Proc. 16th IEEE Int. Conf. Intell. Transport.
  Syst.}, The Hague, the Netherlands, Oct. 2013, pp. 1487--1492.

\bibitem{Bierlaire2013}
M.~Bierlaire, J.~Chen, and J.~Newman, ``A probabilistic map matching method for
  smartphone {GPS} data,'' \emph{Transport. Research Part C: Emerging
  Technol.}, vol.~26, pp. 78--98, Jan. 2013.

\bibitem{Haklay2008}
M.~Haklay and P.~Weber, ``{OpenStreetMap}: User-generated street maps,''
  \emph{IEEE Pervasive Comput.}, vol.~7, no.~4, pp. 12--18, Oct. 2008.

\bibitem{Hashemi2015}
P.~Hashemi and R.~Ali~Abbaspour, ``\BIBforeignlanguage{English}{Assessment of
  logical consistency in {OpenStreetMap} based on the spatial similarity
  concept},'' in \emph{\BIBforeignlanguage{English}{OpenStreetMap in
  GISci.}}\hskip 1em plus 0.5em minus 0.4em\relax Springer, Mar. 2015, pp.
  19--36.

\bibitem{Haklay2010}
M.~Haklay, ``How good is volunteered geographical information? {A comparative
  study of OpenStreetMap}\ and ordnance survey datasets,'' \emph{Environment
  and Planning B: Planning and Design}, vol.~37, no.~4, pp. 682--703, Aug.
  2010.

\bibitem{Hu2013}
S.~Hu, L.~Su \emph{et~al.}, ``{SmartRoad: A} crowd-sourced traffic regulator
  detection and identification system,'' Department of Comput. Sci., University
  of Illinois at Urbana-Champaign, Tech. Rep., 2013.

\bibitem{Bo2013b}
C.~Bo, X.-Y. Li \emph{et~al.}, ``{SmartLoc: Sensing} landmarks silently for
  smartphone based metropolitan localization,'' \emph{CoRR}, vol.
  arXiv:1310.8187, Aug. 2013.

\bibitem{Tan2014}
G.~Tan, M.~Lu \emph{et~al.}, ``Bumping: A bump-aided inertial navigation method
  for indoor vehicles using smartphones,'' \emph{IEEE Trans. Parallel Distrib.
  Syst.}, vol.~25, no.~7, pp. 1670--1680, Jul. 2014.

\bibitem{Aly2016}
H.~Aly, A.~Basalamah, and M.~Youssef, ``Robust and ubiquitous smartphone-based
  lane detection,'' \emph{Pervasive Mobile Comput.}, vol.~26, pp. 35--56, Feb.
  2016.

\bibitem{Godha2005b}
S.~Godha and M.~E. Cannon, ``Development of a {DGPS/MEMS IMU} integrated system
  for navigation in urban canyon conditions,'' in \emph{Proc. Int. Symp. on
  {GPS/GNSS}}, Hong Kong, Dec. 2005.

\bibitem{Niu2010}
X.~Niu, H.~Zhang \emph{et~al.}, ``Using land-vehicle steering constraint to
  improve the heading estimation of {MEMS GPS/INS} georeferencing systems,'' in
  \emph{Proc. Int. Conf. Canadian Geomatics}, Calgary, AB, Jun. 2010.

\bibitem{Marti2014}
E.~Marti, J.~Garcia, and J.~Molina, ``Navigation capabilities of mid-cost
  {GNSS/INS} vs. smartphone: Analysis and comparison in urban navigation
  scenarios,'' in \emph{Proc. 17th Int. Conf. Inf. Fusion}, Salamanca, Spain,
  Jul. 2014.

\bibitem{Zhou2014b}
P.~Zhou, Y.~Zheng, and M.~Li, ``How long to wait? {Predicting} bus arrival time
  with mobile phone based participatory sensing,'' \emph{IEEE Trans. Mobile
  Comput.}, vol.~13, no.~6, pp. 1228--1241, Jun. 2014.

\bibitem{Sohn2006}
T.~Sohn, A.~Varshavsky \emph{et~al.}, ``\BIBforeignlanguage{English}{Mobility
  detection using everyday {GSM} traces},'' in
  \emph{\BIBforeignlanguage{English}{Ubiquitous Comput.}}\hskip 1em plus 0.5em
  minus 0.4em\relax Springer, Sep. 2006, vol. 4206, pp. 212--224.

\bibitem{Anderson2006}
I.~Anderson and H.~Muller, ``Practical activity recognition using {GSM} data,''
  Dep. of Comp. Sci., Bristol, Tech. Rep., 2006.

\bibitem{Bolbol2012}
A.~Bolbol, T.~Cheng \emph{et~al.}, ``Inferring hybrid transportation modes from
  sparse {GPS} data using a moving window {SVM} classification,''
  \emph{Comput., Environment Urban Syst.}, vol.~36, no.~6, pp. 526--537, Nov.
  2012.

\bibitem{Stenneth2012}
L.~Stenneth, K.~Thompson \emph{et~al.}, ``Automated transportation transfer
  detection using {GPS} enabled smartphones,'' in \emph{Proc. 15th IEEE Int.
  Conf. Intell. Transport. Syst.}, Anchorage, AK, Sep. 2012, pp. 802--807.

\bibitem{Hemminki2013}
S.~Hemminki, P.~Nurmi, and S.~Tarkoma, ``Accelerometer-based transportation
  mode detection on smartphones,'' in \emph{Proc. 11th ACM Int. Conf. Embedded
  Netw. Sensor Syst.}, Roma, Italy, Nov. 2013, pp. 13:1--13:14.

\bibitem{Thiagarajan2010}
A.~Thiagarajan, J.~Biagioni \emph{et~al.}, ``Cooperative transit tracking using
  smart-phones,'' in \emph{Proc. 8th ACM Conf. Embedded Netw. Sensor Syst.},
  Zurich, Switzerland, Nov. 2010, pp. 85--98.

\bibitem{Lari2015}
Z.~A. Lari and A.~Golroo, ``Automated transportation mode detection using smart
  phone applications via machine learning: {Case} study mega city of
  {Tehran},'' in \emph{Transport. Research Board 94th Annu. Meeting},
  Washington DC, Jan. 2015.

\bibitem{Assemi2016}
B.~Assemi, H.~Safi \emph{et~al.}, ``Developing and validating a statistical
  model for travel mode identification on smartphones,'' \emph{IEEE Trans.
  Intell. Transport. Syst.}, vol.~17, no.~7, pp. 1920--1931, Jul. 2016.

\bibitem{Montoya2015}
D.~Montoya, S.~Abiteboul, and P.~Senellart, ``{Hup-me: Inferring} and
  reconciling a timeline of user activity from rich smartphone data,'' in
  \emph{Proc. 23rd Int. Conf. Advances in Geographic Inf. Syst.}, Bellevue, WA,
  Nov. 2015, pp. 62:1--62:4.

\bibitem{Eftekhari2016}
H.~R. Eftekhari and M.~Ghatee, ``An inference engine for smartphones to
  preprocess data and detect stationary and transportation modes,''
  \emph{Transport. Research Part C: Emerging Technol.}, vol.~69, pp. 313--327,
  Aug. 2016.

\bibitem{Li2016c}
C.~Li, P.~C. Zegras \emph{et~al.}, ``{FMS-TQ}: combining smartphone and
  {iBeacon} 4 technologies in a transit quality survey,'' in
  \emph{Transportation Research Board 95th Meeting}, Washington, DC, Jan. 2016.

\bibitem{Girling1996}
P.~Girling, Ed., \emph{Automotive Handbook}, 4th~ed.\hskip 1em plus 0.5em minus
  0.4em\relax Robert Bosch GmbH, 1996.

\bibitem{Progressive2012}
``Linking driving behavior to automobile accidents and insurance rates,''
  Progressive, Tech. Rep., Jul. 2012.

\bibitem{Klauer2009}
S.~G. Klauer, T.~A. Dingus \emph{et~al.}, ``Comparing real-world behaviors of
  drivers with high versus low rates of crashes and near-crashes,'' National
  Highway Traffic Safety Administration, Tech. Rep., Feb. 2009.

\bibitem{Yamakado2009}
M.~Yamakado, J.~Takahashi \emph{et~al.}, ``G-vectoring, new vehicle dynamics
  control technology for safe driving,'' \emph{Ind. Syst.}, vol.~58, no.~7, pp.
  346--350, Dec. 2009.

\bibitem{Saiprasert2013b}
C.~Saiprasert, T.~Pholprasit, and W.~Pattara-Atikom, ``Detecting driving events
  using smartphone,'' in \emph{Proc. 20th Intell. Transport. Syst. World
  Congress}, Tokyo, Japan, Oct. 2013.

\bibitem{Saiprasert2014b}
C.~Saiprasert, S.~Thajchayapong \emph{et~al.}, ``Driver behaviour profiling
  using smartphone sensory data in a {V2I} environment,'' in \emph{Proc. IEEE
  Int. Conf. Connected Vehicles Expo}, Vienna, Austria, Nov. 2014, pp.
  552--557.

\bibitem{Saiprasert2015}
C.~Saiprasert, T.~Pholprasit, and S.~Thajchayapong,
  ``\BIBforeignlanguage{English}{Detection of driving events using sensory data
  on smartphone},'' \emph{\BIBforeignlanguage{English}{Int. J. Intell.
  Transpor. Syst. Research}}, Jul. 2015.

\bibitem{Bruwer2015b}
F.~J. Bruwer and M.~J. Booysen, ``Comparison of {GPS and MEMS} support for
  smartphone-based driver behavior monitoring,'' in \emph{Proc. IEEE Symp.
  Comput. Intell.}, Cape Town, South Africa, Dec. 2015, pp. 434--441.

\bibitem{Skog2009}
I.~Skog and P.~Händel, ``In-car positioning and navigation technologies - {A}
  survey,'' \emph{IEEE Trans. Intell. Transport. Syst.}, vol.~10, no.~1, pp.
  4--21, Mar. 2009.

\bibitem{Castignani2015}
G.~Castignani, T.~Derrmann \emph{et~al.}, ``Driver behavior profiling using
  smartphones: A low-cost platform for driver monitoring,'' \emph{IEEE Intell.
  Transport. Syst. Mag.}, vol.~7, no.~1, pp. 91--102, Jan. 2015.

\bibitem{Stoichkov2013}
R.~Stoichkov, ``{Android} smartphone application for driving style
  recognition,'' Dept. Electr. Eng. Inf. Technol., Tech. Rep., Jul. 2013.

\bibitem{Chen2015}
D.~Chen, K.-T. Cho \emph{et~al.}, ``Invisible sensing of vehicle steering with
  smartphones,'' in \emph{Proc. 13th Int. Conf. Mobile Syst., Appl., Services},
  Florence, Italy, May 2015.

\bibitem{Bergasa2014}
L.~Bergasa, D.~Almeria \emph{et~al.}, ``{DriveSafe: An} app for alerting
  inattentive drivers and scoring driving behaviors,'' in \emph{Proc. IEEE
  Intell. Veh. Symp.}, Dearborn, MI, Jun. 2014, pp. 240--245.

\bibitem{Chowdhury2015}
A.~Chowdhury, T.~Banerjee \emph{et~al.}, ``Smartphone based estimation of
  relative risk propensity for inducing good driving behavior,'' in \emph{Proc.
  ACM Int. Conf. Pervasive Ubiquitous Comput. and Proc. ACM Int. Symp. Wearable
  Comput.}, Osaka, Japan, Sep. 2015, pp. 743--751.

\bibitem{Chowdhury2015b}
A.~Chowdhury, T.~Chakravarty \emph{et~al.}, ``Aggregate driver model to enable
  predictable behaviour,'' \emph{J. Physics: Conf. Series}, vol. 633, no.~1,
  Jun. 2015.

\bibitem{Banerjee2016}
T.~Banerjee, A.~Chowdhury, and T.~Chakravarty, ``{MyDrive: Drive} behavior
  analytics method and platform,'' in \emph{Proc. 3rd Workshop Physical
  Analytics}, Singapore, Singapore, Jun. 2016, pp. 7--12.

\bibitem{Vaiana2014}
R.~Vaiana, T.~Iuele \emph{et~al.}, ``Driving behavior and traffic safety: An
  acceleration-based safety evaluation procedure for smartphones,''
  \emph{Modern Appl. Sci.}, vol.~8, no.~1, pp. 88--96, Jan. 2014.

\bibitem{Wahlstrom2014}
J.~Wahlström, I.~Skog, and P.~Händel, ``Risk assessment of vehicle cornering
  events in gnss data driven insurance telematics,'' in \emph{Proc. 17th IEEE
  Int. Conf. Intell. Transport. Syst.}, Qingdao, China, Oct. 2014, pp.
  3132--3137.

\bibitem{Akhtar2014}
N.~Akhtar, K.~Pandey, and S.~Gupta, ``Mobile application for safe driving,'' in
  \emph{Proc. 4th IEEE Int. Conf. Commun. Syst. Netw. Technol.}, Bhopal, India,
  Apr. 2014, pp. 212--216.

\bibitem{Daptardar2015}
S.~Daptardar, V.~Lakshminarayanan \emph{et~al.}, ``Hidden markov model based
  driving event detection and driver profiling from mobile inertial sensor
  data,'' in \emph{SENSORS}, Busan, South Korea, Nov. 2015.

\bibitem{Engelbrecht2014}
J.~Engelbrecht, M.~J. Booysen, and G.-J. van Rooyen, ``Recognition of driving
  manoeuvres using smartphone-based inertial and {GPS} measurement,'' in
  \emph{Proc. 1st Int. Conf. Use of Mobile ICT}, Stellenbosch, South Africa,
  Dec. 2014.

\bibitem{Engelbrecht2015b}
J.~Engelbrecht, M.~J. Booysen \emph{et~al.}, ``{Performance Comparison of
  Dynamic Time Warping (DTW) and a Maximum Likelihood (ML) Classifier in
  Measuring Driver Behavior with Smartphones},'' in \emph{Proc. IEEE Symp.
  Comput. Intell.}, Cape Town, South Africa, Dec 2015, pp. 427--433.

\bibitem{Pholprasit2015}
T.~Pholprasit, W.~Choochaiwattana, and C.~Saiprasert, ``A comparison of driving
  behaviour prediction algorithm using multi-sensory data on a smartphone,'' in
  \emph{Proc. 16th IEEE/ACIS Int. Conf. Software Eng., Artificial Intell.,
  Netw. Parallel/Distributed Comput.}, Takamatsu, Japan, Jun. 2015.

\bibitem{Castignani2013}
G.~Castignani, R.~Frank, and T.~Engel, ``Driver behavior profiling using
  smartphones,'' in \emph{Proc. 16th IEEE Int. Conf. Intell. Transport. Syst.},
  The Hague, the Netherlands, Oct. 2013, pp. 552--557.

\bibitem{Buscarino2014}
A.~Buscarino, L.~Fortuna, and M.~Frasca, ``Driving assistance using
  smartdevices,'' in \emph{Proc. IEEE Int. Symp. Intell. Control}, Antibes,
  France, Oct. 2014, pp. 838--842.

\bibitem{Fazeen2012}
M.~Fazeen, B.~Gozick \emph{et~al.}, ``Safe driving using mobile phones,''
  \emph{IEEE Trans. Intell. Transport. Syst.}, vol.~13, no.~3, pp. 1462--1468,
  Sep. 2012.

\bibitem{Kalra2014b}
N.~Kalra, G.~Chugh, and D.~Bansal, ``Analyzing driving and road events via
  smartphone,'' \emph{Int. J. Comput. Appl.}, vol.~98, no.~12, pp. 5--9, Jul.
  2014.

\bibitem{Chigurupati2012}
S.~Chigurupati, S.~Polavarapu \emph{et~al.}, ``Integrated computing system for
  measuring driver safety index,'' \emph{Int. J. Emerg. Technol. Advanced
  Eng.}, vol.~2, no.~6, Jun. 2012.

\bibitem{Chakravarty2013}
T.~Chakravarty, A.~Ghose \emph{et~al.}, ``{MobiDriveScore: A} system for mobile
  sensor based driving analysis: {A} risk assessment model for improving one's
  driving,'' in \emph{Proc. 7th Int. Conf. Sens. Technol.}, Wellington, New
  Zealand, Dec. 2013, pp. 338--344.

\bibitem{Khedkar2015}
S.~Khedkar, A.~Oswal \emph{et~al.}, ``Driver evaluation system using mobile
  phone and {OBD-II} system,'' \emph{Int. J. Comput. Sci. Inf. Technol.},
  vol.~6, no.~3, pp. 2738--2745, May 2015.

\bibitem{Ylizaliturri-Salcedo2015}
M.~A. Ylizaliturri-Salcedo, M.~Tentori, and J.~A. Garcia-Macias,
  \emph{Ubiquitous Comput. Ambient Intell.. Sensing, Process., Using
  Environmental Inf.}\hskip 1em plus 0.5em minus 0.4em\relax Springer, Dec.
  2015, vol. 9454, ch. Detecting Aggressive Driving Behavior with Participatory
  Sensing, pp. 249--261.

\bibitem{Ouyang2016}
Z.~Ouyang, J.~Niu \emph{et~al.}, ``{Multiwave: A} novel vehicle steering
  pattern detection method based on smartphones,'' in \emph{Proc. IEEE ICC
  Ad-hoc Sensor Netw. Symp.}, Kuala Lumpur, Malaysia, May 2016.

\bibitem{AbuAli2015}
N.~AbuAli, ``Advanced vehicular sensing of road artifacts and driver
  behavior,'' in \emph{Proc. IEEE Symp. Comput. Commun.}, Larnaca, Cyprus, Jul.
  2015, pp. 45--49.

\bibitem{Carmona2015}
J.~Carmona, F.~García \emph{et~al.}, ``Data fusion for driver behaviour
  analysis,'' \emph{Sensors}, vol.~15, no.~10, pp. 25\,968--25\,991, Oct. 2015.

\bibitem{Hosseinioun2015}
S.~V. Hosseinioun, H.~Al-Osman, and A.~E. Saddik, ``Employing sensors and
  services fusion to detect and assess driving events,'' in \emph{Proc. IEEE
  Int. Symp. Multimedia}, Miami, FL, Dec. 2015, pp. 395--398.

\bibitem{Tchankue2013}
P.~Tchankue, J.~Wesson, and D.~Vogts, ``Using machine learning to predict the
  driving context whilst driving,'' in \emph{Proc. South African Institute
  Comput. Sci. Inf. Technol. Conf.}, East London, South Africa, Oct. 2013, pp.
  47--55.

\bibitem{Chaovalit2013}
P.~Chaovalit, C.~Saiprasert, and T.~Pholprasit, ``A method for driving event
  detection using {SAX} on smartphone sensors,'' in \emph{Proc. 13th Int. Conf.
  ITS Telecommun.}, Tampere, Finland, Nov. 2013, pp. 450--455.

\bibitem{Castignani2015b}
G.~Castignani, T.~Derrmann \emph{et~al.}, ``Validation study of risky event
  classification using driving pattern factors,'' in \emph{Proc. IEEE Symp.
  Commun. Vehicular Technol. in the Benelux}, Luxembourg-Kirchberg, Luxembourg,
  Nov. 2015.

\bibitem{Bhoyar2013}
V.~Bhoyar, P.~Lata \emph{et~al.}, ``Symbian based rash driving detection
  system,'' \emph{Int. J. Emerging Trends \& Technol in Comput. Sci.}, vol.~2,
  no.~2, pp. 124--126, Mar. 2013.

\bibitem{Antoniou2014}
C.~Antoniou, V.~Papathanasopoulou \emph{et~al.}, ``Classification of driving
  characteristics using smartphone sensor data,'' in \emph{Proc. 3rd Symp.
  European Association for Research in Transport.}, Leeds, UK, Sep. 2014.

\bibitem{Meseguer2013}
J.~Meseguer, C.~Calafate \emph{et~al.}, ``{DrivingStyles: A} smartphone
  application to assess driver behavior,'' in \emph{Proc. IEEE Symp. Comput.
  Commun.}, Split, Croatia, Jul. 2013, pp. 535--540.

\bibitem{Chen2015b}
Z.~Chen, J.~Yu \emph{et~al.}, ``D3: Abnormal driving behaviors detection and
  identification using smartphone sensors,'' in \emph{Proc. IEEE Int. Conf.
  Sensing, Commun. Netw.}, Seattle, WA, Jun. 2015, pp. 524--532.

\bibitem{Tecimer2013}
A.~A. Tecimer, Z.~C. Taysi \emph{et~al.}, ``Assessment of vehicular
  transportation quality via smartphones,'' \emph{Turkish J. Elect. Eng. \&
  Comput. Sci.}, vol.~23, pp. 2161--2170, Aug. 2013.

\bibitem{Ly2013}
M.~V. Ly, S.~Martin, and M.~Trivedi, ``Driver classification and driving style
  recognition using inertial sensors,'' in \emph{Proc. 6th. IEEE Symp. Intell.
  Vehicles}, Gold Coast City, Australia, Jun. 2013, pp. 1040--1045.

\bibitem{Lee2004}
S.~E. Lee, E.~C.~B. Olsen, and W.~W. Wierwille, ``A comprehensive examination
  of naturalistic lane-changes,'' US Department of Transport., Tech. Rep., Mar.
  2004.

\bibitem{Joubert2016}
J.~W. Joubert, D.~de~Beer, and N.~de~Koker, ``Combining accelerometer data and
  contextual variables to evaluate the risk of driver behaviour,''
  \emph{Transport. Research Part F: Traffic Psychology and Behaviour}, vol. 41,
  Part A, pp. 80--96, Aug. 2016.

\bibitem{Toledo2008}
T.~Toledo, O.~Musicant, and T.~Lotan, ``In-vehicle data recorders for
  monitoring and feedback on driver's behavior,'' \emph{Transport. Research
  Part C: Emerging Technol.}, vol.~16, no.~3, pp. 320--331, Jun. 2008.

\bibitem{Whipple2009}
J.~Whipple, W.~Arensman, and M.~Boler, ``A public safety application of
  {GPS}-enabled smartphones and the {Android} operating system,'' in
  \emph{Proc. IEEE Int. Conf. Systems, Man and Cybernetics}, San Antonio, TX,
  Oct. 2009, pp. 2059--2061.

\bibitem{Chakravarty2014}
T.~Chakravarty, A.~Chowdhury \emph{et~al.}, ``Statistical analysis of
  road-vehicle-driver interaction as an enabler to designing behavioral
  models,'' \emph{Int. J. Model. Simul. Sci. Comput.}, vol.~5, Oct. 2014.

\bibitem{Lindfors2016}
M.~Lindfors, G.~Hendeby \emph{et~al.}, ``Vehicle speed tracking using chassis
  vibrations,'' in \emph{Proc. IEEE Intell. Vehicles Symp.}, Göteborg, Sweden,
  Jun. 2016, pp. 214--219.

\bibitem{Handel2009}
P.~Händel, ``Discounted least-squares gearshift detection using accelerometer
  data,'' \emph{IEEE Trans. Instrum. Meas.}, vol.~58, no.~12, pp. 3953--3958,
  Dec. 2009.

\bibitem{Singh2012}
S.~Singh, S.~Nelakuditi \emph{et~al.}, ``Your smartphone can watch the road and
  you: Mobile assistant for inattentive drivers,'' in \emph{Proc. 13th ACM Int.
  Symp. Mobile Ad Hoc Netw. Comput.}, Hilton Head, SC, Jun. 2012, pp. 261--262.

\bibitem{You2013}
C.-W. You, N.~D. Lane \emph{et~al.}, ``{CarSafe App: Alerting} drowsy and
  distracted drivers using dual cameras on smartphones,'' in \emph{Proc. 11th
  Int. Conf. Mobile Syst., Appl., Services}, Taipei, Taiwan, Jun. 2013, pp.
  13--26.

\bibitem{Douangphachanh2014thesis}
V.~Douangphachanh, ``The development of a simple method for networkwide road
  surface roughness condition estimation and monitoring using smartphone
  sensors,'' Ph.D. dissertation, Tokyo Metropolitan University, Sep. 2014.

\bibitem{Seraj2016}
F.~Seraj, B.~J. van~der Zwaag \emph{et~al.}, \emph{Big Data Analytics in the
  Social and Ubiquitous Context}.\hskip 1em plus 0.5em minus 0.4em\relax
  Springer, Jan. 2016, vol. 9546, ch. RoADS: A Road Pavement Monitoring System
  for Anomaly Detection Using Smart Phones, pp. 128--146.

\bibitem{Strazdins2011}
G.~Strazdins, A.~Mednis \emph{et~al.}, ``Towards vehicular sensor networks with
  {Android} smartphones for road surface monitoring,'' in \emph{Proc. 2nd Int.
  Workshop Netw. Cooperating Objects}, Chicago, IL, Apr. 2011.

\bibitem{Mednis2011}
A.~Mednis, G.~Strazdins \emph{et~al.}, ``Real time pothole detection using
  {Android} smartphones with accelerometers,'' in \emph{Proc. Int. Conf.
  Distrib. Comput. Sensor Syst. Workshops}, Barcelona, Spain, Jun. 2011.

\bibitem{Ghose2012}
A.~Ghose, P.~Biswas \emph{et~al.}, ``Road condition monitoring and alert
  application: Using in-vehicle smartphone as internet-connected sensor,'' in
  \emph{Proc. IEEE Int. Conf. Pervasive Comput. Commun. Workshops}, Lugano,
  Switzerland, Mar. 2012, pp. 489--491.

\bibitem{Astarita2012}
V.~Astarita, M.~V. Caruso \emph{et~al.}, ``A mobile application for road
  surface quality control,'' in \emph{Proc. 15th Meeting {EURO} Working Group
  on Transport.}, Paris, France, Sep. 2012, pp. 1135--1144.

\bibitem{Darawade2016}
K.~Darawade, P.~Karmare \emph{et~al.}, ``Estimation of road surface roughness
  condition from android smartphone sensors,'' \emph{Int. J. Recent Trends in
  Eng. \& Research}, vol.~2, no.~3, pp. 339--346, Mar. 2016.

\bibitem{Mahajan2015b}
D.~V. Mahajan and T.~Dange, ``Analysis of road smoothness based on
  smartphones,'' \emph{Int. J. Innovative Research in Comput. Commun. Eng.},
  vol.~3, no.~6, pp. 5201--5206, Jun. 2015.

\bibitem{Toke2016}
P.~S. Toke, S.~Maheshwari \emph{et~al.}, ``Road quality \& valley complexity
  analysis using android application,'' \emph{Int. J. Advanced Research Comput.
  Commun. Eng.}, vol.~5, no.~5, pp. 732--736, May 2016.

\bibitem{Kulkarni2016}
K.~Kulkarni, K.~V. Prashant \emph{et~al.}, ``Predicting road anomalies using
  sensors in smartphones,'' \emph{Imperial J. Interdisciplinary Research},
  vol.~2, no.~6, pp. 219--222, 2016.

\bibitem{Lanjewar2016}
B.~Lanjewar, R.~Sagar \emph{et~al.}, ``Road bump and intensity detection using
  smartphone sensors,'' \emph{Int. J. Innovative Research Comput. Commun.
  Eng.}, vol.~4, no.~5, pp. 9185--9192, May 2016.

\bibitem{Tonde2015}
V.~P. Tonde, A.~Jadhav \emph{et~al.}, ``Road quality and ghats complexity
  analysis using android sensors,'' \emph{Int. J. Advanced Research Comput.
  Commun. Eng.}, vol.~4, no.~3, pp. 101--104, Mar. 2015.

\bibitem{Perttunen2011}
M.~Perttunen, O.~Mazhelis \emph{et~al.},
  ``\BIBforeignlanguage{English}{Distributed road surface condition monitoring
  using mobile phones},'' in \emph{\BIBforeignlanguage{English}{Ubiquitous
  Intell. Comput.}}, Sep. 2011, vol. 6905, pp. 64--78.

\bibitem{Mednis2010}
A.~Mednis, G.~Strazdins \emph{et~al.}, \emph{Netw. Digit. Technol.}, Jan. 2010,
  vol.~88, ch. {RoadMic: Road} surface monitoring using vehicular sensor
  networks with microphones, pp. 417--429.

\bibitem{Rajamohan2015}
D.~Rajamohan, B.~Gannu, and K.~S. Rajan, ``{MAARGHA: A} prototype system for
  road condition and surface type estimation by fusing multi-sensor data,''
  \emph{ISPRS Int. J. Geo-Inf.}, vol.~4, no.~3, pp. 1225--1245, Jul. 2015.

\bibitem{Orhan2013}
F.~Orhan and P.~E. Eren, ``Road hazard detection and sharing with multimodal
  sensor analysis on smartphones,'' in \emph{Proc. 7th Int. Conf. Next
  Generation Mobile Apps, Services Technol.}, Prague, Czech Republic, Sep.
  2013, pp. 56--61.

\bibitem{Wang2015}
H.-W. Wang, C.-H. Chen \emph{et~al.}, ``A real-time pothole detection approach
  for intelligent transportation system,'' \emph{Mathematical Problems in
  Eng.}, Aug. 2015.

\bibitem{Vittorio2014}
A.~Vittorio, V.~Rosolino \emph{et~al.}, ``Automated sensing system for
  monitoring of road surface quality by mobile devices,'' \emph{Procedia -
  Social and Behavioral Sci.}, vol. 111, pp. 242--251, Feb. 2014.

\bibitem{Seraj2015}
F.~Seraj, N.~Meratnia \emph{et~al.}, ``A smartphone based method to enhance
  road pavement anomaly detection by analyzing the driver behavior,'' in
  \emph{Proc. ACM Int. Conf. Pervasive Ubiquitous Comput.}, Osaka, Japan, Sep.
  2015, pp. 1169--1177.

\bibitem{Jain2012}
M.~Jain, N.~I. S.~B. Ajeet Pal~Singh, JIIT, and S.~Kaul, ``Speed-breaker early
  warning system,'' in \emph{Proc. 6th USENIX/ACM Workshop Netw. Syst. for
  Developing Regions}, Boston, MA, Jun. 2012.

\bibitem{Alessandroni2014}
G.~Alessandroni, L.~C. Klopfenstein \emph{et~al.}, ``{SmartRoadSense:
  Collaborative} road surface condition monitoring,'' in \emph{Proc. 8th Int.
  Conf. Mobile Ubiquitous Comput., Syst., Services Technol.}, Rome, Italy, Aug.
  2014, pp. 210--215.

\bibitem{Tai2010}
Y.-c. Tai, C.-w. Chan, and J.~Y.-j. Hsu, ``Automatic road anomaly detection
  using smart mobile device,'' in \emph{Proc. Conf. Technol. Appl. Artificial
  Intell.}, Hsinchu, Taiwan, Nov. 2010.

\bibitem{Forslof2015}
L.~Forslöf and H.~Jones, ``{Roadroid: Continuous} road condition monitoring
  with smart phones,'' \emph{J. Civil Eng. Architecture}, vol.~9, pp. 485--496,
  2015.

\bibitem{Mohamed2015}
A.~Mohamed, M.~M.~M. Fouad \emph{et~al.}, \emph{Advances in Intelligent Systems
  and Computing}.\hskip 1em plus 0.5em minus 0.4em\relax Springer, Sep. 2014,
  vol. 323, ch. RoadMonitor: An Intelligent Road Surface Condition Monitoring
  System, pp. 377--387.

\bibitem{Mahajan2015}
D.~V. Mahajan and T.~Dange, ``Estimation of road roughness condition by using
  sensors in smartphones,'' \emph{Int. J. Comput. Eng. \& Technol.}, vol.~6,
  no.~7, pp. 41--49, Jul. 2015.

\bibitem{Skog2006}
I.~Skog, A.~Schumacher, and P.~Händel, ``A versatile {PC}-based platform for
  inertial navigation,'' in \emph{Proc. Nordic Signal Process. Symp.},
  Reykjavik, Iceland, Jun. 2006, pp. 262--265.

\bibitem{Skog2006b}
I.~Skog and P.~Händel, ``Calibration of a {MEMS} inertial measurement unit,''
  in \emph{Proc. IMEKO XVIII World Congress}, Rio de Janerio, Brazil, Sep.
  2006.

\bibitem{Handel2010}
P.~Händel, B.~Enstedt, and M.~Ohlsson, ``Combating the effect of chassis squat
  in vehicle performance calculations by accelerometer measurements,''
  \emph{Meas.}, vol.~43, no.~4, pp. 483--488, May 2010.

\end{thebibliography}

\begin{IEEEbiography}[{\includegraphics[width=1in,height=1.25in,clip,keepaspectratio]{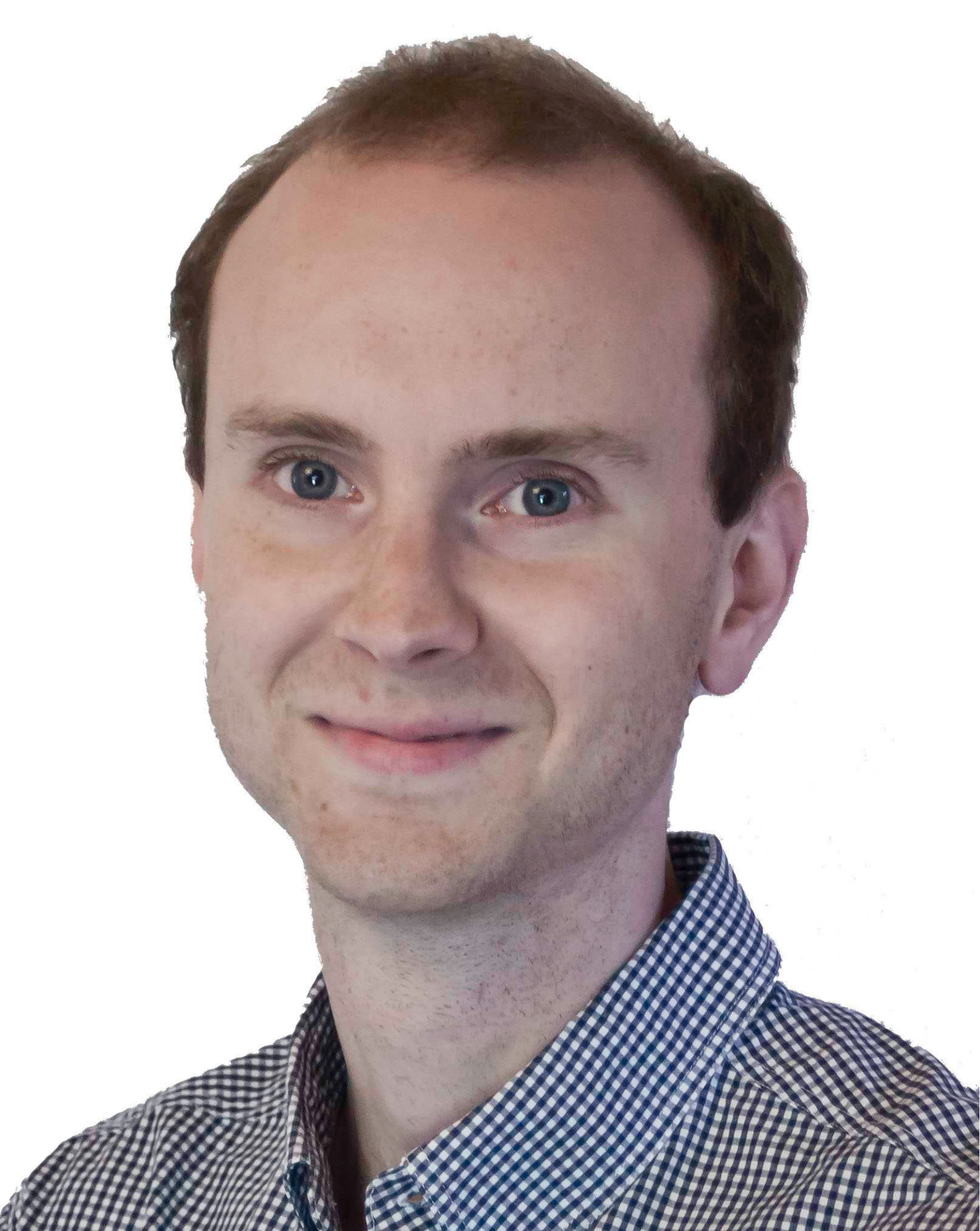}}]
{Johan Wahlstr\"{o}m} received his MSc degree in Engineering Physics from the KTH Royal Institute of Technology, Stockholm, Sweden, in 2014. He subsequently joined the Signal Processing Department at KTH, working towards his PhD. His main research topic is smartphone-based automotive navigation. In 2015, he received a scholarship from the Sweden-America foundation and spent six months at Washington University, St. Louis, USA.
\end{IEEEbiography}

\begin{IEEEbiography}[{\includegraphics[width=1in,height=1.25in,clip,keepaspectratio]{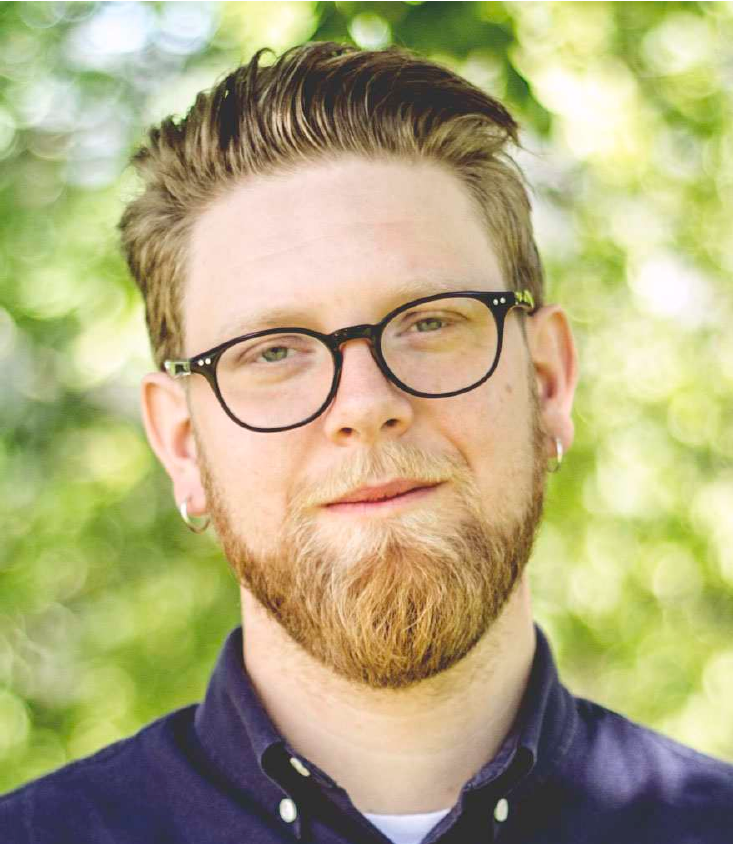}}]
{Isaac Skog}(S'09-M'10) received the BSc and MSc degrees in Electrical Engineering from the KTH Royal Institute of Technology, Stockholm, Sweden, in 2003 and 2005, respectively. In 2010, he received the Ph.D. degree in Signal Processing with a thesis on low-cost navigation systems. In 2009, he spent 5 months at the Mobile Multi-Sensor System
research team, University of Calgary, Canada, as a visiting scholar and in 2011 he spent 4 months at the Indian Institute of Science (IISc), Bangalore, India, as a visiting scholar. He is currently a Researcher at KTH coordinating the KTH Insurance Telematics Lab. He was a recipient of a Best Survey Paper Award by the IEEE Intelligent Transportation Systems Society in 2013.
\end{IEEEbiography}

\begin{IEEEbiography}[{\includegraphics[width=1in,height=1.25in,clip,keepaspectratio]{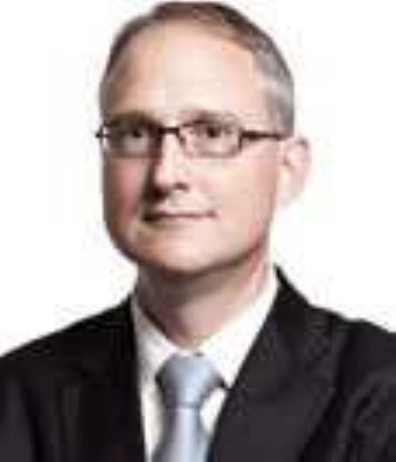}}]
{Peter H\"{a}ndel}(S'88-M'94-SM'98) received a Ph.D. degree from Uppsala University, Uppsala, Sweden, in 1993. From 1987 to 1993, he was with Uppsala University. From 1993 to 1997, he was with Ericsson AB, Kista, Sweden. From 1996 to 1997, he was a Visiting Scholar with the Tampere University of Technology, Tampere, Finland. Since 1997, he has been with the KTH Royal Institute of Technology, Stockholm, Sweden, where he is currently a Professor of Signal Processing and the Head of the Department of Signal Processing. From 2000 to 2006, he held an adjunct position at the Swedish Defence Research Agency. He has been a Guest Professor at the Indian Institute of Science (IISc), Bangalore, India, and at the University of G\"avle, Sweden. He is a co-founder of Movelo AB. Dr. H\"andel has served as an associate editor for the IEEE TRANSACTIONS ON SIGNAL PROCESSING. He was a recipient of a Best Survey Paper Award by the IEEE Intelligent Transportation Systems Society in 2013.
\end{IEEEbiography}









\end{document}